\documentclass[final,leqno,onefignum,onetabnum]{siamltex1213}

\usepackage{geometry}
\usepackage{graphicx}		
\usepackage{amssymb}
\usepackage{amsmath}
\usepackage{bm}
\newcommand{\ud}{\textrm{d}}
\newcommand{\dd}[2]{\frac{\ud #1}{\ud #2}}
\newcommand{\df}[2]{\frac{\partial #1}{\partial #2}}
\newcommand{\con}{\bm{q}}

\newcommand{\ee}{\textrm{e}}

\newcommand{\imh}{{i - \frac{1}{2}}}
\newcommand{\iph}{{i + \frac{1}{2}}}

\newcommand{\fl}{\bm{f}}
\newcommand{\gl}{\bm{g}}

\newcommand{\nfl}{\hat{\fl}}
\newcommand{\ofl}{\hat{f}}
\newcommand{\bT}{\bar{T}}
\newcommand{\numfl}{\hat{F}}

\newcommand{\bs}{\bm{s}}
\newcommand{\avg}[1]{\bar{#1}}

\newcommand{\half}{\frac{1}{2}}

\newcommand{\cpoly}{\mathbb{P}}
\newcommand{\tpoly}{\mathbb{Q}}

\newcommand{\mesh}{\mathcal{T}_h}
\newcommand{\kref}{\hat{K}}
\newcommand{\basis}{\varphi}
\newcommand{\ip}[1]{\left( #1 \right)}

\newcommand{\minmod}[1]{\mbox{minmod}\left( #1 \right)}

\newcommand{\pot}{\Phi}

\newtheorem{remark}{Remark}

\title{Well-balanced nodal discontinuous Galerkin method for Euler equations with gravity}
\author{Praveen Chandrashekar\footnotemark[2] \ and Markus Zenk\footnotemark[3]}

\begin{document}

\maketitle

\renewcommand{\thefootnote}{\fnsymbol{footnote}}
\footnotetext[2]{TIFR Center for Applicable Mathematics, Bangalore, India ({\tt praveen@tifrbng.res.in})}
\footnotetext[3]{Dept. of Mathematics, University of W\"urzburg, W\"urzburg, Germany ({\tt markus.zenk@gmx.de})}

\slugger{sisc}{xxxx}{xx}{x}{x--x}

\begin{abstract}
We present a well-balanced nodal discontinuous Galerkin (DG) scheme for compressible Euler equations with gravity. The DG scheme makes use of discontinuous Lagrange basis functions supported at Gauss-Lobatto-Legendre (GLL) nodes together with GLL quadrature using the same nodes. The well-balanced property is achieved by a specific form of source term discretization that depends on the nature of the hydrostatic solution, together with the GLL nodes for quadrature of the source term. The scheme is able to preserve isothermal and polytropic stationary solutions upto machine precision on any mesh composed of quadrilateral cells and for any gravitational potential. It is applied on several examples to demonstrate its well-balanced property and the improved resolution of small perturbations around the stationary solution.
\end{abstract}


\begin{AMS}\end{AMS}

\pagestyle{myheadings}
\thispagestyle{plain}
\markboth{Well-balanced DG scheme for Euler equations with gravity}{Chandrashekar \& Zenk}
%----------------------------------------------------------------------------
\section{Introduction}
The Euler equations in the presence of a gravitational field are an important mathematical model arising in atmospheric flows and astrophysical applications. Due to the presence of gravitational force, these equations have non-trivial stationary solutions, usually refered to as {\em hydrostatic} solutions. These stationary solutions are of interest in themselves and particularly their stability to small perturbations. Many atmospheric phenoma are small perturbations around the hydrostatic solution. The accurate computation of these small perturbations about the hydrostatic solution is hence important in many applications. In general, there are many other mathematical models involving source terms which exhibit non-trivial stationary solutions. An important class of such models with source terms are the shallow water equations arising in river and ocean modeling. Any numerical scheme which preserves the hydrostatic solution  on any mesh is said to be {\em well-balanced}. 

Stationary solutions are obtained by the precise balance of fluxes and source terms. Schemes which are not well-balanced will not be able to achieve this precise balance and may give rise to large numerical errors close to the stationary solutions, especially on coarse meshes~\cite{Xing:2006:HOW:1140818.1140826,doi:10.1137/140984373}. In order to obtain reliable solutions, such schemes would require very fine meshes, which may be impractical in realistic simulations in three dimensions. Well-balanced schemes on the other hand yield accurate solutions even on coarse meshes and are capable of resolving small perturbations around the stationary solution.

In the finite difference and finite volume approach, there are many well-balanced schemes available in the literature for the Euler equations with gravity, see e.g.~\cite{Leveque:2011:WPF:2004328.2004371, Xing:2013:HOW:2434597.2434626, Kappeli:2014:WSE:2567016.2567396, doi:10.1137/140984373}. However well-balanced DG schemes are not as well developed for the Euler equations. Xing and Shu~\cite{Xing:2006:HOW:1140818.1140826} have proposed a well-balanced DG scheme for the shallow-water equations where the hydrostatic solution is characterized by a quadratic invariant, see also~\cite{FLD:FLD1674,Xing2014536}. In the case of Euler equations with gravity, the invariants are not simple polynomials which makes it difficult to preserve them in a DG scheme where numerical quadrature has to be used. The hydrostatic solution is determined by a non-linear ODE whose solutions cannot be written down explicitly except in some simple settings like ideal gas model, isothermal or polytropic gas, etc.  To illustrate the difficulty, consider a scalar conservation law with source term $q_t + f(q)_x = s(q)$ which has a stationary solution $q_e$.  The DG scheme has the form
\[
\dd{}{t}\ip{q_h, \phi_h}_h + a_h(q_h, \phi_h) = (s_{h}(q_h), \phi_h)_h, \qquad \forall \phi_h \in V_h^k
\]
where $\ip{\cdot,\cdot}_h$ is some approximation of the $L^2$ inner product by quadratures, $a_h$ is a bilinear form approximated by quadrature and $s_h$ is an approximation of the source term. Let $q_{e,h}$ be an approximation to the stationary solution, e.g., obtained by interpolation or projection of $q_e$ on $V_h^k$. For the above scheme to be well-balanced, we require that
\[
a_h(q_{e,h}, \phi_h) = (s_{h}(q_{e,h}), \phi_h)_h, \qquad \forall \phi_h \in V_h^k
\]
Depending on the degree $k$, this contains many equations that need to be satisfied at the same time. The non-linearity of the PDE means that the quadratures may not be exact and we cannot use integration by parts to prove well-balanced property. 

Li and Xing~\cite{LiXing2015} have proposed a well-balanced DG scheme using orthogonal basis functions for Euler equations with gravity for isothermal hydrostatic solutions. The well-balanced property is achieved by a re-writing of the source terms together with an integration by parts, and using a Lax-Friedrich type numerical flux with a modified viscosity. Since orthogonal basis functions are used, the initial condition is projected onto the finite element space.

In this work, we propose a well-balanced DG scheme for isothermal and polytropic hydrostatic solutions under the ideal gas assumption. The scheme is based on nodal Lagrange basis functions using Gauss-Lobatto-Legendre points on arbitrary quadrilateral cells in 2-D. The same GLL points are also used for quadrature in the weak formulation of the DG scheme. The source term is re-written based on whether we are near isothermal or polytropic hydrostatic solution and then discretized using the GLL points. For continuous isothermal and polytropic hydrostatic solutions, the scheme is well-balanced for {\em any} consistent numerical flux function. The scheme is also well-balanced for isothermal hydrostatic solutions in which density might be discontinuous provided we use a numerical flux function which is exact for stationary contact discontinuities, like the Roe or HLLC flux, and the initial discontinuity in density coincides with the cell boundaries. For discontinuous solutions, a non-linear TVD limiter is necessary to avoid unphysical oscillations. The limiter might destroy the well-balanced property but this is easily solved by preventing the application of the limiter in case the solution residual in any cell is zero (close to machine precision), which would be the case for a hydrostatic solution.

The rest of the paper is organized as follows. In section~(\ref{sec:1d}) we introduce the 1-D Euler equations, explain the hydrostatic solutions, introduce the DG scheme and prove its well-balanced property. In section~(\ref{sec:2d}) we perform the same steps for the 2-D Euler equations and explain the limiter in section~(\ref{sec:limiter}). Numerical results are shown in section~(\ref{sec:res}) to demonstrate the well-balanced property and the accurate resolution of perturbations around hydrostatic solutions. Finally we end the paper with a summary and conclusions.
%----------------------------------------------------------------------------
\section{1-D Euler equations with gravity}
\label{sec:1d}
Consider the system of compressible Euler equations in one dimension which models conservation of mass, momentum and energy. These equations are given by
\begin{eqnarray*}
\df{\rho}{t} + \df{}{x}(\rho u) &=& 0 \\
\df{}{t}(\rho u) + \df{}{x}(p+\rho u^2) &=& -\rho \df{\pot}{x} \\
\df{E}{t} + \df{}{x}(E + p)u &=& -\rho u \df{\pot}{x}
\end{eqnarray*}
Here $\rho$ is the density, $u$ is the velocity, $p$ is the pressure, $E$ is the energy per unit volume excluding the gravitational energy and $\pot$ is the gravitational potential. The pressure is given by
\[
p = (\gamma-1) \left[ E - \frac{1}{2}\rho u^2 \right], \qquad \gamma = \frac{c_p}{c_v} > 1
\]
where $\gamma$ is the ratio of specific heats at constant pressure and volume, which is taken to be constant. We can write the above set of coupled equations in a compact notation as
\[
\df{\con}{t} + \df{\fl}{x} = -\begin{bmatrix}
0 \\ \rho \\ \rho u \end{bmatrix} \df{\pot}{x}
\]
where $\con$ is the set of conserved variables and $\fl$ is the corresponding flux vector. In the case of a self-gravitating system, the gravitational potential $\pot$ is governed by a Poisson-type equation. In the present work, we will consider the case of a static gravitational potential, which is assumed to be given as a function of the spatial coordinates.
%----------------------------------------------------------------------------
\subsection{Hydrostatic states}
Consider the hydrostatic stationary solution, i.e., the state with zero velocity, $u_e = 0$. In this case, the mass and energy fluxes are identically zero and their conservation equations are automatically satisfied. The momentum equation becomes an ordinary differential equation given by
\begin{equation}
\dd{p_e}{x} = -\rho_e \dd{\pot}{x}
\label{eq:hydro}
\end{equation}
We will assume the ideal gas equation of state
\begin{equation}
p = \rho R T
\label{eq:ieos}
\end{equation}
where $R$ is the gas constant.
%----------------------------------------------------------------------------
\subsubsection{Isothermal solution}
If the temperature is constant, $T= T_e = \const$, we can integrate the stationary momentum equation (\ref{eq:hydro}) together with (\ref{eq:ieos}) to obtain
\begin{equation}
p_e(x) \exp\left(\frac{\pot(x)}{RT_e}\right) = \const
\label{eq:isot}
\end{equation}
Since the potential is obtained as the solution of a Poisson equation, it is reasonable to assume that it is a continuous function of the spatial variable, which implies that the pressure in the hydrostatic state is a continuous function. In general, we can consider a hydrostatic solution where the density is discontinuous, e.g., two regions of different but constant temperature. Even in this case, the hydrostatic pressure would be continuous.
%----------------------------------------------------------------------------
\subsubsection{Polytropic solution}
For a gas in a polytropic equilibrium, the condition is characterized by
\begin{equation}
p \rho^{-\nu} = \alpha = \const
\label{eq:eospoly}
\end{equation}
for some constant $\nu > 1$. Using this in the hydrostatic equation~(\ref{eq:hydro}) we obtain
\[
\frac{\alpha \nu}{\nu-1} \rho_e^{\nu-1}(x) + \pot(x) = \beta = \const
\]
%----------------------------------------------------------------------------
\subsection{Mesh and basis functions}
Consider a partition of the domain into disjoint cells $C_i = (x_\imh,x_\iph)$ with the cell size being $\Delta x_i = x_\iph - x_\imh$. We will approximate the solution inside each cell by a polynomial of degree $N$ and we will refer to this as a $Q_N$ polynomial. To construct the basis functions inside each cell $C_i$ we map it to a reference cell, say $\hat{C}=[0,1]$, the mapping being given by
\begin{equation}
x = \xi \Delta x_i + x_\imh, \qquad \xi \in [0,1]
\label{eq:map1d}
\end{equation}
On this reference cell, let $\xi_j$, $0 \le j \le N$ be the Gauss-Lobatto-Legendre nodes. The nodal Lagrange basis functions using these GLL points are denoted by $\ell_j(\xi)$ with the interpolation property
\[
\ell_j(\xi_k) = \delta_{jk}, \qquad 0 \le j,k \le N
\]
The basis functions are then taken as
\[
\phi_j(x) = \ell_j(\xi), \qquad 0 \le j \le N
\]
where $x$ and $\xi$ are related by~(\ref{eq:map1d}). To compute the derivatives of the shape functions $\phi_j$ we apply the chain rule of differentiation
\[
\dd{}{x}\phi_j(x) = \ell_j'(\xi) \dd{\xi}{x} = \frac{1}{\Delta x_i} \ell_j'(\xi)
\]
Moreover, let $x_j \in C_i$ denote the physical locations of the GLL points, i.e., $x_j = \xi_j \Delta x_i + x_\imh$, $0 \le j \le N$.
%----------------------------------------------------------------------------
\subsection{Semi-discrete DG scheme in 1-D}
In order to explain the DG scheme, we will let $q$ denote one of the components of $\con$, and consider the single conservation law with source term
\[
\df{q}{t} + \df{f}{x} = s
\]
We will approximate the solution inside cell $C_i$ by the polynomial of degree $N$ which can be written as
\[
q_h(x,t) = \sum_{j=0}^N q_j(t) \phi_j(x)
\]
The coefficients $q_j$ in the above expansion are the degrees of freedom which determine the solution. In actual computations the above function is evaluated on the reference cell. Similarly we will approximate the flux inside the cell by a polynomial of degree $N$ given by
\[
f_h(x,t) = \sum_{j=0}^N f(q_h(x_j,t)) \phi_j(x) = \sum_{j=0}^N f_j(t) \phi_j(x)
\]
In the weak formulation of the DG scheme, we need to perform some quadrature to approximate the integrals. We will use Gauss-Lobatto-Legendre quadrature which uses the same nodes as used in the solution representation, i.e.,
\[
\int_{C_i} \phi(x) \psi(x) \ud x \approx \ip{\phi, \psi}_h = \ip{\phi, \psi}_{N,C_i} = \Delta x_i \sum_{q=0}^N \omega_q \phi(x_q) \psi(x_q)
\]
where the GLL weights $\omega_q$ correspond to the reference interval $[0,1]$. The semi-discrete DG scheme is given by
\begin{equation}
\begin{aligned}
\dd{}{t}\ip{q_h,\phi_j}_h + \ip{ \partial_x f_h, \phi_j}_h + & [\ofl_\iph - f_h(x_\iph^-)] \phi_j(x_\iph^-)  \\
- & [\ofl_\imh - f_h(x_\imh^+)] \phi_j(x_\imh^+) = \ip{s_h, \phi_j}_h, \qquad 0 \le j \le N
\end{aligned}
\label{eq:dg1d}
\end{equation}
where $\ofl_\iph = \ofl(q_\iph^-, q_\iph^+)$ is a numerical flux function. This equation will be solved using a Runge-Kutta scheme. The mass matrix in the above scheme is diagonal which can be inverted explicitly. The above type of DG scheme is also refered to as a quadrature-free scheme, see~\cite{atkinsshu1998}, and sometimes as a spectral element method. Note that we have performed an additional integration by parts to obtain the flux divergence term $\partial_x f_h$ leading to what is known as the strong form of the DG scheme~\cite{hesthavenbook}. For a detailed discussion on different forms of the nodal DG method, we refer the reader to~\cite{koprivagassner2010}. 
%----------------------------------------------------------------------------
\begin{remark}
In fact, the quadrature in the DG scheme is trivial and the scheme~(\ref{eq:dg1d}) can be written in simplified form as
\begin{equation*}
\begin{aligned}
\omega_j \Delta x_i \dd{q_j}{t} + \partial_x f_h(x_j,y_j) \omega_j \Delta x_i + & [\ofl_\iph - f_h(x_\iph^-)] \phi_j(x_\iph^-)  \\
- & [\ofl_\imh - f_h(x_\imh^+)] \phi_j(x_\imh^+) = s_h(x_j) \omega_j \Delta x_i, \qquad 0 \le j \le N
\end{aligned}
\label{eq:dg1db}
\end{equation*}
In the actual code, the above simplified form is used, however in the proofs, we will still write the scheme in the general form~(\ref{eq:dg1d}) since it has a compact notation.
\end{remark}
%----------------------------------------------------------------------------
\subsection{Numerical flux function}
The numerical flux function $\nfl(\con^-,\con^+)$ must be consistent in the sense that $\nfl(\con,\con) = \fl(\con)$. There are many different choices for these fluxes for the Euler equations based on exact or approximate Riemann solvers~\cite{godlewski-raviart-2,torobook}. Some flux functions also satisfy the property of exactly resolving a stationary contact discontinuity. For the two stationary states $\con^- = [\rho^-, \ 0, \ E]^\top$ and $\con^+ = [\rho^+, \ 0, \ E]^\top$ with $E=p/(\gamma-1)$, the numerical flux satisfies
\begin{equation}
\nfl(\con^-, \con^+) = [0, \ p, \ 0]^\top
\label{eq:cprop}
\end{equation}
We will refer to this as the {\em contact property}. Some examples of numerical fluxes having the contact property are the Roe scheme~\cite{Roe1981357} and the HLLC scheme~\cite{hllc}. This property is necessary to prove the well-balanced property of the scheme for hydrostatic solutions containing a discontinuity in the density. If the hydrostatic solution is continuous then any consistent numerical flux function is sufficient to obtain well-balanced property, as proved in the theorem below.
%----------------------------------------------------------------------------
\subsection{Approximation of source term}
If the potential is known to us as an explicit function of the spatial coordinate, we can compute the source term exactly by taking its derivative, but this does not lead to a well balanced scheme. In order to achieve the well-balanced property we will find it useful to approximate the spatial derivative of the potential in a form similar to the flux derivative in the above DG scheme. 
%----------------------------------------------------------------------------
\subsubsection{Isothermal case} 
Let $\bT_i$ be the temperature corresponding to the cell average value in cell $C_i$. We can write the source term in the momentum equation as~\cite{Xing:2013:HOW:2434597.2434626}
\[
s(x) = -\rho \df{\pot}{x} = \rho R \bT_i \exp\left(\frac{\pot}{R\bT_i}\right) \df{}{x} \exp\left(-\frac{\pot}{R\bT_i}\right)
\]
Using the above form, the source term is approximated as follows
\begin{equation}
s_h(x) = \rho_h(x) R \bT_i \exp\left(\frac{\pot(x)}{R\bT_i}\right) \df{}{x} \sum_{j=0}^N \exp\left(-\frac{\pot(x_j)}{R\bT_i}\right) \phi_j(x), \qquad x \in C_i
\label{eq:src1d}
\end{equation}
Note that we use the same approximation for the source term as for the fluxes which is key to the well-balanced property. The source term in the energy equation will be approximated as $(\rho u)_h s_h/\rho_h$.
%----------------------------------------------------------------------------
\subsubsection{Polytropic case}
Define the function $H(x)$ inside each cell $C_i$ as
\[
H(x) = \frac{\nu}{\nu-1} \ln\left( \frac{\nu-1}{\nu \alpha_i} (\beta_i - \pot(x)) \right), \qquad x \in C_i
\]
where $\alpha_i$, $\beta_i$ are constants to be chosen. The source term in the momentum equation can be written as
\[
s(x) = -\rho \df{\pot}{x} = \frac{\nu-1}{\nu} \rho (\beta_i - \pot(x)) \exp(-H(x)) \df{}{x} \exp(H(x)), \qquad x \in C_i
\]
Using the above form, the source term is approximated as
\begin{equation}
s_h(x) = \frac{\nu-1}{\nu} \rho_h(x) (\beta_i - \pot(x)) \exp(-H(x)) \df{}{x} \sum_{j=0}^N \exp(H(x_j)) \phi_j(x), \quad x \in C_i
\label{eq:src1dpoly}
\end{equation}
The parameter $\beta_i$ is chosen as
\[
\beta_i = \max_{0 \le j \le N} \left[ \frac{\nu}{\nu-1} \frac{p_j}{\rho_j} + \pot(x_j) \right], \qquad \alpha_i = p_{j^*} \rho_{j^*}^{-\nu}
\]
where $j^*$ is the index of the GLL point where the maximum is attained. With this choice the function $H(x)$ is well defined at all the GLL nodes.
%-------------------------------------------------------------------------------------
\begin{theorem}
Let the initial condition be obtained by interpolating the hydrostatic solution corresponding to a continuous gravitational potential $\pot$. Then the  DG scheme (\ref{eq:dg1d}) together with the source term approximation given by (\ref{eq:src1d}) or (\ref{eq:src1dpoly}) preserves the initial condition under any time integration scheme.
\end{theorem}

{\em Proof}: Since the hydrostatic solution is interpolated at the GLL points, the initial condition is exactly equal to the hydrostatic solution at the GLL points. We will first show that the boundary flux terms in~(\ref{eq:dg1d}) vanish in the hydrostatic case. Firstly, assume that the hydrostatic solution (isothermal or polytropic) is continuous. Since we interpolate the hydrostatic solution at the GLL points and we have GLL points located at the element boundaries, the finite element approximation of the initial condition is also continuous across the element boundaries. Hence by consistency of the numerical flux, we have
\[
\ofl_\iph - f_h(x_\iph^-) = 0, \qquad \ofl_\imh - f_h(x_\imh^+) = 0
\]
If the isothermal hydrostatic solution has discontinuous density, we will assume that the discontinuity exactly coincides with some cell boundary. If we use a numerical flux with the contact property then the above conditions are again satisfied. Thus the boundary terms vanish from the DG scheme.

The velocity being zero for the hydrostatic solution, the flux $f_h(x)=0$ in the density and energy equation, so that density and energy remain constant with time. It remains to check the momentum equation.  The flux $f_h$ has the form
\[
f_h(x,t) = \sum_{j=0}^N p_j(t) \phi_j(x)
\]
where $p_j$ is the pressure at the GLL point $x_j$. If the initial condition is isothermal, then $\bT_i = T_e = \const$. Now the source term evaluated at any GLL node $x_k$ is given by
\begin{eqnarray*}
s_h(x_k) &=& \rho_h(x_k) R T_e \exp\left(\frac{\pot(x_k)}{RT_e}\right) \sum_{j=0}^N \exp\left(-\frac{\pot(x_j)}{RT_e}\right) \df{}{x} \phi_j(x_k) \\
&=& p_k \exp\left(\frac{\pot(x_k)}{RT_e}\right) \sum_{j=0}^N \exp\left(-\frac{\pot(x_j)}{RT_e}\right) \df{}{x} \phi_j(x_k) \\
&=& \sum_{j=0}^N p_k \exp\left(\frac{\pot(x_k)}{RT_e}\right) \exp\left(-\frac{\pot(x_j)}{RT_e}\right) \df{}{x} \phi_j(x_k) \\
&=& \sum_{j=0}^N p_j \exp\left(\frac{\pot(x_j)}{RT_e}\right) \exp\left(-\frac{\pot(x_j)}{RT_e}\right) \df{}{x} \phi_j(x_k) \\
&=& \sum_{j=0}^N p_j \df{}{x} \phi_j(x_k) \\
&=& \df{}{x} f_h(x_k)
\end{eqnarray*}
Since $\partial_x f_h = s_h$ at all the GLL nodes, we can conclude that $\ip{ \partial_x f_h, \phi_j}_h = \ip{s_h, \phi_j}_h$ and hence the scheme is well-balanced for the momentum equation also.

If the initial condition is polytropic with exponent $\nu$ then $\alpha_i = \alpha$ and $\beta_i = \beta$ are constants. At any GLL node $x_k$, we have $H(x_k) = \ln \rho_k^\nu$. The source term at any GLL node $x_k$ can be written as
\begin{eqnarray*}
s_h(x_k) &=& \frac{\nu-1}{\nu} \rho_h(x_k) (\beta - \pot(x_k)) \exp(-H(x_k)) \sum_{j=0}^N \exp(H(x_j)) \df{}{x} \phi_j(x_k) \\
&=& \frac{\nu-1}{\nu} \rho_k (\beta - \pot(x_k)) \rho_k^{-\nu} \sum_{j=0}^N \rho_j^\nu \df{}{x} \phi_j(x_k) \\
&=& \alpha \sum_{j=0}^N \rho_j^\nu \df{}{x} \phi_j(x_k) = \sum_{j=0}^N p_j \df{}{x} \phi_j(x_k) \\
&=& \partial_x f_h(x_k)
\end{eqnarray*}
Hence in this case also, the DG scheme is well-balanced.
%----------------------------------------------------------------------------
\paragraph{Remark} The use of GLL nodes was important in the above proof. The boundary flux terms vanish since the GLL nodes ensure that the interpolation of the hydrostatic solution on the mesh is continuous across the elements. The entire scheme makes use of only the solution at the GLL nodes which is exact in the hydrostatic case and helps us to satisfy the well-balanced property.
%----------------------------------------------------------------------------
\section{2-D Euler equations with gravity}
\label{sec:2d}
The Euler equations in 2-D are given by the following set of four coupled conservation laws for mass, momentum and energy
\[
\df{\con}{t} + \df{\fl}{x} + \df{\gl}{y} = \bs
\]
where $\con$ is the vector of conserved variables, $(\fl,\gl)$ is the flux vector and $\bs$ is the source term, given by
\[
\con = \begin{bmatrix}
\rho \\
\rho u \\
\rho v \\
E \end{bmatrix}, \qquad \fl = \begin{bmatrix}
\rho u \\
p + \rho u^2 \\
\rho u v \\
(E+p)u \end{bmatrix}, \qquad \gl = \begin{bmatrix}
\rho v \\
\rho u v \\
p + \rho v^2 \\
(E+p)v \end{bmatrix}, \qquad \bs = \begin{bmatrix}
0 \\
-\rho \df{\pot}{x} \\
-\rho \df{\pot}{y} \\
-\left( \rho u \df{\pot}{x} + \rho v \df{\pot}{y} \right)
\end{bmatrix}
\]
%----------------------------------------------------------------------------
\subsection{Hydrostatic solution}
In the hydrostatic state we have $u_e = v_e = 0$ so that the mass and energy equations are identically satisfied. The momentum equation can be written as
\[
\nabla p_e = - \rho_e \nabla\pot
\]
Assuming the ideal gas equation of state~(\ref{eq:ieos}) and a constant temperature $T = T_e = \const$, we get $\ud p_e = - \rho_e \ud\pot = -\frac{p_e}{RT_e} \ud\pot$. 
Integrating this equations gives the condition
\begin{equation}
p_e(x,y) \exp\left( \frac{\pot(x,y)}{RT_e} \right) = \const
\label{eq:hyrdo2d}
\end{equation}
In the case of a polytropic solution satisfying~(\ref{eq:eospoly}) we obtain
\[
\frac{\alpha \nu}{\nu-1} \rho_e^{\nu-1}(x,y) + \pot(x,y) = \beta = \const
\]
We will exploit the above properties of the hydrostatic state to construct the well-balanced schemes.
%----------------------------------------------------------------------------
\subsection{Mesh and basis functions}
\begin{figure}
\begin{center}
\includegraphics[width=0.8\textwidth]{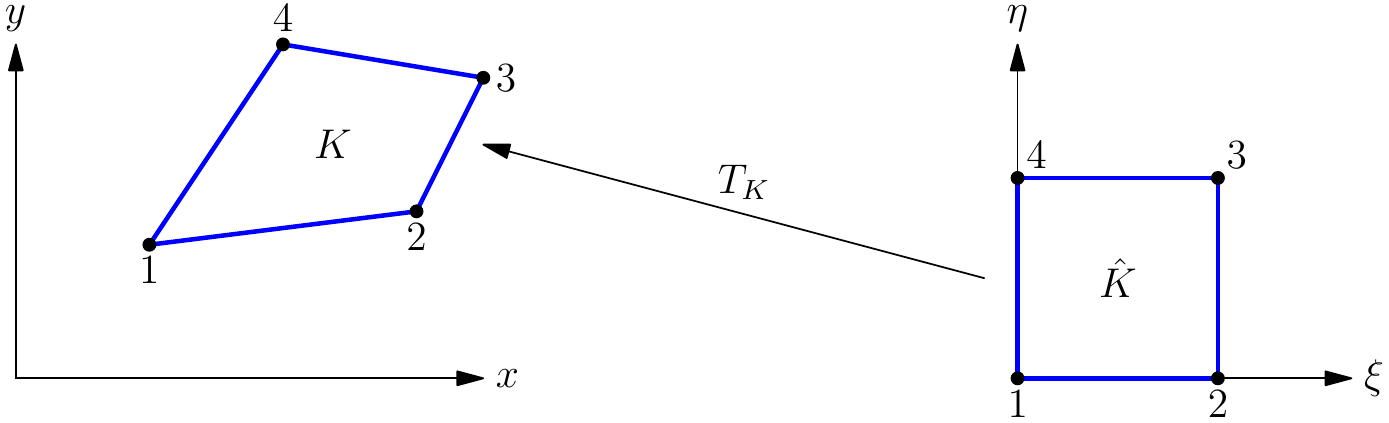}
\caption{A quadrilateral cell $K$ and reference cell $\kref$}
\label{fig:kref}
\end{center}
\end{figure}
\begin{figure}
\begin{center}
\includegraphics[width=0.8\textwidth]{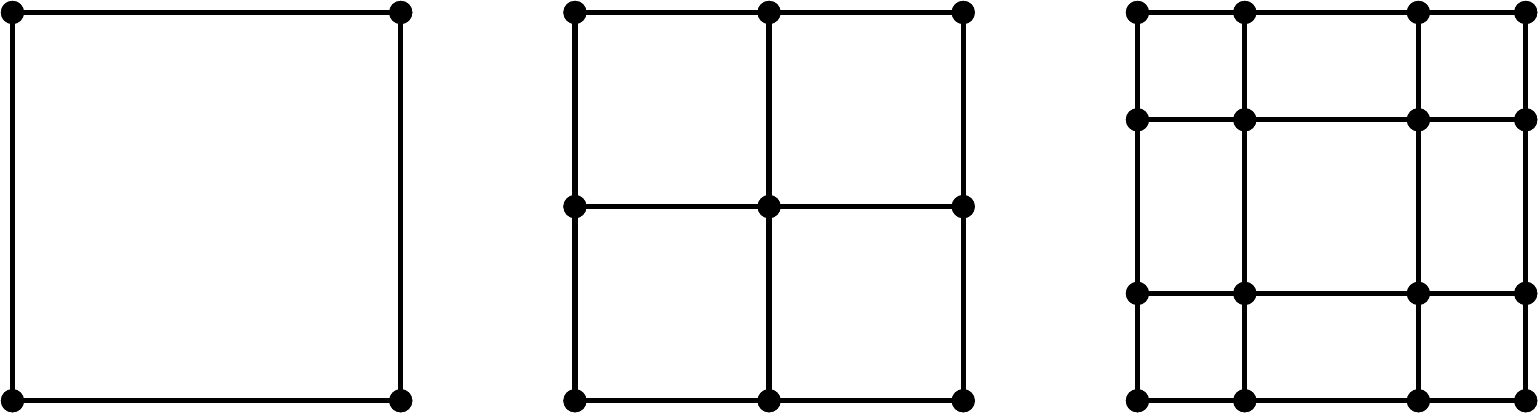}
\caption{GLL points for degree $N=1,2,3$}
\label{fig:gll}
\end{center}
\end{figure}

Consider a partition of the domain into a mesh $\mesh$ of disjoint quadrilateral cells $K$ and let $\kref = [0,1] \times [0,1]$ be the reference cell. We assume that the mesh is conforming, i.e., there are no hanging nodes. Let $T_K : \kref \to K$ be the mapping from the reference cell $\kref$ to the physical cell $K$, see figure~\ref{fig:kref}. If $\xi_r \in [0,1]$, $0 \le r \le N$ are the GLL points, then consider their tensor product $(\xi_r, \xi_s)$, $0 \le r,s \le N$. This is illustrated in figure~(\ref{fig:gll}) for $N=1,2,3$. Assume that there is a one dimensional indexing of these points as $i=1,2,\ldots,M$ where $M=(N+1)^2$. The basis functions in cell $K$ for the space $Q_N$ of tensor product polynomials of degree $N$ are of the form
\[
\basis_i^K(x,y) = \ell_r(\xi) \ell_s(\eta), \qquad (x,y) = T_K(\xi,\eta), \qquad i=1,2,\ldots,M
\]
The computation of flux derivatives like $\partial_x f_h$ requires derivatives of the basis functions which are evaluated by applying the chain rule, e.g.,
\[
\df{}{x}\phi_i^K(x,y) = \ell_r'(\xi) \ell_s(\eta) \df{\xi}{x} + \ell_r(\xi) \ell_s'(\eta) \df{\eta}{x}
\]
The derivatives $\df{\xi}{x}$ etc. are computed from the mapping $T_K$ between the reference cell $\kref$ and the actual cell $K$. This map is affine only if $K$ is a parallelogram and is non-affine for more general cell shapes.
%----------------------------------------------------------------------------
\subsection{Semi-discrete DG scheme}
To explain the DG scheme in two dimensions, we will consider a single conservation law with source term of the form
\[
\df{q}{t} + \df{f}{x} + \df{g}{x} = s
\]
The solution inside cell $K$ is represented as
\[
q_h(x,y,t) = \sum_{i=1}^M q_i^K(t) \basis_i^K(x,y)
\]
In actual computations, the above expression would be evaluated on the reference element.  Similarly, we approximate the fluxes inside the cell by the following interpolation at the same GLL nodes
\[
f_h(x,y) = \sum_{i=1}^M f(q_h(x_i, y_i)) \phi_i^K(x,y) = \sum_{i=1}^M f_i \phi_i^K(x,y)
\]
and
\[
g_h(x,y) = \sum_{i=1}^M g(q_h(x_i, y_i)) \phi_i^K(x,y) = \sum_{i=1}^M g_i \phi_i^K(x,y)
\]
Define the quadrature on element $K$ using GLL points
\[
\ip{\phi, \psi}_K = \sum_{r=0}^N \sum_{s=0}^N \phi(\xi_r, \xi_s) \psi(\xi_r, \xi_s) \omega_r \omega_s |J_K(\xi_r,\xi_s)|
\]
where $\omega_r$ are the GLL quadrature weights and $J_K$ is the Jacobian of the map $T_K$. We also need the quadrature on the faces of the cell $K$. This is approximated as
\[
\ip{\phi, \psi}_{\partial K} = \sum_{e \in \partial K} \ip{\phi, \psi}_e
\]
where $(\cdot,\cdot)_e$ are one dimensional quadrature rules using the subset of GLL nodes located on the boundary of the cell. The semi-discrete DG scheme is given by
\begin{equation}
\begin{aligned}
\dd{}{t}\ip{q_h, \phi_i^K}_K + \ip{\partial_x f_h, \phi_i^K}_K + & \ip{\partial_y g_h, \phi_i^K}_K \\
+ &  \ip{\numfl_h - F_h^-, \phi_i^K}_{\partial K} = \ip{s_h, \phi_i^K}, \qquad 1 \le i \le M
\end{aligned}
\label{eq:dg2d}
\end{equation}
where $\numfl_h = \numfl(q^-_h, q^+_h, n)$ is a numerical flux function in the direction of the unit outward normal vector $n=(n_x,n_y)$ to the boundary of the cell $\partial K$, and $F_h^-$ is the flux in the direction of $n$ given by $F_h^- = f_h^- n_x + g_h^- n_y$ which is evaluated using the solution inside the element $K$. As discussed in the 1-D case, the above scheme can also be written in a simplified form.%-----------------------------------------------------------------------------------
\subsection{Approximation of source term}
We rewrite the source term as in the 1-D problem starting with the case of an isothermal solution. Let $\bar{T}_K$ be the temperature corresponding to the cell average value in cell $K$. The source term in the $x$-momentum equation can be re-written as
\[
s = -\rho \df{\pot}{x} = \rho R \bT_K \exp\left(\frac{\pot}{R\bT_K}\right) \df{}{x} \exp\left(-\frac{\pot}{R\bT_K}\right)
\]
To be consistent with the discretization of the flux derivative term in the hydrostatic situation we discretize the above form of the source term by
\begin{equation}
s_h(x,y) = \rho_h(x,y) R \bT_K \exp\left(\frac{\pot(x,y)}{R\bT_K}\right) \df{}{x} \sum_{j=1}^M \exp\left(-\frac{\pot(x_j,y_j)}{R\bT_K}\right) \phi_j^K(x,y)
\label{eq:src2d}
\end{equation}
A similar expression is used for the source term in the $y$-momentum equation. 

In the case of polytropic solution, define the function $H(x,y)$ inside each cell $K$ as
\[
H(x,y) = \frac{\nu}{\nu-1} \ln\left( \frac{\nu-1}{\nu \alpha_K} (\beta_K - \pot(x,y)) \right), \qquad (x,y) \in K
\]
where $\alpha_K$, $\beta_K$ are constants to be chosen. The source term in the $x$-momentum equation can be written as
\[
s(x,y) = -\rho \df{\pot}{x} = \frac{\nu-1}{\nu} \rho (\beta_K - \pot(x,y)) \exp(-H(x,y)) \df{}{x} \exp(H(x,y)), \qquad (x,y) \in K
\]
Using the above form, the source term for $(x,y) \in K$ is approximated as
\begin{equation}
s_h(x,y) = \frac{\nu-1}{\nu} \rho_h(x,y) (\beta_K - \pot(x,y)) \exp(-H(x,y)) \df{}{x} \sum_{j=1}^M \exp(H(x_j,y_j)) \phi_j(x,y)
\label{eq:src2dpoly}
\end{equation}
The parameter $\beta_K$ is chosen as
\[
\beta_K = \max_{1 \le j \le M} \left[ \frac{\nu}{\nu-1} \frac{p_j}{\rho_j} + \pot(x_j,y_j) \right], \qquad \alpha_K = p_{j^*} \rho_{j^*}^{-\nu}
\]
where $j^*$ is the index of the GLL point where the maximum is attained. With this choice the function $H(x,y)$ is well defined at all the GLL nodes in the cell $K$.

If we denote the source terms in the momentum equation as $(s_h^x, s_h^y)$, then the source term in the energy equation is given by $ \frac{1}{\rho_h}[ (\rho u)_h s_h^x + (\rho v)_h s_h^y]$. This completes the specification of the DG scheme.
%-------------------------------------------------------------------------------------
\begin{theorem}
Let the initial condition be obtained by interpolating the  hydrostatic solution corresponding to a continuous gravitational potential $\pot$. Then the DG scheme (\ref{eq:dg2d}) together with the source term approximation given by (\ref{eq:src2d})  or (\ref{eq:src2dpoly}) preserves the initial condition under any time integration scheme.
\end{theorem}\\
\underline{Proof}: The interpolation of the hydrostatic solution ensures that the solution $q_h$ and the fluxes $f_h$, $g_h$ are equal to the hydrostatic solution at all the GLL points. Firstly, assuming that the hydrostatic solution is continuous, the finite element approximation $q_h$ of the initial condition is continuous on the element boundaries. Due to consistency of the numerical flux, we get
\[
\numfl_h = F_h \quad \textrm{at all GLL points on } \partial K
\]
If the hydrostatic density is discontinuous, then assuming that the discontinuity surface coincides with the cell boundaries, the above condition is satisfied if the numerical flux has the contact property. Thus in both cases, the boundary terms are zero.

In the hydrostatic solution, the velocity is zero and hence the mass and energy equations are well-balanced since $f_h = g_h = 0$ in these equations. Now consider the $x$-momentum equation. The fluxes at the GLL points have the form
\[
f_j = p_j, \qquad g_j = 0, \qquad 1 \le j \le M
\]
and hence
\[
\partial_x f_h(x,y) = \sum_{j=1}^M p_j \partial_x \phi_j^K(x,y), \qquad \partial_x g_h(x,y) = 0
\]
For the initial condition which is isothermal and hydrostatic, the source term at any GLL point can be written as
\begin{eqnarray*}
s_h(x_i,y_i) &=& \rho_h(x_i,y_i) R T_e \exp\left(\frac{\pot(x_i,y_i)}{RT_e}\right) \df{}{x} \sum_{j=1}^M \exp\left(-\frac{\pot(x_j,y_j)}{RT_e}\right) \phi_j^K(x_i,y_i) \\
&=& p_i \exp\left(\frac{\pot(x_i,y_i)}{RT_e}\right)  \sum_{j=1}^M \exp\left(-\frac{\pot(x_j,y_j)}{RT_e}\right) \df{}{x} \phi_j^K(x_i,y_i) \\
&=& \sum_{j=1}^M  p_i \exp\left(\frac{\pot(x_i,y_i)}{RT_e}\right) \exp\left(-\frac{\pot(x_j,y_j)}{RT_e}\right) \df{}{x} \phi_j^K(x_i,y_i)  \\
&=& \sum_{j=1}^M  p_j \exp\left(\frac{\pot(x_j,y_j)}{RT_e}\right) \exp\left(-\frac{\pot(x_j,y_j)}{RT_e}\right) \df{}{x} \phi_j^K(x_i,y_i) \\
&=& \sum_{j=1}^M  p_j \df{}{x} \phi_j^K(x_i,y_i) \\
&=& \partial_x f_h(x_i,y_i)
\end{eqnarray*}
where we have made use of~(\ref{eq:hyrdo2d}). Hence at all the GLL points we have $\partial_x f_h = s_h$ which is enough to conclude that
\[
\ip{\partial_x f_h, \phi_i^K}_K = \ip{s_h, \phi_i^K}_K
\]
This proves that the $x$-momentum equation is well-balanced. A similar proof shows that the $y$-momentum equation is also well-balanced. The case of polytropic hydrostatic solution can also be proved along similar lines. Thus the DG scheme is well-balanced for isothermal and polytropic hydrostatic solutions.
%-------------------------------------------------------------------------------------
\section{Limiter}
\label{sec:limiter}
The computation of discontinuous solutions by a high order accurate scheme leads to unphysical oscillations which can spoil the accuracy of the scheme and may even lead to a break-down of the computations. In order to control the oscillations, we can use a non-linear TVD limiter~\cite{Cockburn:1998:RDG:287244.287254} which is applied as a post processing step. The limiter reduces the degree of the polynomial solution if it detects the presence of oscillations in the linear part of the solution. The standard TVD limiter is given for the case of complete polynomial solutions belonging to $\cpoly_k$~\cite{Cockburn:1998:RDG:287244.287254} but in our work we use tensor product polynomials $\tpoly_k$. For such solutions, we construct a limiter for Cartesian meshes as follows. In each cell the average gradient is computed
\[
\overline{\nabla q}_h = \frac{1}{|K|} \int_K \nabla q_h \ud x
\]
This gradient is compared with forward and backward differences of the cell average quantities
\[
\partial_x q_h^{(m)} = \minmod{ \overline{\partial_x q}_h, \beta \frac{\Delta^-_x \avg{q}_h}{\Delta x}, \beta \frac{\Delta^+_x \avg{q}_h}{\Delta x} }, \qquad \partial_y q_h^{(m)} = \minmod{ \overline{\partial_y q}_h, \beta \frac{\Delta^-_y \avg{q}_h}{\Delta y}, \beta \frac{\Delta^+_y \avg{q}_h}{\Delta y} }
\]
where $\Delta_x^\pm$, $\Delta_y^\pm$ are backward and forward difference operators in the $x$ and $y$ directions, while $\beta \in [1,2]$. In all the computations, we use $\beta=2$ which corresponds to MC limiter of Van Leer and leads to more accurate solutions. On unstructured meshes, we cannot define backward and forward differences. Hence we adopt the minmax limiter as developed for finite volume methods~\cite{BarJ89}. The average gradient is computed in each cell as described above. This is used to define an affine function in the cell, which is evaluated at the centers of each face. We then check that these values are bounded between the minimum and maximum of the neighbouring cell average values. If any of these face values falls outside the range, then the gradient is scaled to ensure that values do not exceed the local minimum and maximum values.

As described above, the limiter is applied component-wise to the conserved variables. However it is beneficial to apply the limiter to characteristic variables~\cite{Cockburn:1998:RDG:287244.287254}. The average gradient is transformed by multiplying with the matrix of left eigenvectors. The limiter is applied as above and transformed back to the original variables by multiplying with the matrix of right eigenvectors. 

The application of limiter can change the solution in a cell even if the solution is hydrostatic and smooth, especially at extrema. In order to preserve the well-balanced property of the scheme, we apply the limiter in a cell only if the $L^2$ norm of the cell residual is greater than a small number and in the computations we use a tolerance of $10^{-12}$. In all the numerical tests, we find that this is sufficient to prevent the application of limiter when the solution is in hydrostatic state.

%-------------------------------------------------------------------------------------
\section{Numerical results}
\label{sec:res}
The DG code is written in C++ using the {\tt deal.II} finite element library~\cite{BangerthHartmannKanschat2007} and all computations are performed in double precision. All the grids used in the computations are generated using the open source grid generator {\tt Gmsh}~\cite{NME:NME2579}. The time integration is performed using a strong stability preserving Runge-Kutta scheme, see~\cite{Shu1988439}. For the ODE $\dd{q}{t} = R(t,q)$, the second order scheme is given by
\begin{eqnarray*}
q^{(1)} &=& q^n + \Delta t \cdot R(t_n, q^n) \\
q^{(2)} &=& \frac{1}{2} q^n + \frac{1}{2}[q^{(1)} + \Delta t \cdot R(t_n+\Delta t,q^{(1)})] \\
q^{n+1} &=& q^{(2)}
\end{eqnarray*}
and the third order scheme is
\begin{eqnarray*}
q^{(1)} &=& q^n + \Delta t R(t_n, q^n) \\
q^{(2)} &=& \frac{3}{4} q^n + \frac{1}{4}[q^{(1)} + \Delta t \cdot R(t_n+\Delta t,q^{(1)})] \\
q^{(3)} &=& \frac{1}{3} q^n + \frac{2}{3}[q^{(2)} + \Delta t \cdot R(t_n+\Delta t/2,q^{(2)})] \\
q^{n+1} &=& q^{(3)}
\end{eqnarray*}
In all the computations we take the gas constant $R=1$ and $\gamma =1.4$.
%-------------------------------------------------------------------------------------
\subsection{1-D hydrostatic solution}

\subsubsection{Isothermal case}
We first perform numerical test of well-balanced property for one dimensional solutions. The potential is taken to be either $\Phi = x$ or $\Phi = \sin(2\pi x)$. The corresponding states for density and pressure are given by $\rho = p = \exp(-\pot(x))$. Starting with the  hydrostatic solution as the initial condition, the solution is updated upto a time of $t=0.1$ and the $L^2$ norm of the difference in final solution and the initial condition is computed on different mesh sizes. The computations are performed on the unit square with a Cartesian mesh of different cell sizes. This is shown in table~(\ref{tab:1dlin$Q_1$}) and (\ref{tab:1dlin$Q_2$}) for the second and third order schemes for the potential $\Phi = x$, and in tables (\ref{tab:1dsin$Q_1$}) and (\ref{tab:1dsin$Q_2$}) for the potential $\Phi = \sin(2\pi x)$. In all cases we see that the difference is small and of the order of machine precision.
\begin{table}
\begin{center}
\begin{tabular}{|c|c|c|c|c|}
\hline
Mesh & $\rho u$ & $\rho v$ & $\rho$ & $E$ \\
\hline
25x25  & 1.03822e-13  & 6.68114e-15  & 2.72604e-14 & 9.53913e-14  \\
50x50  & 1.04783e-13  & 5.92391e-15  & 2.67559e-14 & 9.36725e-14  \\
100x100 & 1.05019e-13 &  5.6383e-15  & 2.66323e-14 & 9.34503e-14  \\
200x200 & 1.05088e-13 & 5.54862e-15  & 2.66601e-14 & 9.33861e-14  \\
\hline
\end{tabular}
\caption{Well-balanced test for isothermal case on Cartesian mesh using $Q_1$ polynomials and potential $\pot = x$}
\label{tab:1dlin$Q_1$}
\end{center}
\end{table}

\begin{table}
\begin{center}
\begin{tabular}{|c|c|c|c|c|}
\hline
Mesh & $\rho u$ & $\rho v$ & $\rho$ & $E$ \\
\hline
25x25  & 1.04518e-13 & 7.29936e-15 & 2.7548e-14 & 9.64205e-14  \\
50x50  & 1.04983e-13 & 6.03994e-15 & 2.69317e-14 & 9.43158e-14  \\
100x100 & 1.05069e-13 & 5.68612e-15 & 2.69998e-14 & 9.39126e-14  \\
200x200 & 1.05089e-13 & 5.69125e-15 & 2.68828e-14 & 9.462e-14 \\
\hline
\end{tabular}
\caption{Well-balanced test for isothermal case on Cartesian mesh using $Q_2$ polynomials and potential $\pot =x$}
\label{tab:1dlin$Q_2$}
\end{center}
\end{table}

\begin{table}
\begin{center}
\begin{tabular}{|c|c|c|c|c|}
\hline
Mesh & $\rho u$ & $\rho v$ & $\rho$ & $E$ \\
\hline
25x25  & 9.23424e-13 & 1.16432e-13 & 2.31405e-13 & 8.16645e-13  \\
50x50  & 9.36459e-13 & 1.04921e-13 & 2.28315e-13 & 8.04602e-13  \\
100x100 & 9.39613e-13 & 1.00384e-13 & 2.28001e-13 & 8.03005e-13  \\
200x200 & 9.40422e-13 & 9.89098e-14 & 2.2792e-13 & 8.02653e-13 \\
\hline
\end{tabular}
\caption{Well-balanced test for isothermal case on Cartesian mesh using $Q_1$ polynomials and $\Phi = \sin(2\pi x)$}
\label{tab:1dsin$Q_1$}
\end{center}
\end{table}

\begin{table}
\begin{center}
\begin{tabular}{|c|c|c|c|c|}
\hline
Mesh & $\rho u$ & $\rho v$ & $\rho$ & $E$ \\
\hline
25x25 & 9.34536e-13 & 1.32134e-13 & 2.35173e-13 & 8.30316e-13  \\
50x50 & 9.39556e-13 & 1.08172e-13 & 2.29908e-13 & 8.10055e-13  \\
100x100 & 9.40442e-13 & 1.00923e-13 & 2.28538e-13 & 8.04638e-13  \\
200x200 & 9.40668e-13 & 9.90613e-14 & 2.28051e-13 & 8.0357e-13 \\
\hline
\end{tabular}
\caption{Well-balanced test for isothermal case on Cartesian mesh using $Q_2$  polynomials and $\Phi = \sin(2\pi x)$}
\label{tab:1dsin$Q_2$}
\end{center}
\end{table}

\begin{figure}
\begin{center}
\begin{tabular}{cc}
\includegraphics[width=0.48\textwidth]{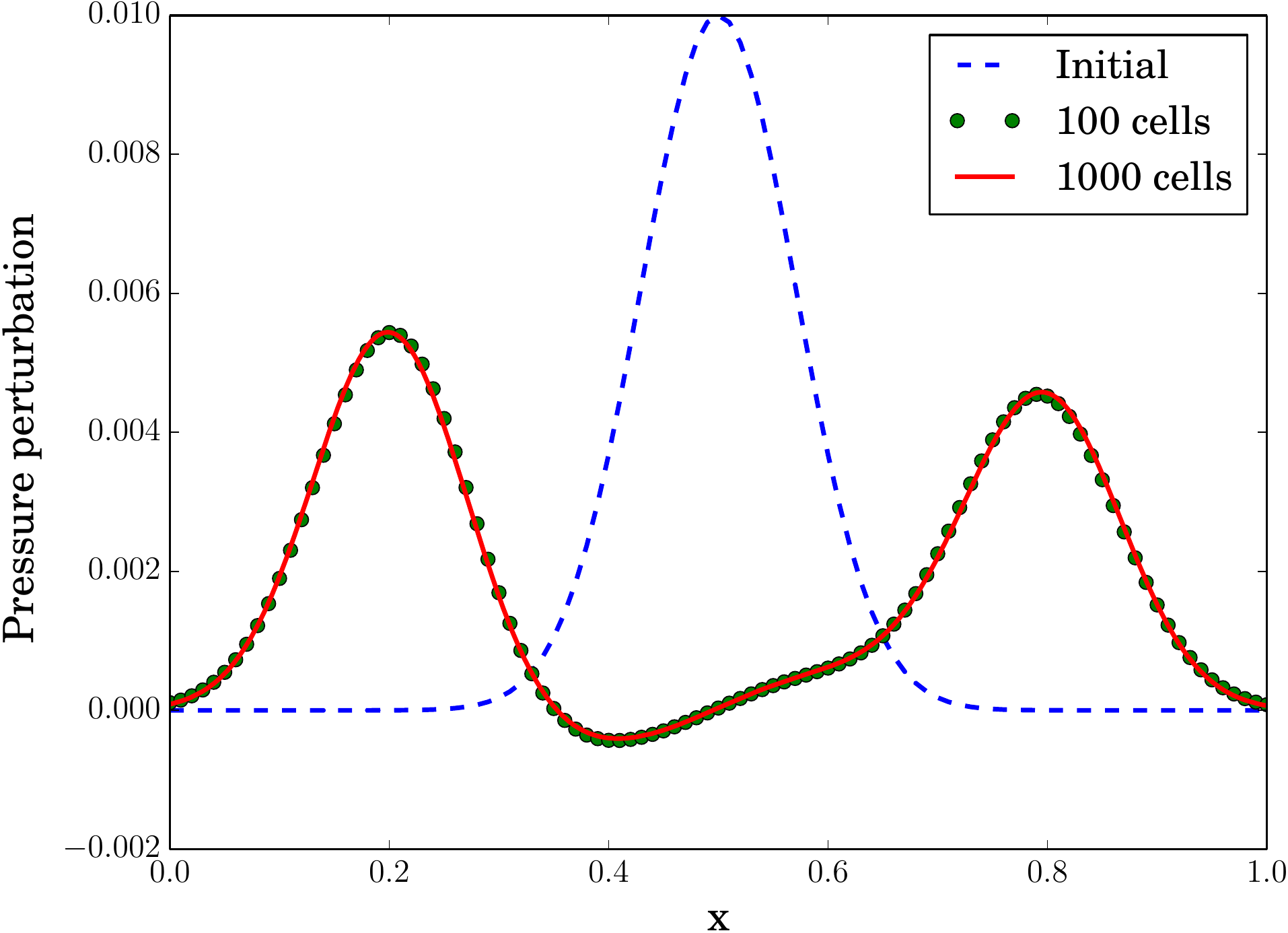} &
\includegraphics[width=0.48\textwidth]{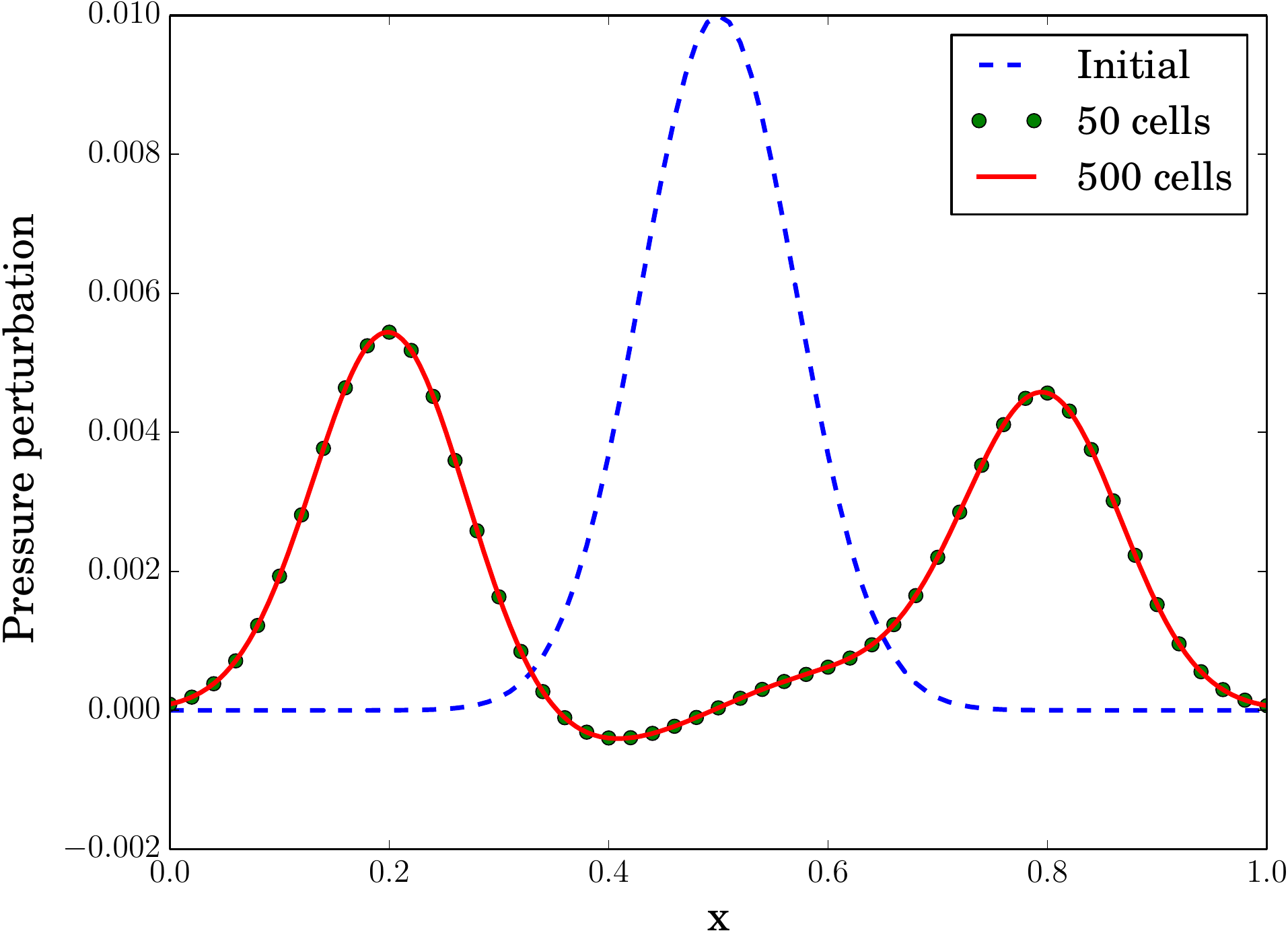} \\
(a) & (b)
\end{tabular}
\caption{Evolution of small perturbations in the isothermal case for $\eta = 10^{-2}$: (a) $Q_1$ (b) $Q_2$}
\label{fig:pert1}
\end{center}
\end{figure}

\begin{figure}
\begin{center}
\begin{tabular}{cc}
\includegraphics[width=0.48\textwidth]{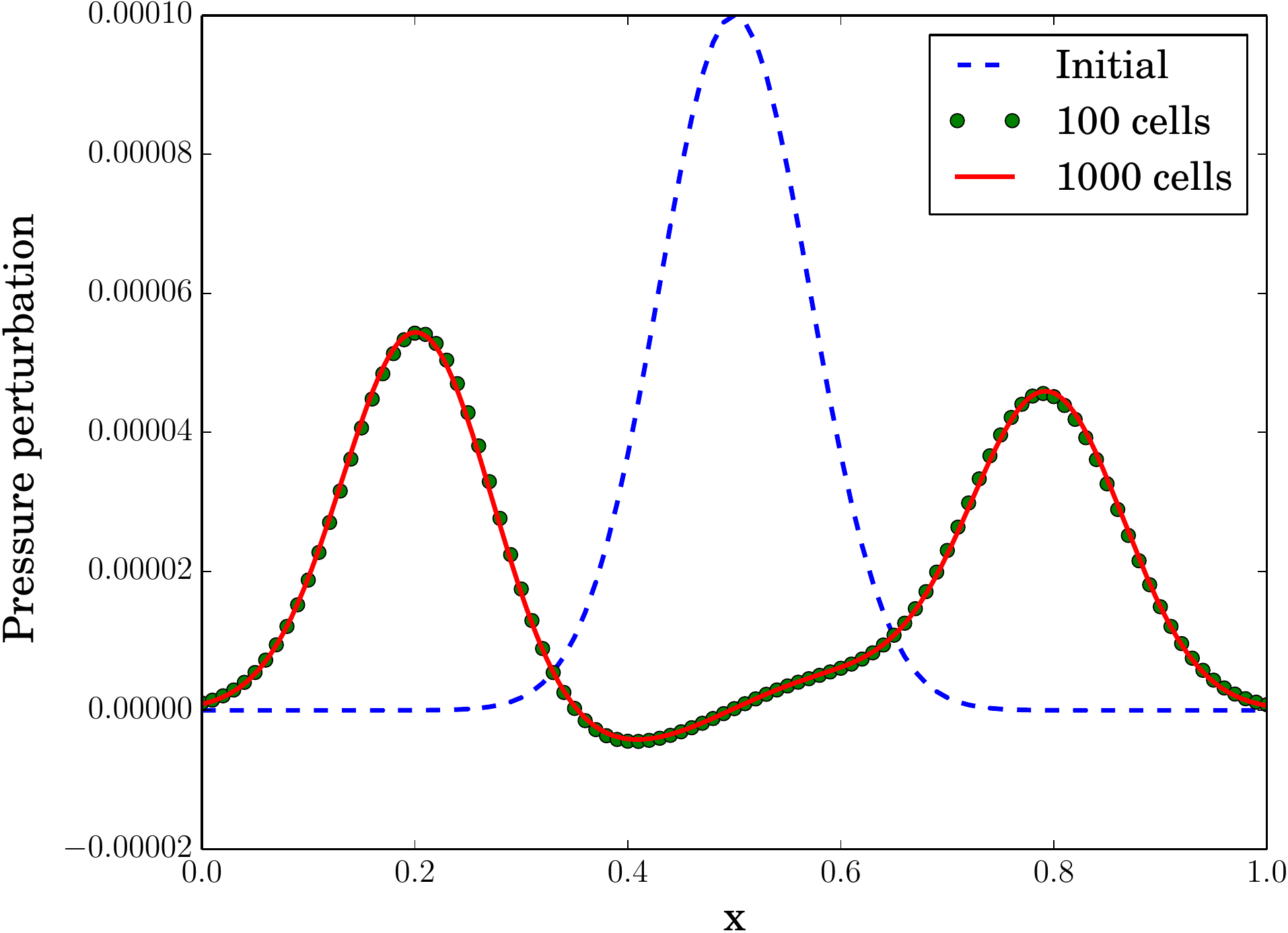} &
\includegraphics[width=0.48\textwidth]{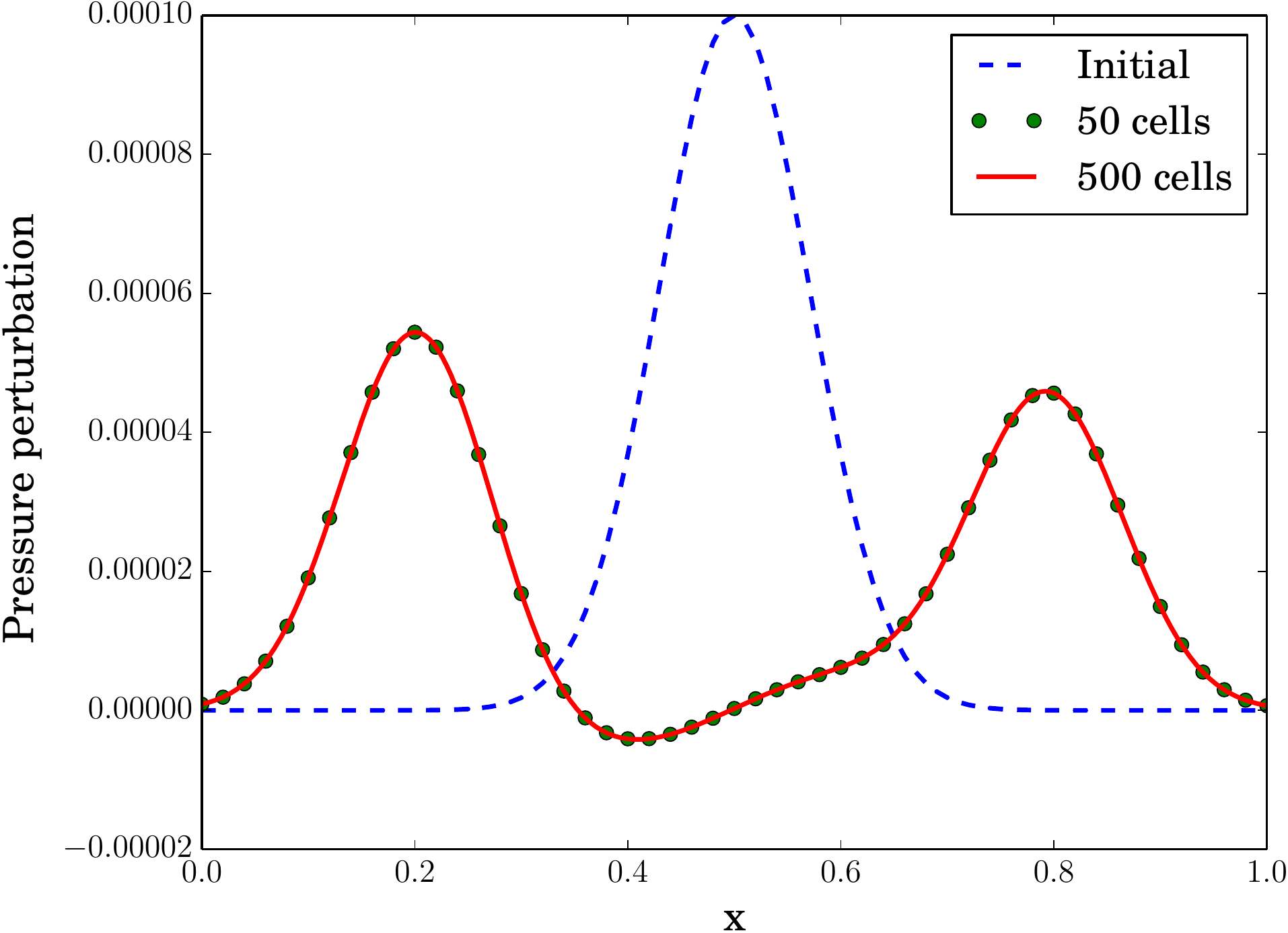} \\
(a) & (b)
\end{tabular}
\caption{Evolution of small perturbations in the isothermal case for $\eta = 10^{-4}$: (a) $Q_1$ (b) $Q_2$}
\label{fig:pert2}
\end{center}
\end{figure}

We next add a perturbation to the initial hydrostatic solution. The perturbation is added to the pressure as follows
\[
p = \exp(-x) + \eta \exp(-100(x-1/2)^2)
\]
where the perturbation amplitude $\eta=10^{-2}$ or $\eta=10^{-4}$. The initial perturbation breaks into two waves which propagate in the negative and positive directions. The solution at time $t=0.25$ is shown in figure~(\ref{fig:pert1}) and (\ref{fig:pert2}) for the two amplitudes and different mesh sizes and polynomial degree. We observe that in all cases, the well-balanced scheme is able to predict the perturbation solution without any spurious oscillations. Even the coarse grid solutions show good accuracy as compared with the fine grid solutions.

\begin{figure}
\begin{center}
\begin{tabular}{cc}
\includegraphics[width=0.48\textwidth]{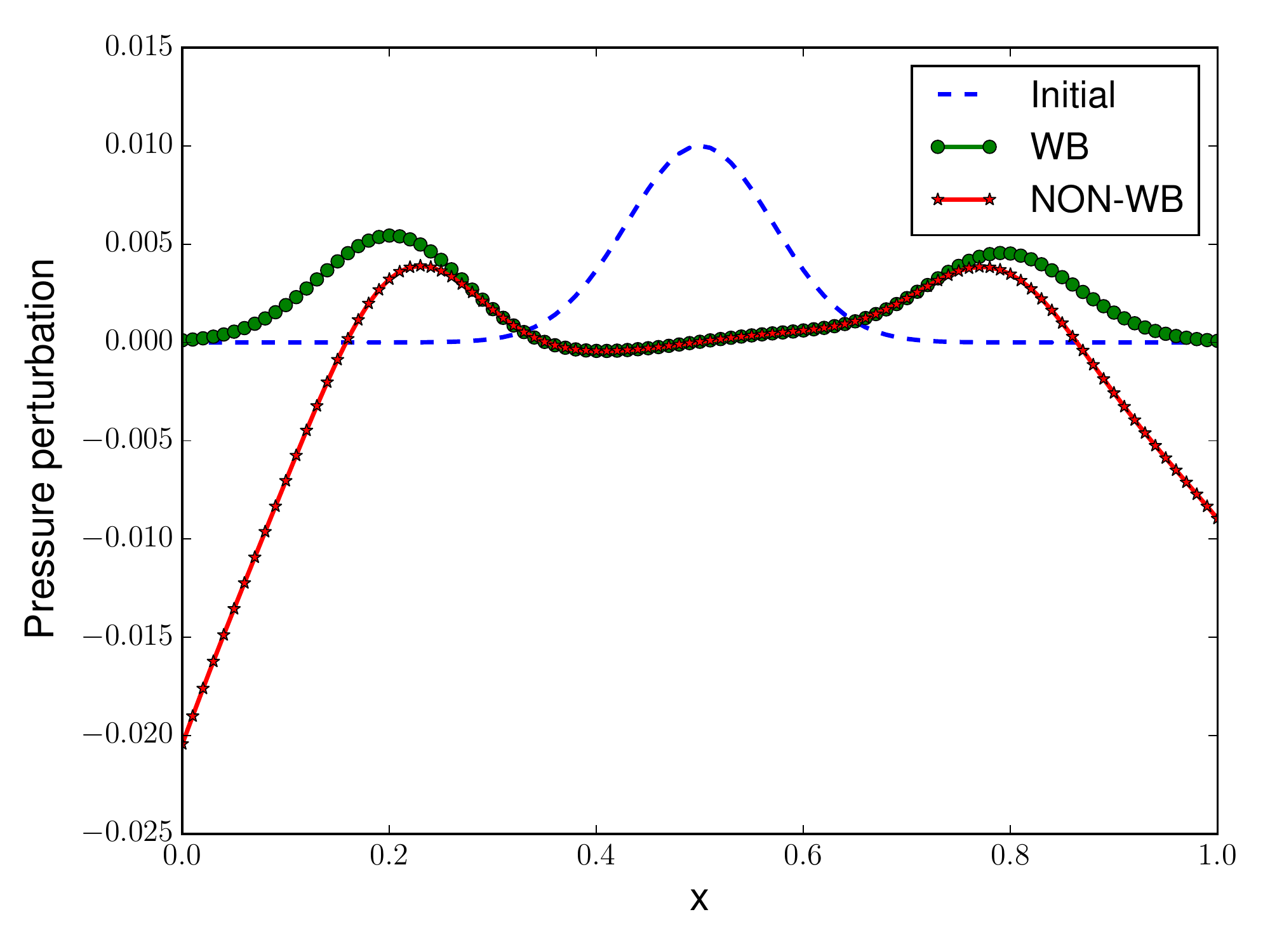} &
\includegraphics[width=0.48\textwidth]{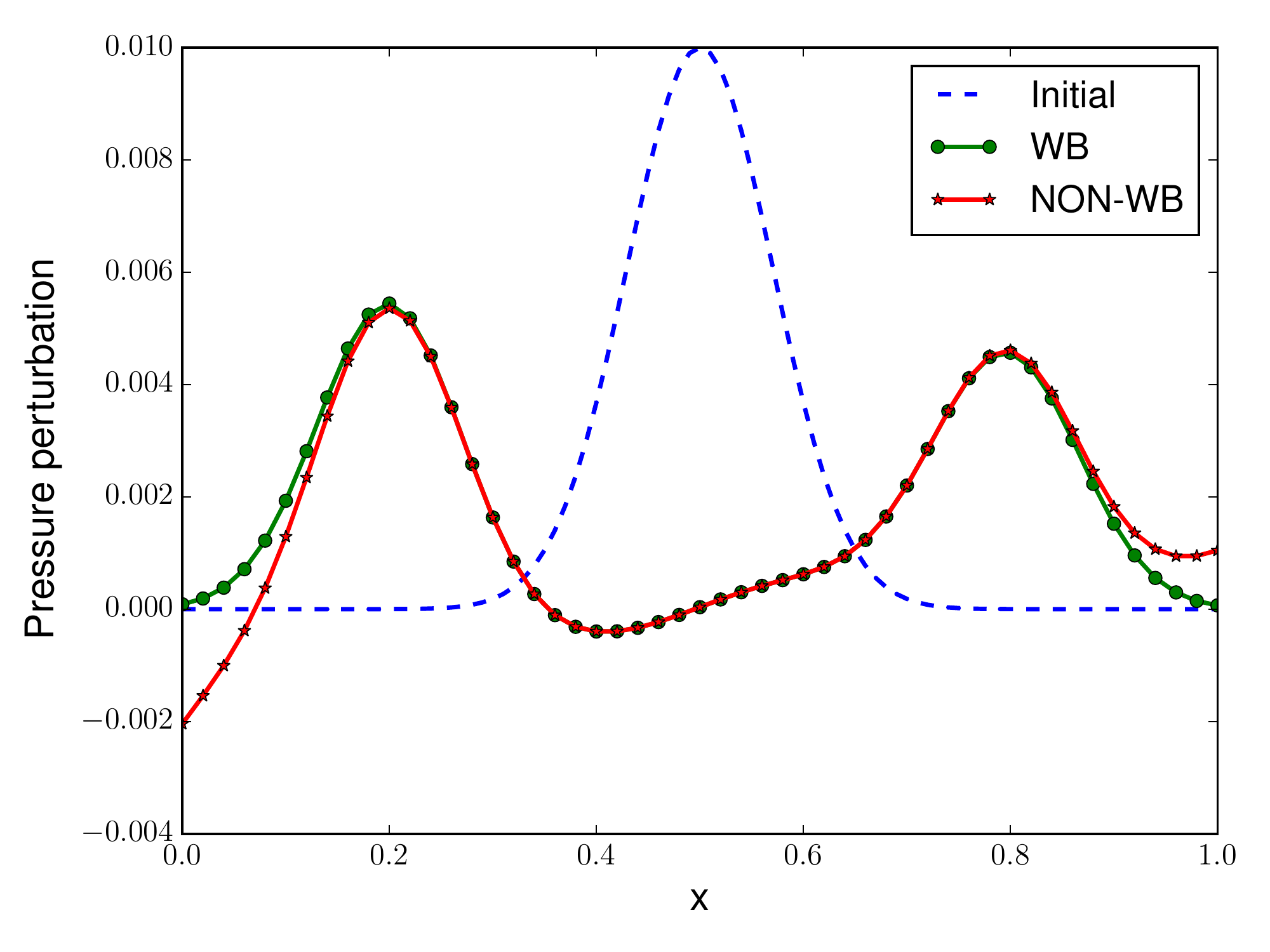} \\
(a) & (b)
\end{tabular}
\caption{Comparison of well-balanced (WB) and non well-balanced (NON-WB) schemes for evolution of small perturbations in the isothermal case for $\eta = 10^{-2}$: (a) $Q_1$, 100 cells (b) $Q_2$, 50 cells}
\label{fig:pert1wbnwb}
\end{center}
\end{figure}

\begin{figure}
\begin{center}
\begin{tabular}{cc}
\includegraphics[width=0.48\textwidth]{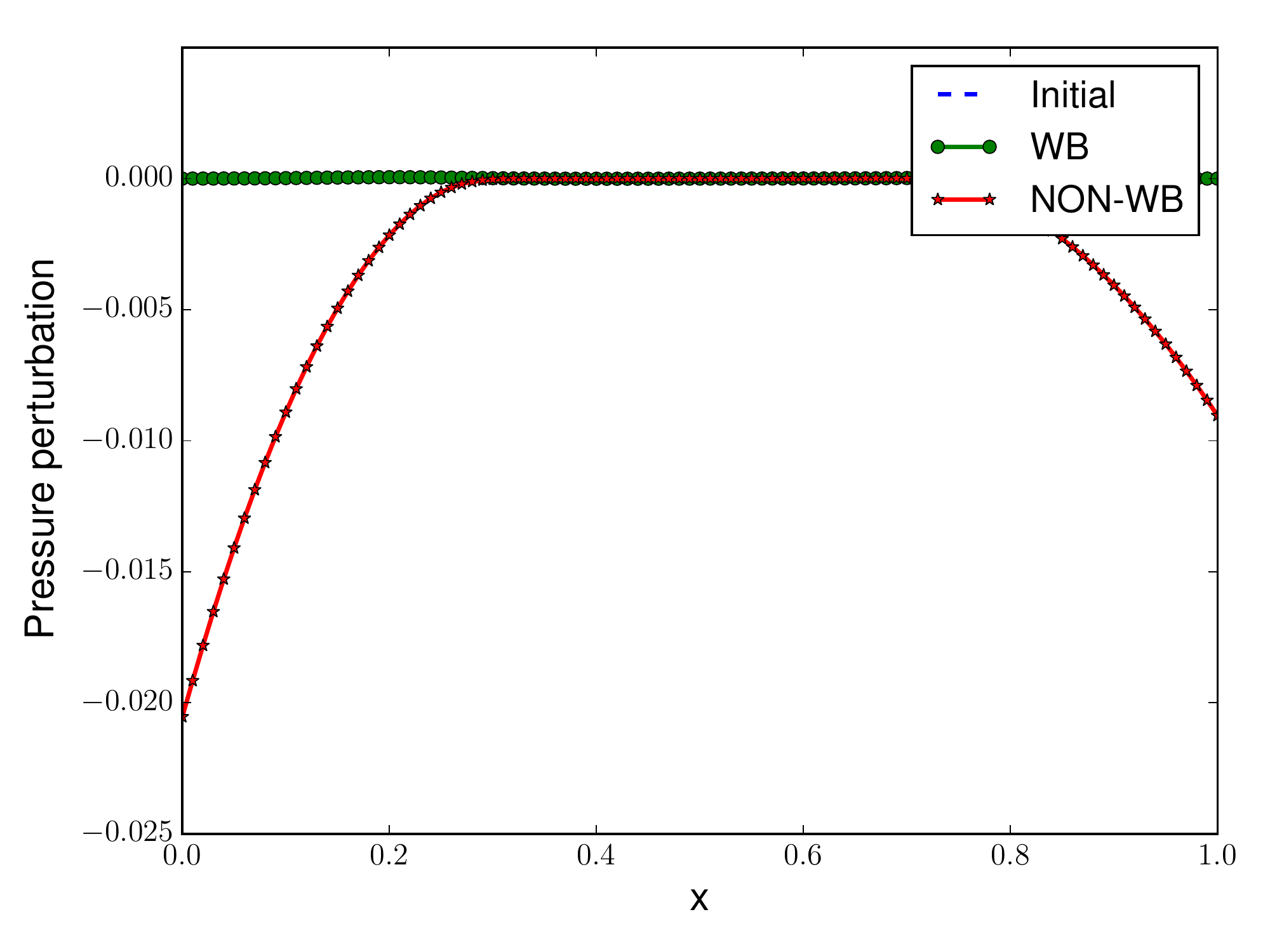} &
\includegraphics[width=0.48\textwidth]{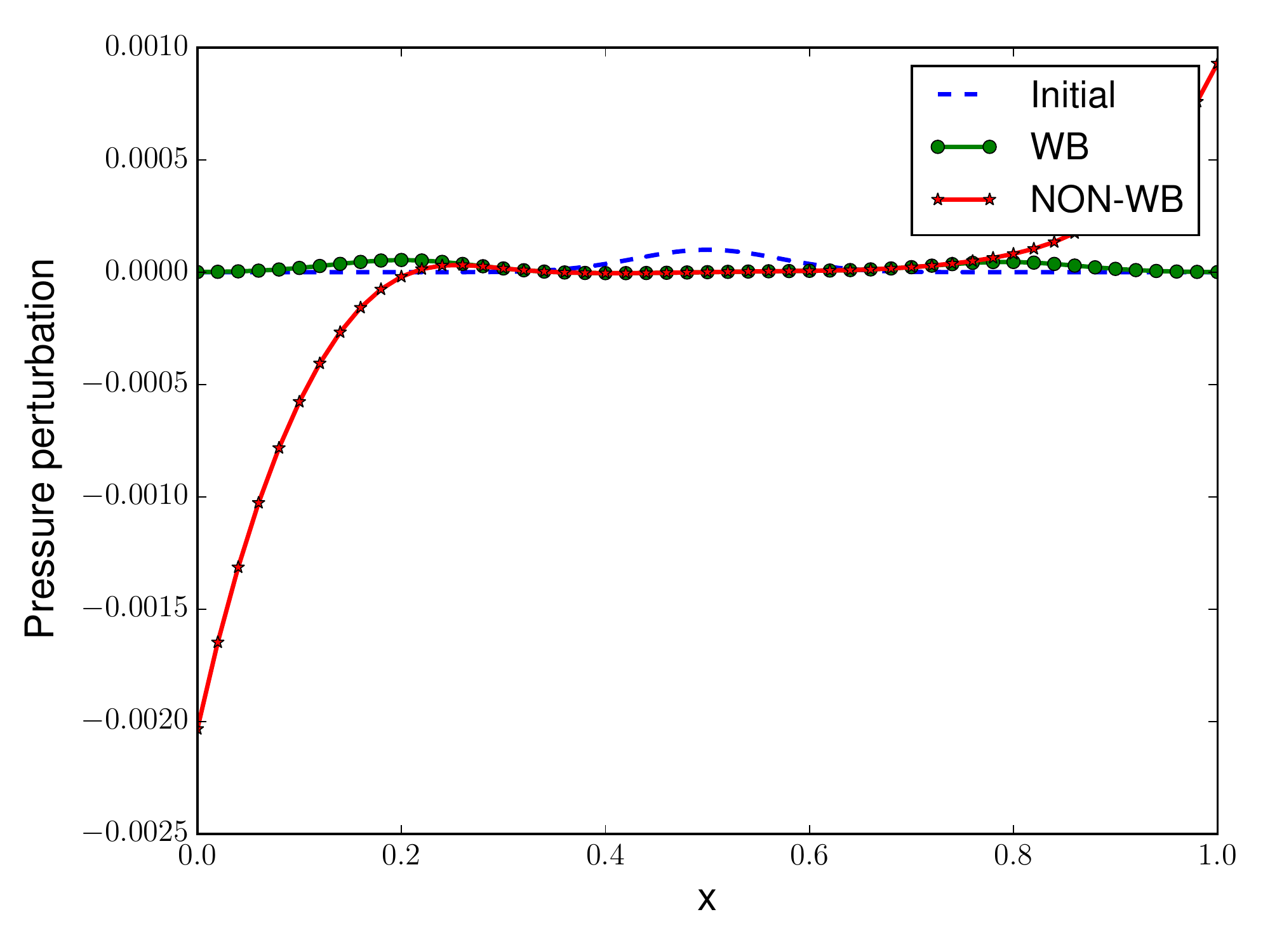} \\
(a) & (b)
\end{tabular}
\caption{Comparison of well-balanced (WB) and non well-balanced (NON-WB) schemes for evolution of small perturbations in the isothermal case for $\eta = 10^{-4}$: (a) $Q_1$, 100 cells (b) $Q_2$, 50 cells}
\label{fig:pert2wbnwb}
\end{center}
\end{figure}

To show the advantage of well-balanced scheme, we compute the above solutions using a scheme that is not well balanced. The source terms in the non well-balanced scheme are computed using the exact value of the gravitational force $\nabla \pot$. The resulting perturbation pressures are shown in figures~(\ref{fig:pert1wbnwb}), (\ref{fig:pert2wbnwb}). The non well-balanced scheme has much larger error in the perturbation pressure compared to the well-balanced scheme. Figure~(\ref{fig:pert1wbnwb}) shows that using a higher order scheme ($Q_2$) reduces the error in the non-well-balanced scheme. However if the perturbation level is smaller, as in figure~(\ref{fig:pert2wbnwb}), then the error in non well-balanced scheme is still too high compared to well-balanced scheme. 

%-------------------------------------------------------------------------------------

\subsubsection{Polytropic case}
\label{sec:poly}
We next consider a polytropic hydrostatic solution with potential $\pot = x$ and state given by
\[
\rho = \left( \rho_0^{\nu-1} - \alpha \frac{\nu-1}{\nu} x \right)^{1/(\nu-1)}, \qquad p = \alpha \rho^{\nu}
\]
where $\rho_0=1$, $\alpha=1$ and $\nu=1.2$. We compute the solution upto a final time of $t=0.1$ units and show the error in solution in table~(\ref{tab:1dlin$Q_1$poly}) and (\ref{tab:1dlin$Q_2$poly}) for second and third order schemes respectively. We see that the scheme preserves the initial condition close to machine precision.

\begin{table}
\begin{center}
\begin{tabular}{|c|c|c|c|c|}
\hline
Mesh & $\rho u$ & $\rho v$ & $\rho$ & $E$ \\
\hline
25x25   & 1.07749e-13  & 6.67302e-15 &  2.89086e-14 &  9.38882e-14 \\
50x50   & 1.08760e-13  & 5.87560e-15 &  2.82644e-14 &  9.17933e-14 \\
100x100 & 1.07487e-13  & 5.29355e-15 &  4.91244e-14 &  9.62769e-14 \\
200x200 & 1.09086e-13  & 5.70042e-15 &  2.81423e-14 &  9.14467e-14 \\
\hline
\end{tabular}
\caption{Well-balanced test for polytropic case on Cartesian mesh using $Q_1$ polynomials and potential $\pot = x$}
\label{tab:1dlin$Q_1$poly}
\end{center}
\end{table}

\begin{table}
\begin{center}
\begin{tabular}{|c|c|c|c|c|}
\hline
Mesh & $\rho u$ & $\rho v$ & $\rho$ & $E$ \\
\hline
25x25   & 1.08483e-13  & 7.35525e-15  & 2.92146e-14 &  9.48770e-14 \\
50x50   & 1.08977e-13  & 6.03037e-15  & 2.85168e-14 &  9.25119e-14 \\
100x100 & 1.09071e-13  & 5.75922e-15  & 2.86190e-14 &  9.23044e-14 \\
200x200 & 1.09107e-13  & 6.11341e-15  & 2.91511e-14 &  9.39398e-14 \\
\hline
\end{tabular}
\caption{Well-balanced test for polytropic case on Cartesian mesh using $Q_2$ polynomials and potential $\pot = x$}
\label{tab:1dlin$Q_2$poly}
\end{center}
\end{table}

\begin{figure}
\begin{center}
\begin{tabular}{cc}
\includegraphics[width=0.48\textwidth]{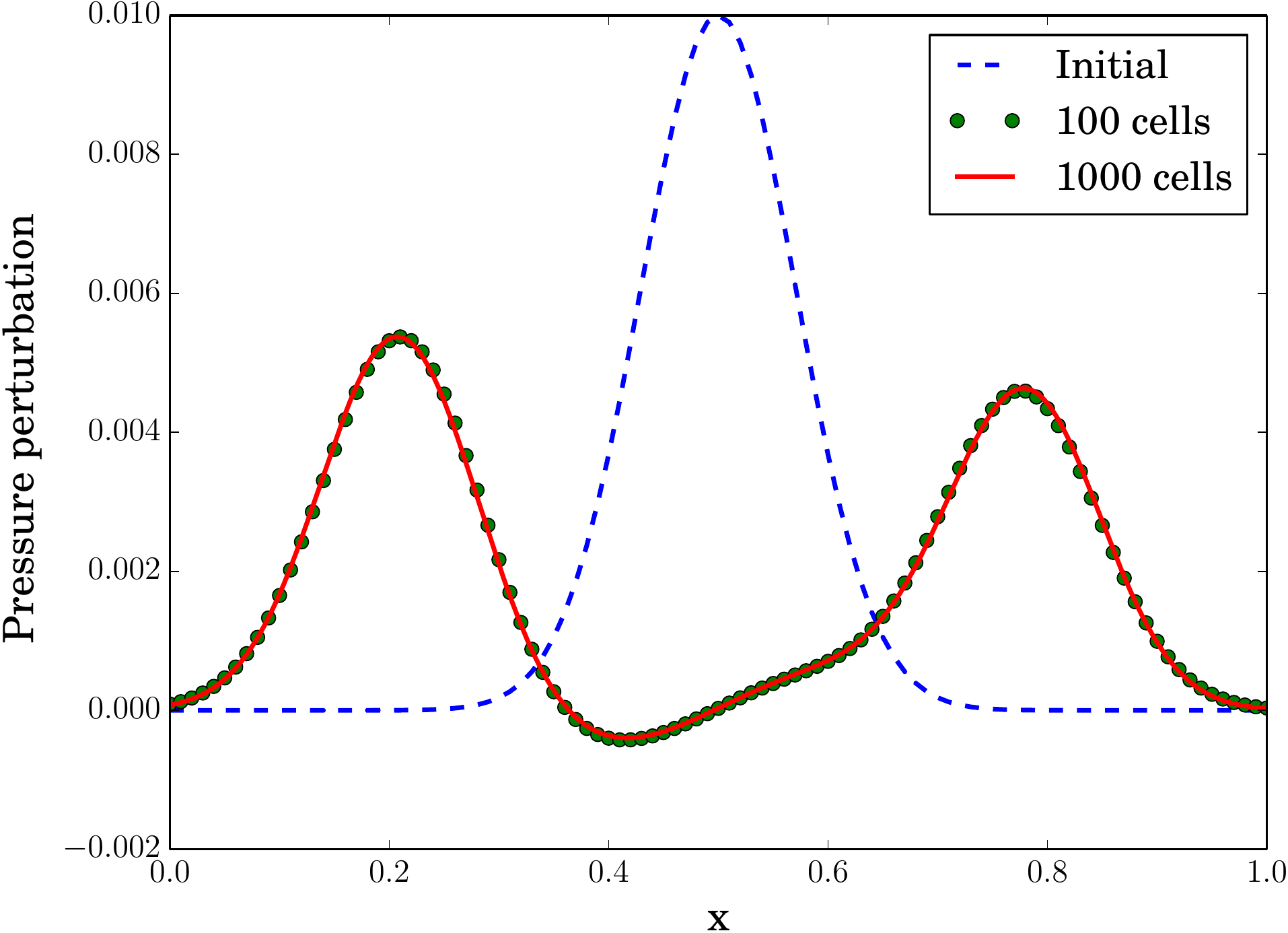} &
\includegraphics[width=0.48\textwidth]{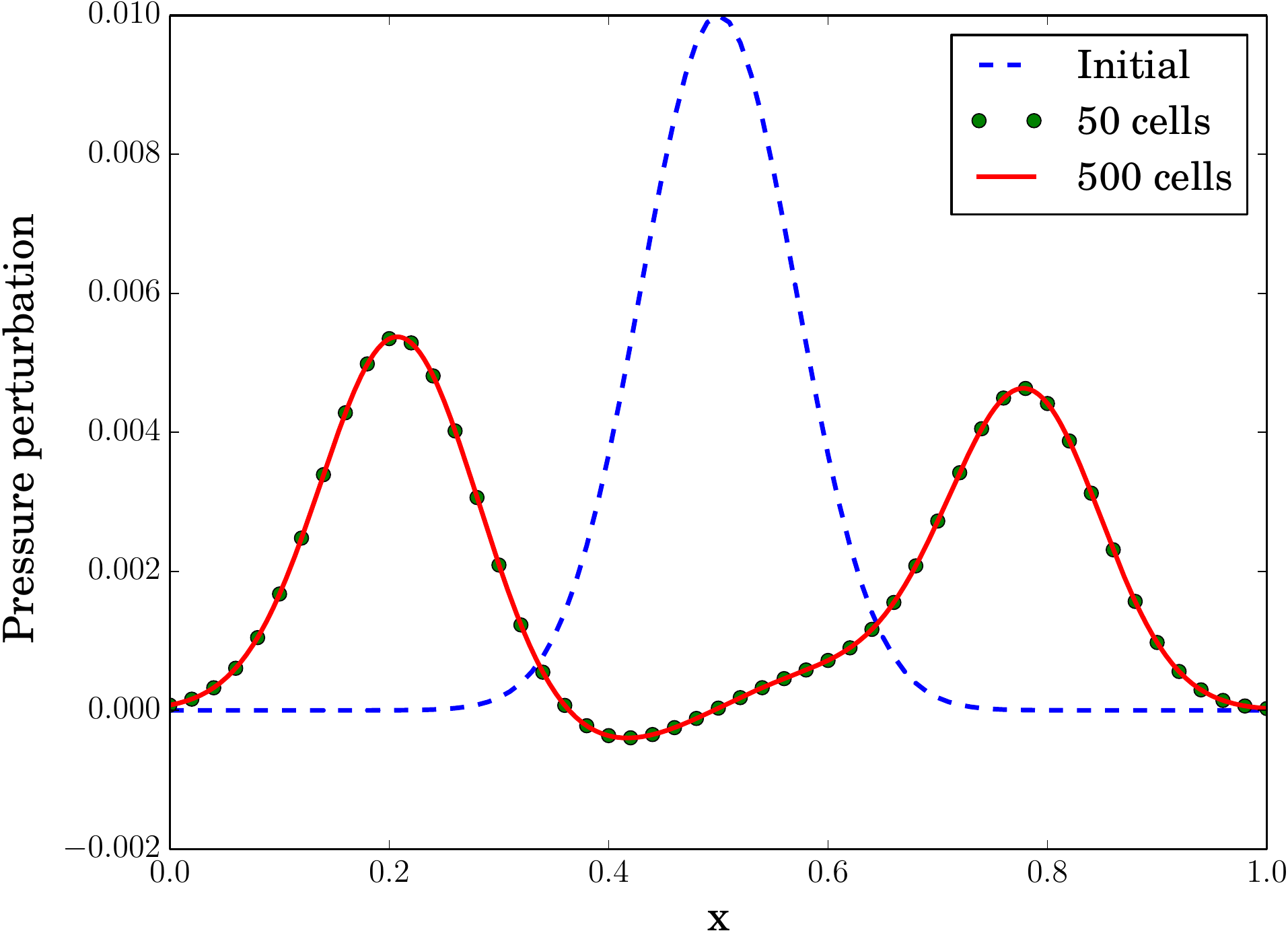} \\
(a) & (b)
\end{tabular}
\caption{Evolution of small perturbations in the polytropic case for $\eta = 10^{-2}$: (a) $Q_1$ (b) $Q_2$}
\label{fig:polypert1}
\end{center}
\end{figure}

\begin{figure}
\begin{center}
\begin{tabular}{cc}
\includegraphics[width=0.48\textwidth]{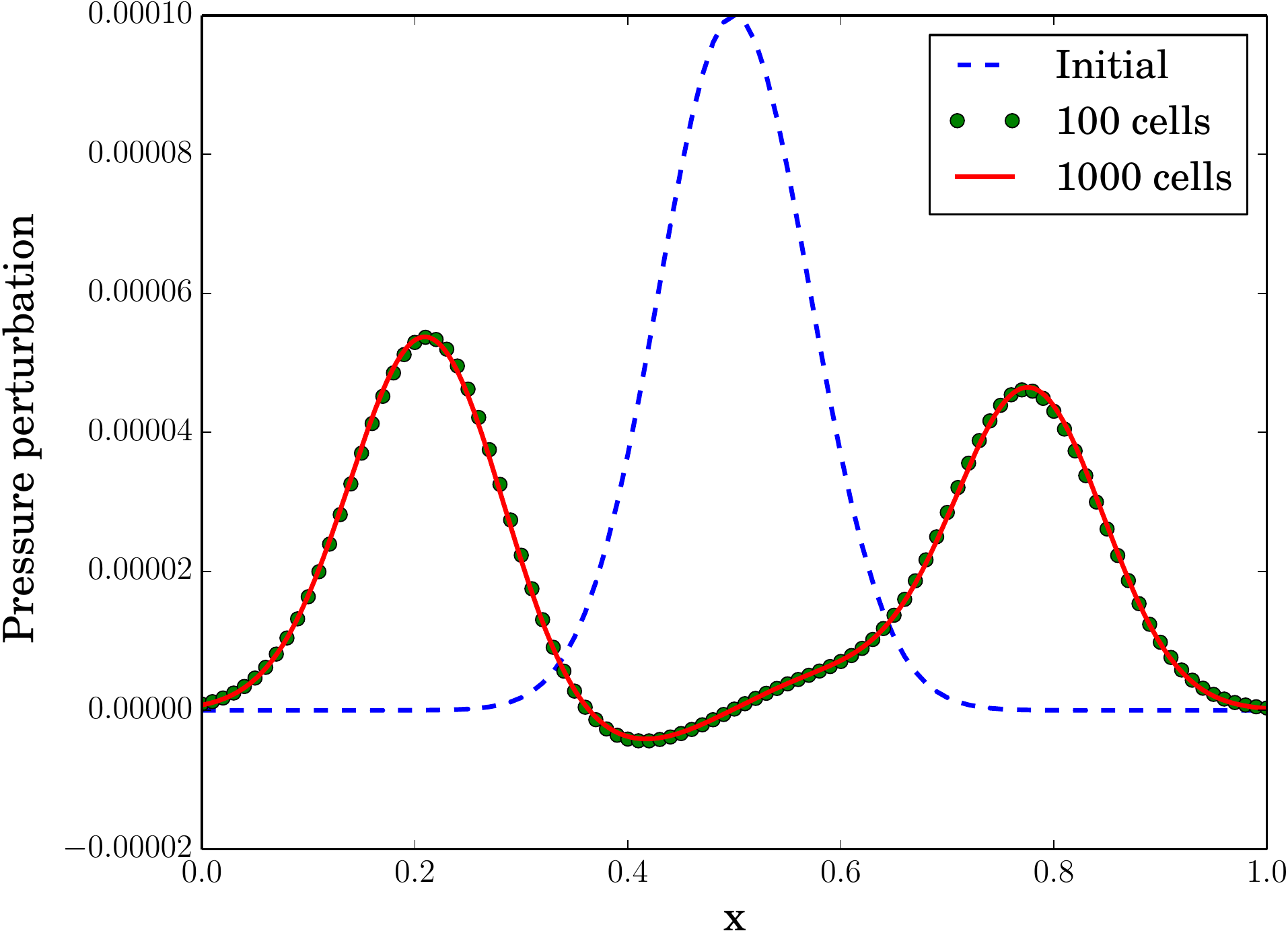} &
\includegraphics[width=0.48\textwidth]{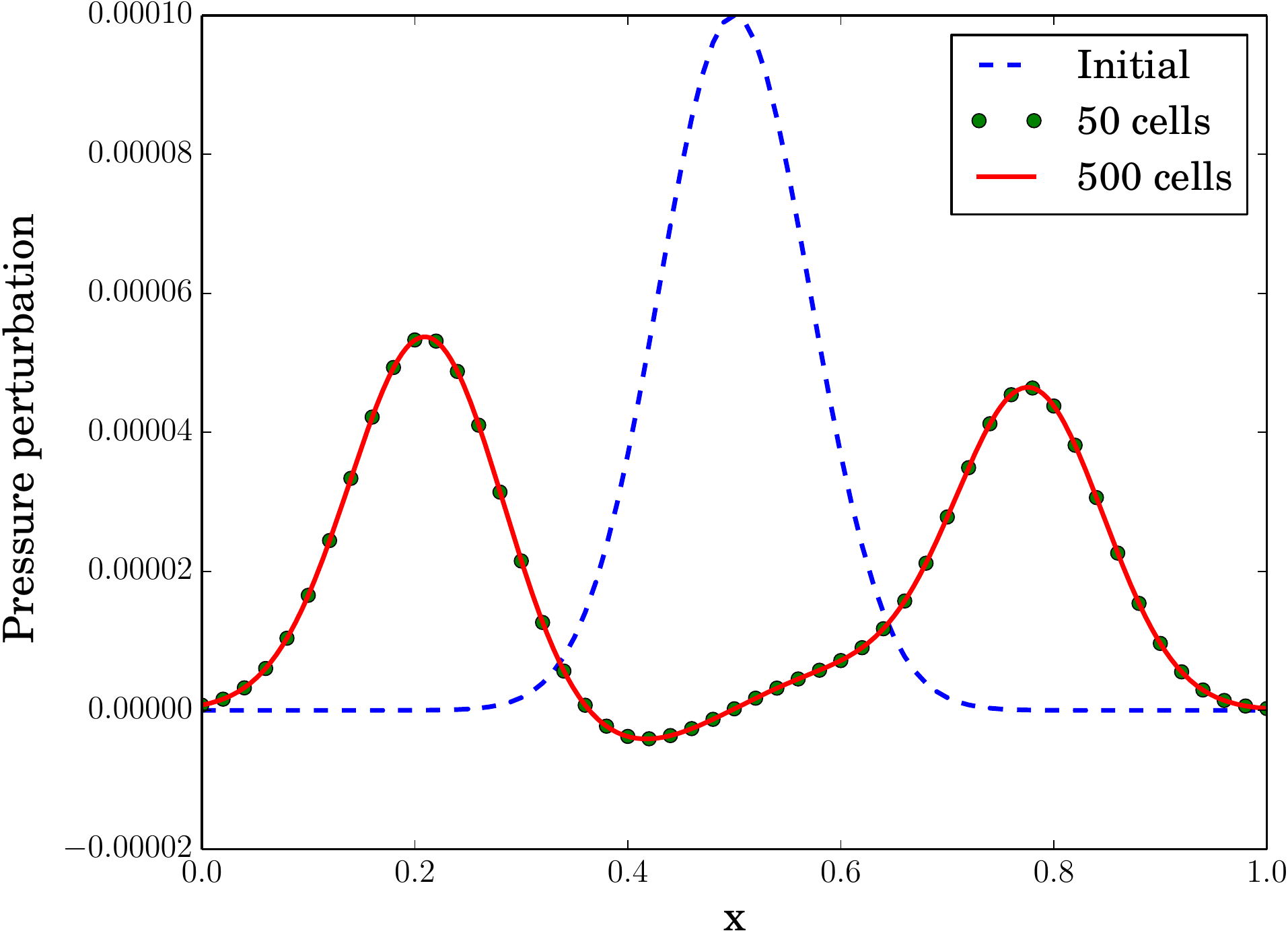} \\
(a) & (b)
\end{tabular}
\caption{Evolution of small perturbations in the polytropic case for $\eta = 10^{-4}$: (a) $Q_1$ (b) $Q_2$}
\label{fig:polypert2}
\end{center}
\end{figure}

To study the evolution of small perturbations, we consider an initial condition given by
\[
\rho = \left( \rho_0^{\nu-1} - \alpha \frac{\nu-1}{\nu} x \right)^{1/(\nu-1)}, \qquad p = \alpha \rho^{\nu} + \eta \exp(-100(x-1/2)^2)
\]
This is advanced upto a time of $t=0.25$ and the perturbation pressure is shown in figures~(\ref{fig:polypert1}) and (\ref{fig:polypert2}) for $\eta=10^{-2}$ and $\eta = 10^{-4}$ respectively. The perturbations are resolved without any spurious oscillations and even the coarse mesh gives solutions comparable to those on the finer mesh.

%-------------------------------------------------------------------------------------
\subsection{Order of accuracy study when scheme is not well-balanced}
In this test, we start with the polytropic hydrostatic solution from section~(\ref{sec:poly}) as initial condition in two dimensions with gravity acting vertically, and solve it with the isothermal well-balanced scheme. This scheme will not be able to exactly preserve the polytropic solution. Since the exact solution is the polytropic hydrostatic solution, we can compute the $L^2$ norm of the error. The error is computed on different grid sizes and polynomials degrees, and the results are shown in table~(\ref{tab:polyisoq1}), (\ref{tab:polyisoq2}). The horizontal momentum error is close to machine zero since gravity acts on in the vertical direction. The other quantities are not exactly preserved but their errors converge at the expected rate, i.e., we get second order accuracy with $Q_1$ polynomials and third order accuracy with $Q_2$ polynomials.
\begin{table}
\begin{center}
\begin{tabular}{|c|c|c|c|c|c|c|c|}
\hline
Mesh & $\rho u$ & \multicolumn{2}{|c|}{$\rho v$} & \multicolumn{2}{|c|}{$\rho$} & \multicolumn{2}{|c|}{$E$} \\
\hline 
Size  & Error & Error & Rate & Error & Rate & Error & Rate \\
\hline
25x25  & 6.22039e-14  & 1.39945e-05  &- & 5.03134e-07  &- & 1.50727e-06 & -\\
50x50  & 6.27891e-14  & 3.51615e-06  &1.99 & 1.71697e-07  &1.55 & 4.03669e-07 & 1.90\\
100x100 & 6.29344e-14  & 8.79605e-07  &1.99 & 4.91080e-08  &1.80 & 1.08737e-07 & 1.89\\
200x200 & 6.29710e-14  & 2.19966e-07  &1.99 & 1.30477e-08  &1.91 & 2.83352e-08 & 1.94 \\
\hline
\end{tabular}
\caption{Polytropic hydrostatic solution solved with isothermal well-balanced scheme on Cartesian mesh using $Q_1$ polynomials and potential $\pot = x$}
\label{tab:polyisoq1}
\end{center}
\end{table}

\begin{table}
\begin{center}
\begin{tabular}{|c|c|c|c|c|c|c|c|}
\hline
Mesh & $\rho u$ & \multicolumn{2}{|c|}{$\rho v$} & \multicolumn{2}{|c|}{$\rho$} & \multicolumn{2}{|c|}{$E$} \\
\hline 
Size  & Error & Error & Rate & Error & Rate & Error & Rate \\
\hline
25x25 & 6.26353e-14  & 1.03474e-07  &-& 1.17234e-07 &-&  3.80288e-07 & - \\
50x50  & 6.24467e-14  & 1.29041e-08  &3.00& 1.46356e-08  &3.00& 4.74617e-08 & 3.00 \\
100x100 & 6.29693e-14  & 1.61142e-09  &3.00& 1.82873e-09  &3.00& 5.92946e-09  & 3.00 \\
200x200 & 6.29792e-14  & 2.01344e-10  &3.00& 2.28559e-10  &3.00& 7.41017e-10 & 3.00 \\
\hline
\end{tabular}
\caption{Polytropic hydrostatic solution solved with isothermal well-balanced scheme on Cartesian mesh using $Q_2$ polynomials and potential $\pot = x$}
\label{tab:polyisoq2}
\end{center}
\end{table}
%-------------------------------------------------------------------------------------
\subsection{2-D hydrostatic solution - I}
\begin{figure}
\begin{center}
\begin{tabular}{cc}
\includegraphics[width=0.48\textwidth]{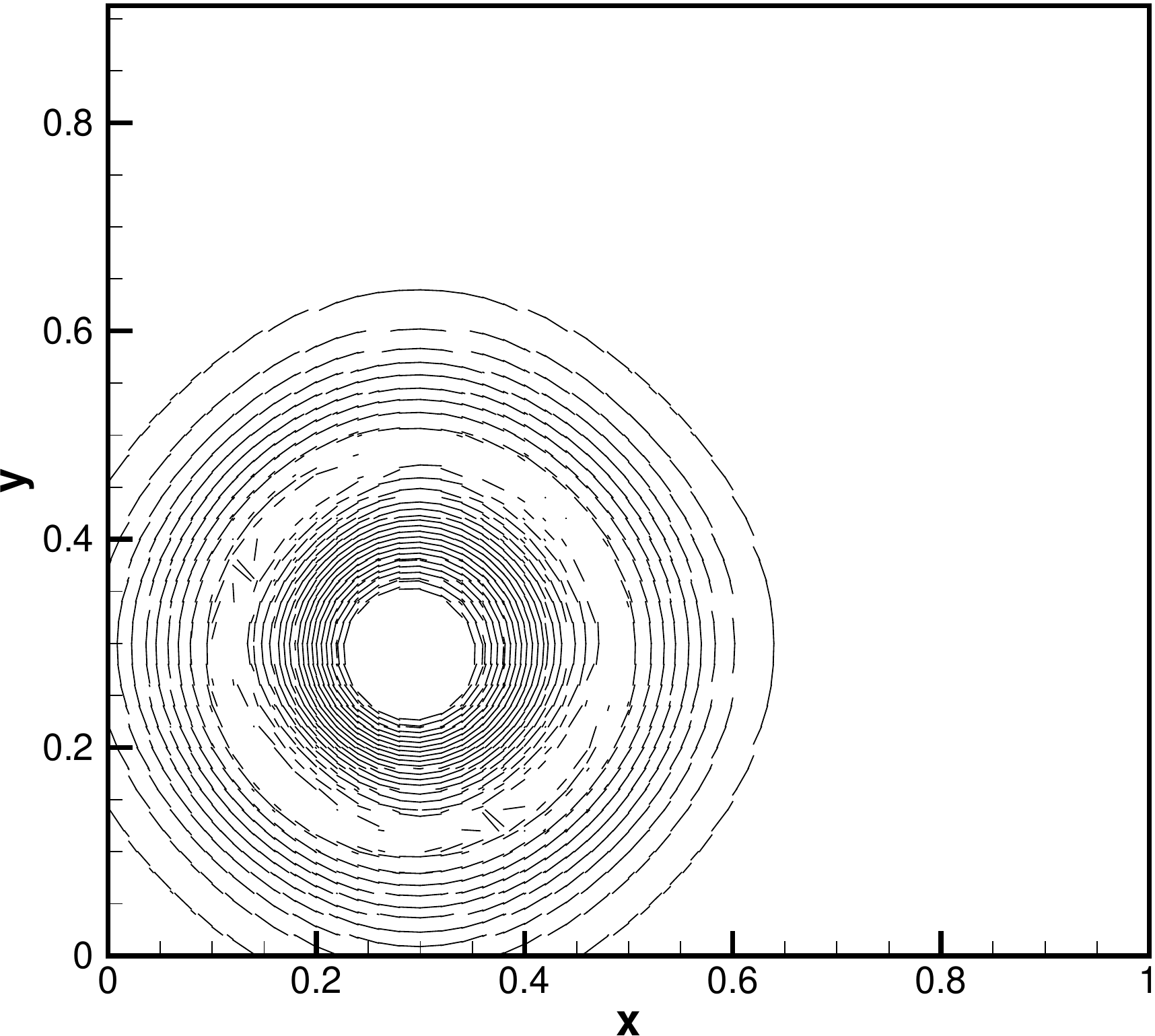} &
\includegraphics[width=0.48\textwidth]{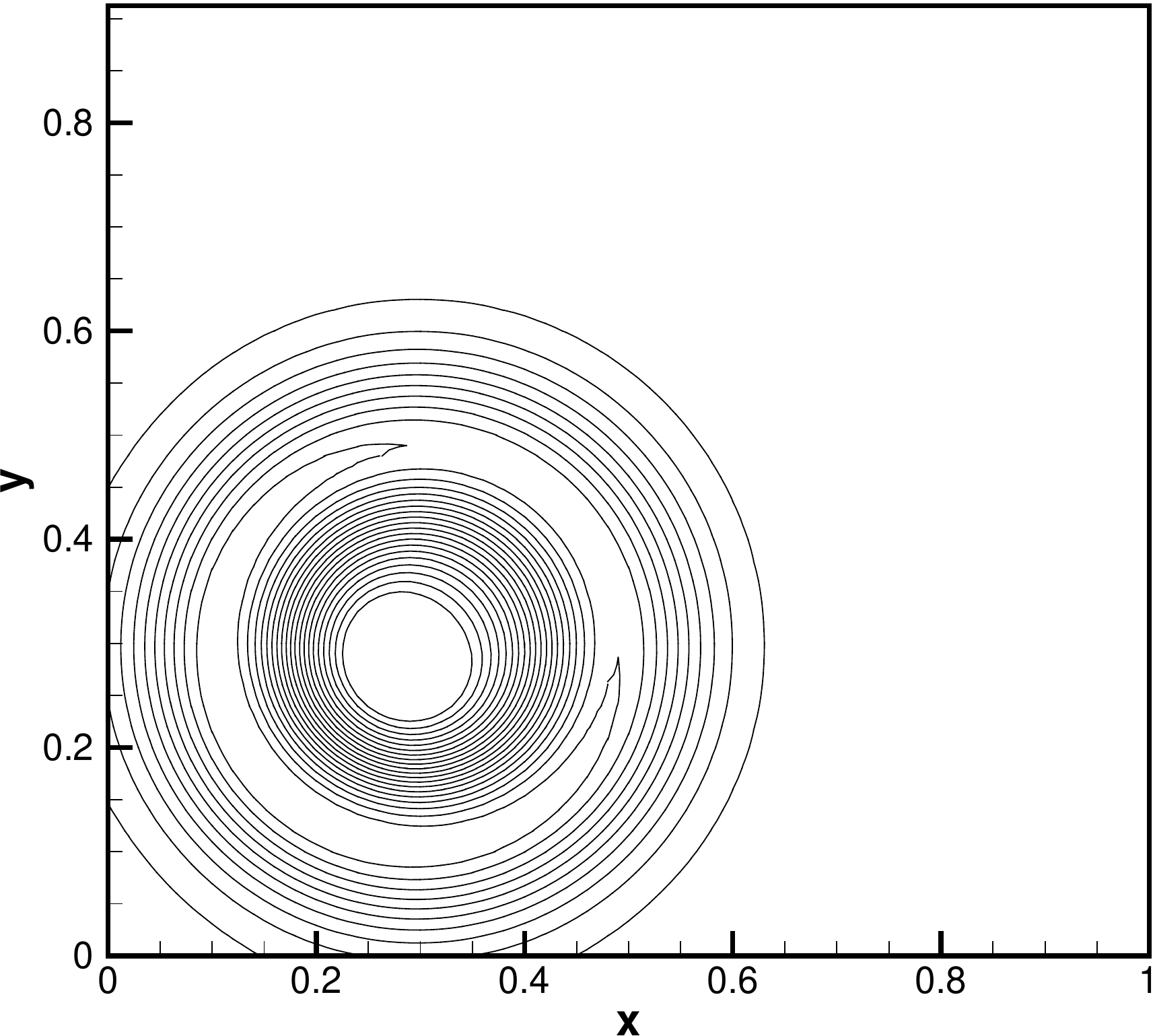}\\
(a) & (b)
\end{tabular}
\caption{Pressure perturbation on $50 \times 50$ mesh at time $t=0.15$ (a) $Q_1$ (b) $Q_2$. Showing 20 contours lines between $-0.0002$ to $+0.0002$}
\label{fig:iso2da}
\end{center}
\end{figure}

\begin{figure}
\begin{center}
\begin{tabular}{cc}
\includegraphics[width=0.48\textwidth]{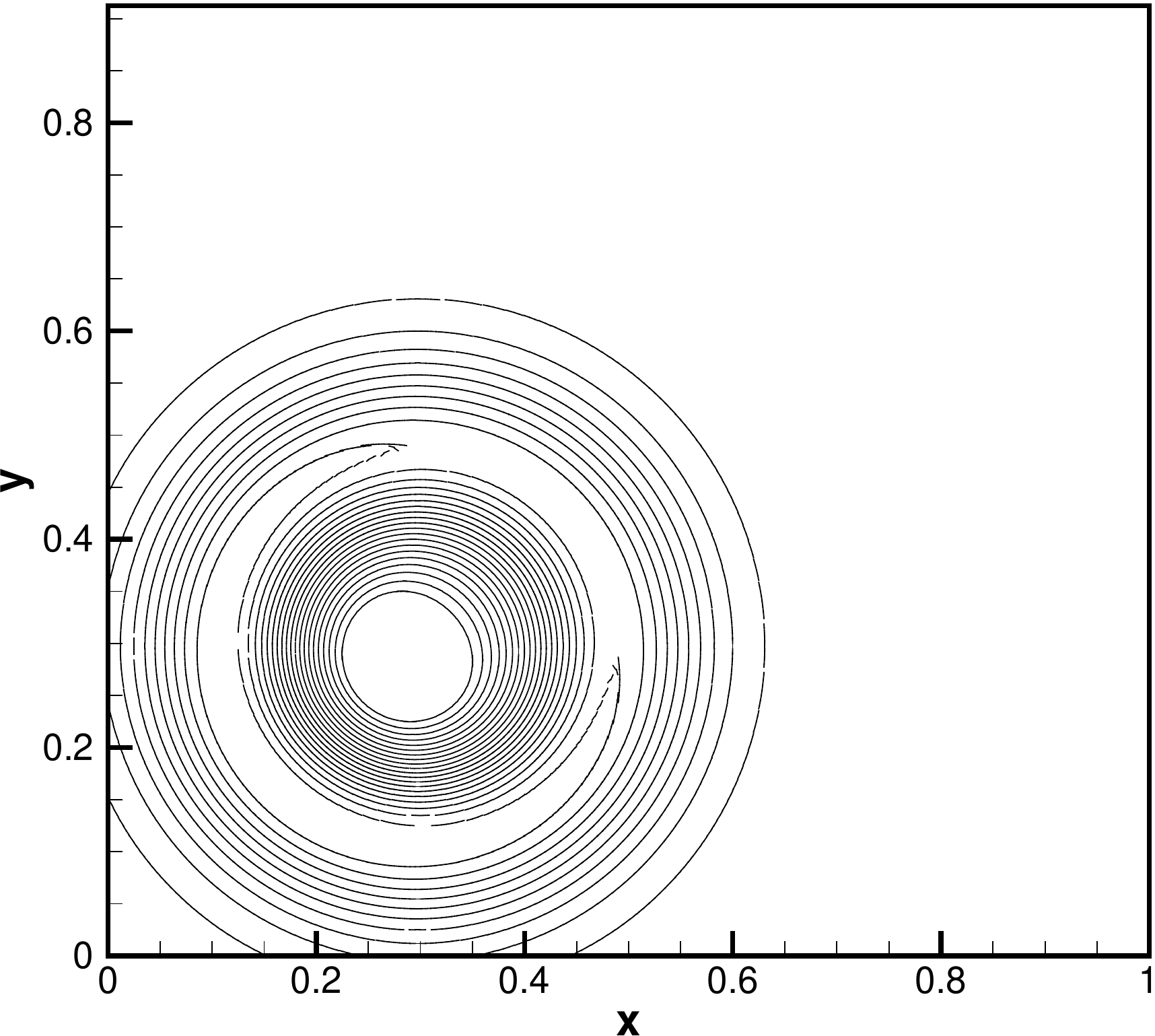} &
\includegraphics[width=0.48\textwidth]{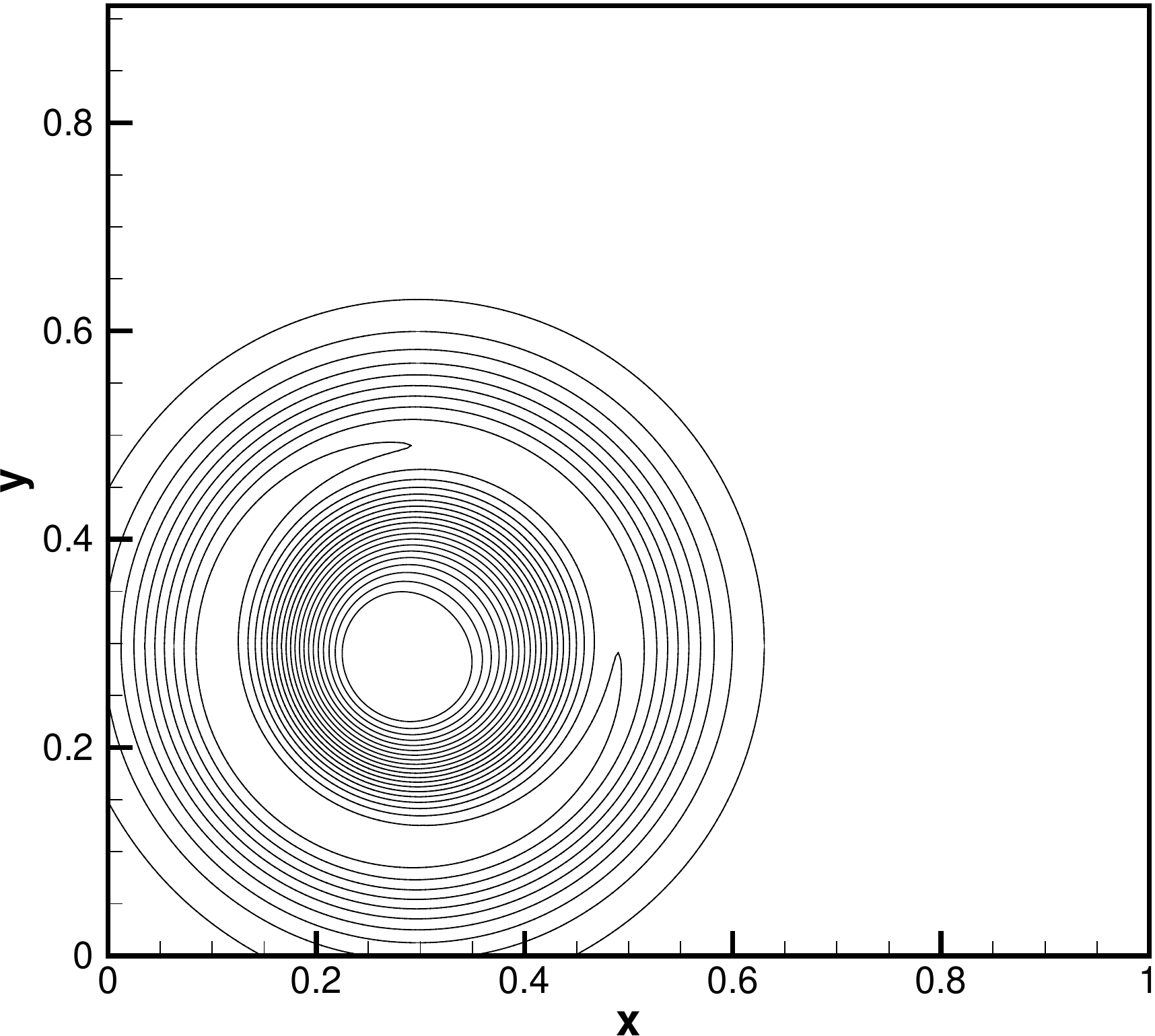}\\
(a) & (b)
\end{tabular}
\caption{Pressure perturbation on $200 \times 200$ mesh at time $t=0.15$ (a) $Q_1$ (b) $Q_2$. Showing 20 contours lines between $-0.0002$ to $+0.0002$}
\label{fig:iso2db}
\end{center}
\end{figure}

This test case is taken from~\cite{Xing:2013:HOW:2434597.2434626}. For the potential $\pot = x+y$, the hydrostatic solution is given by
\[
\rho = \rho_0 \exp\left( - \frac{\rho_0 g}{p_0}(x+y) \right), \qquad p = p_0 \exp\left( - \frac{\rho_0 g}{p_0}(x+y) \right), \qquad u=v= 0
\]
For the parameters appearing in the above equations, we take the values $\rho_0=1$, $p_0=1$ and $g=1$. With the above initial conditions, the DG scheme maintains the solution at the same values upto machine precision as shown in table~(\ref{tab:iso2d}).
\begin{table}
\begin{center}
\begin{tabular}{|l|c|c|c|c|}
\hline
   & $\rho u$ & $\rho v$ & $\rho$ & $E$ \\
\hline
$Q_1$, $25\times 25$  &  9.85926e-14 & 9.85855e-14 & 5.32357e-14 & 1.55361e-13  \\
$Q_1$, $50\times 50$  &  9.94493e-14 & 9.94451e-14 & 5.37084e-14 & 1.56669e-13  \\
$Q_1$, $100\times 100$ & 9.96481e-14 & 9.96474e-14 & 5.38404e-14 & 1.57062e-13  \\
\hline
$Q_2$, $25\times 25$  &  9.9256e-14 & 9.92682e-14 & 5.39863e-14 & 1.57435e-13  \\
$Q_2$, $50\times 50$  &  9.961e-14 & 9.96538e-14 & 5.41091e-14 & 1.57521e-13  \\
$Q_2$, $100\times 100$ & 9.95889e-14 & 9.97907e-14 & 5.43145e-14 & 1.57728e-13 \\
\hline
\end{tabular}
\caption{Well-balanced test for 2-D isothermal hydrostatic solution on Cartesian meshes}
\label{tab:iso2d}
\end{center}
\end{table}
Next we add a small perturbation to the pressure so that the initial pressure is given by
\[
p = p_0 \exp\left( - \frac{\rho_0 g}{p_0}(x+y) \right) + \eta \exp\left( -100\frac{\rho_0 g}{p_0}[ (x-0.3)^2 + (y-0.3)^2] \right)
\]
while the remaining quantities are as before. In this case, the initial condition is not in equilibrium and it evolves with time. Due to the gravitational force, the pressure perturbation is advected towards the lower left corner. We see in figures~(\ref{fig:iso2da}), (\ref{fig:iso2db}) that the well-balanced scheme is able to predict the evolution of the perturbation without any spurious disturbances. The $Q_1$ solutions become more accurate as the mesh becomes finer while the $Q_2$ solutions are visually quite accurate even on the coarse mesh.

%-------------------------------------------------------------------------------------
\subsection{2-D hydrostatic solution - II}
We consider an initial condition with two constant temperatures
\[
T = \begin{cases}
T_l & y < 0 \\
T_u & y > 0
\end{cases}
\]
Taking the potential to be $\pot = y$, the hydrostatic pressure and density are given by
\[
p = \begin{cases}
p_0 \ee^{-y/RT_l}, & y \le 0 \\
p_0 \ee^{-y/RT_u}, & y > 0
\end{cases}, \qquad \rho = \begin{cases}
p/RT_l, & y < 0 \\
p/RT_u, & y > 0 
\end{cases}
\]
While the pressure is continuous at $y=0$, the density experiences a jump due to the jump in temperature. The computational domain is taken to be $[-0.25, +0.25] \times [-1,1]$ which is covered with a Cartesian mesh and we use the Roe numerical flux. We consider two configurations based on the temperature distribution and compute the solution until time $t=0.1$ units.

\noindent
\underline{Case 1}: Here we choose $T_l = 1$, $T_u = 2$. This corresponds to lighter fluid on top of heavier fluid which is a stable configuration. The DG scheme preserves the initial condition upto machine precision as seen in table~(\ref{tab:rt1}).

\noindent
\underline{Case 2}: Here we choose $T_l = 2$, $T_u = 1$. This corresponds to heavier fluid on top of lighter fluid which is an unstable configuration. The DG scheme preserves the initial condition upto machine precision  as seen in table~(\ref{tab:rt2}).

\begin{table}
\begin{center}
\begin{tabular}{|c|c|c|c|c|}
\hline
   & $\rho u$ & $\rho v$ & $\rho$ & $E$ \\
\hline
$Q_1$, 25x100 & 5.13108e-13 & 1.10971e-13 & 2.56836e-13 & 8.91784e-13  \\
$Q_1$, 50x200 & 5.15744e-13 & 1.12234e-13 & 2.84309e-13 & 8.97389e-13  \\
$Q_2$, 25x100 & 5.15726e-13 & 1.1341e-13 & 3.22713e-13 & 9.01183e-13  \\
$Q_2$, 50x200 & 5.16397e-13 & 1.13707e-13 & 3.93623e-13 & 9.02503e-13  \\
\hline
\end{tabular}
\caption{Well-balanced test for Rayleigh-Taylor problem in Case 1, $T_l=1$, $T_u=2$}
\label{tab:rt1}
\end{center}
\end{table}

\begin{table}
\begin{center}
\begin{tabular}{|c|c|c|c|c|}
\hline
   & $\rho u$ & $\rho v$ & $\rho$ & $E$ \\
\hline
$Q_1$, 25x100 & 3.487e-13 & 1.0989e-13 & 1.98949e-13 & 7.42265e-13  \\
$Q_1$, 50x200 & 3.50153e-13 & 1.1121e-13 & 2.34991e-13 & 7.4747e-13  \\
$Q_2$, 25x100 & 3.50149e-13 & 1.12159e-13 & 2.80645e-13 & 7.5104e-13  \\
$Q_2$, 50x200 & 3.50548e-13 & 1.12533e-13 & 3.63806e-13 & 7.52151e-13 \\
\hline
\end{tabular}
\caption{Well-balanced test for Rayleigh-Taylor problem in Case 2, $T_l=2$, $T_u=1$}
\label{tab:rt2}
\end{center}
\end{table}

\begin{figure}
\begin{center}
\begin{tabular}{cccc}
\includegraphics[width=0.24\textwidth]{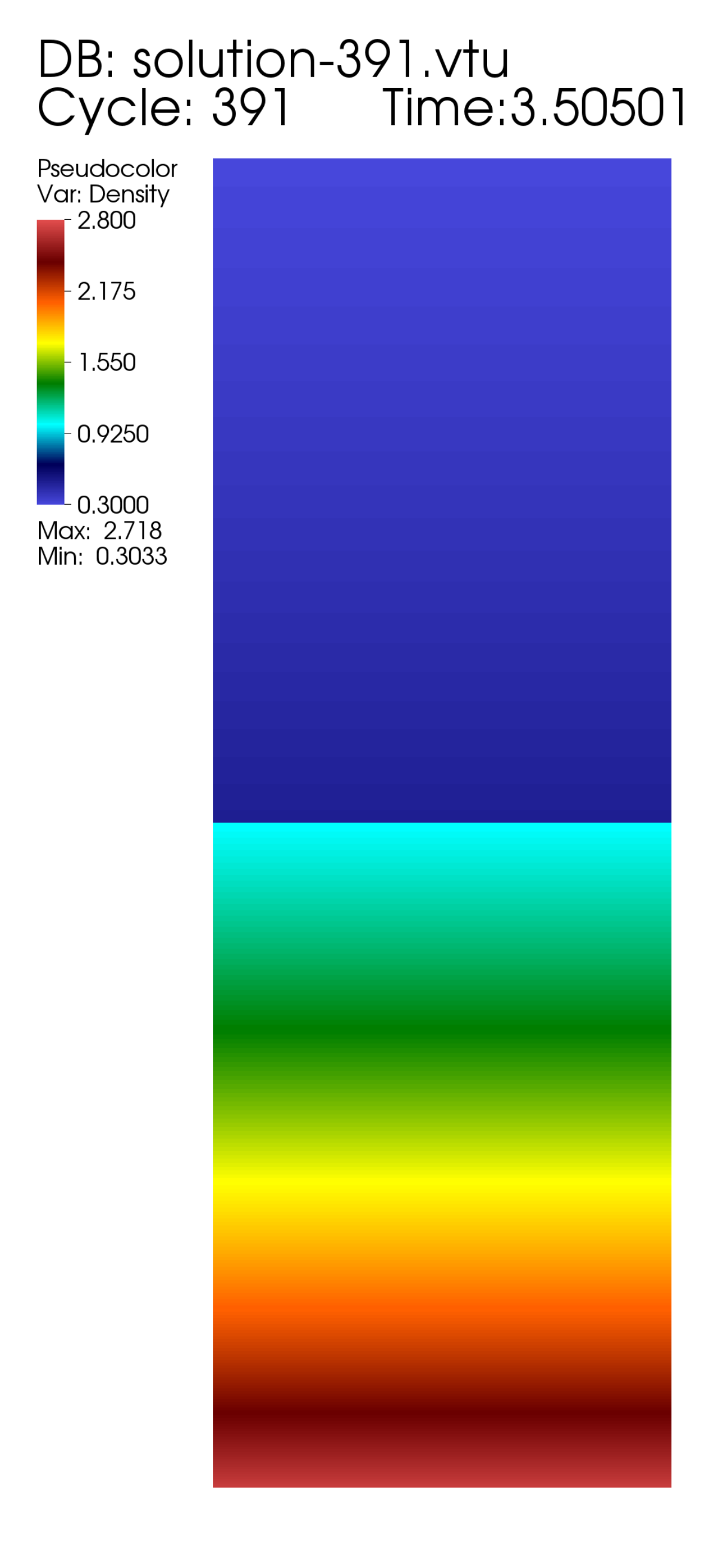} &
\includegraphics[width=0.24\textwidth]{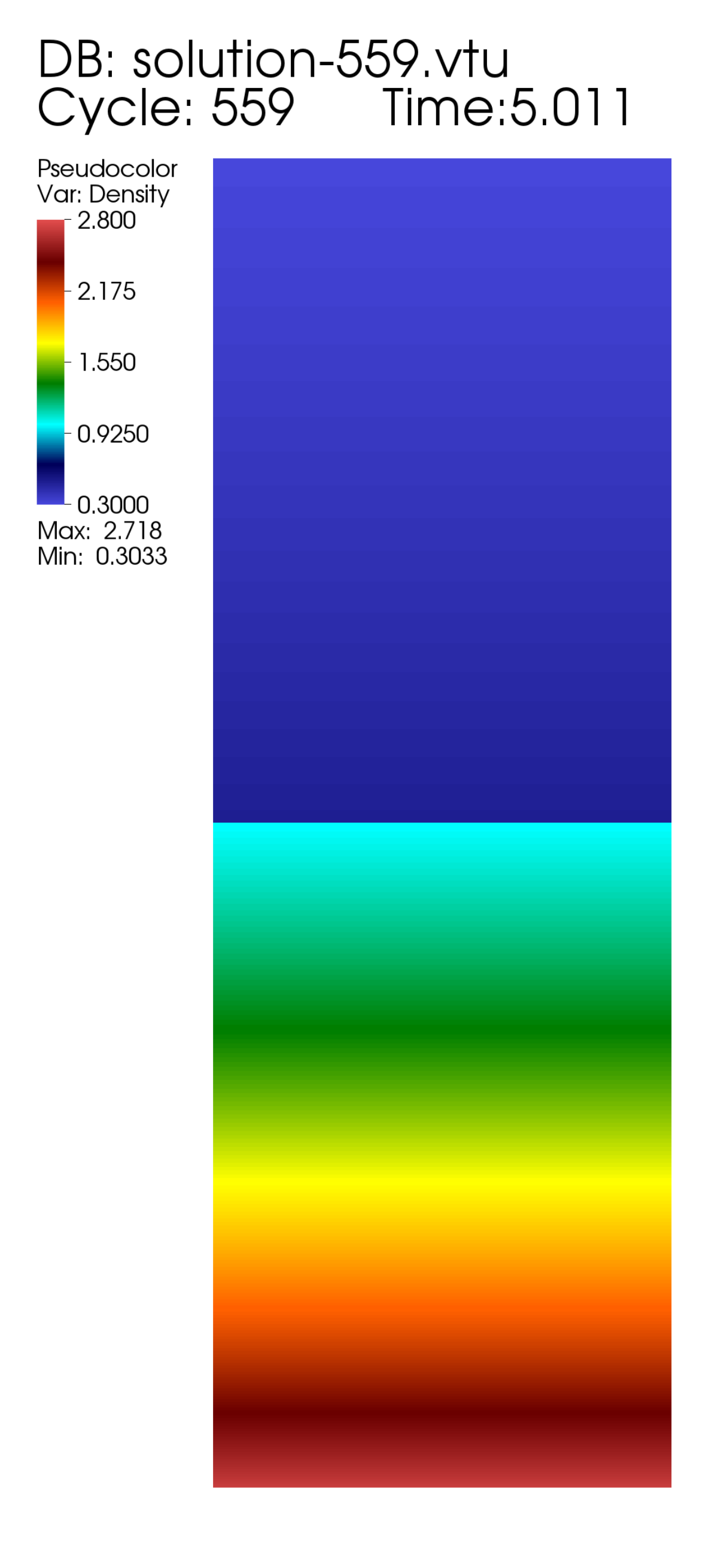} &
\includegraphics[width=0.24\textwidth]{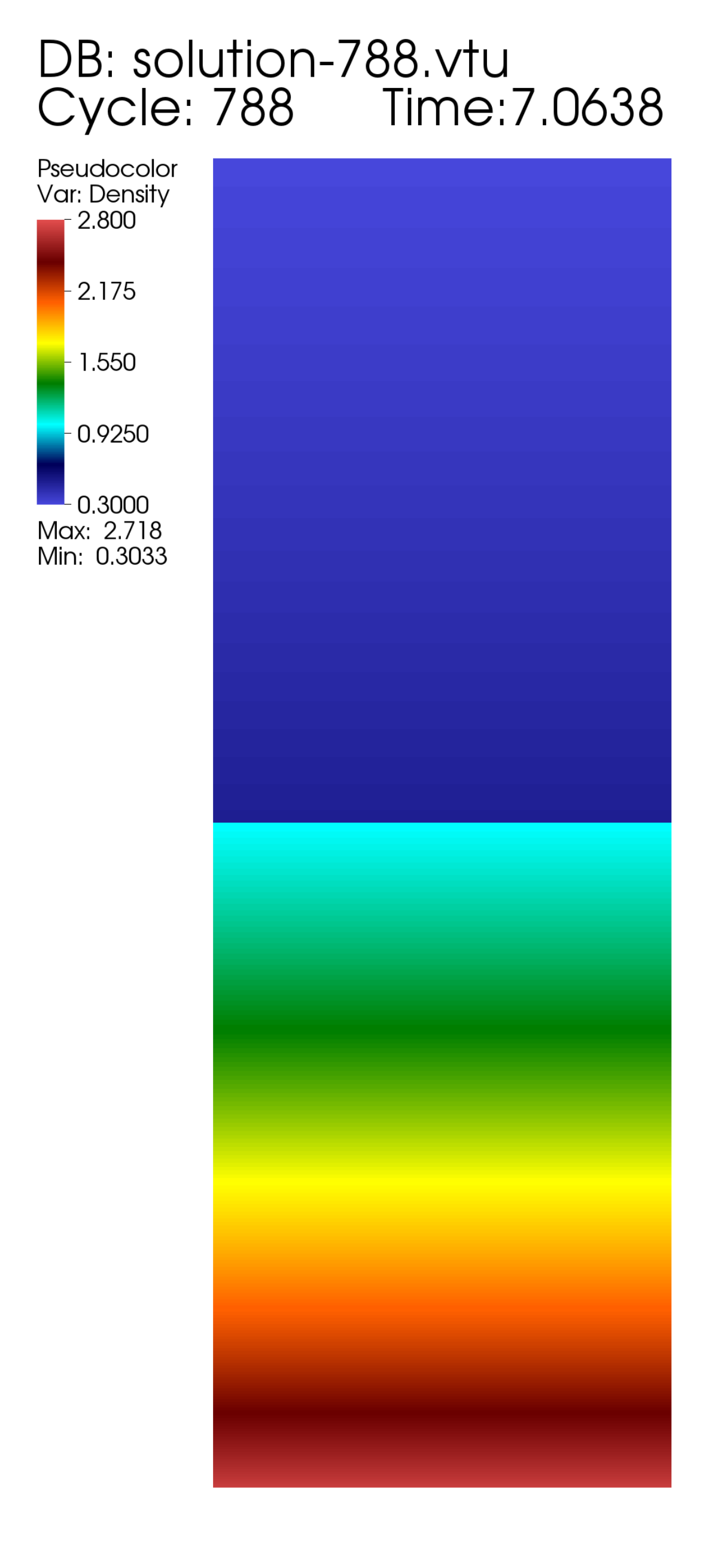} &
\includegraphics[width=0.24\textwidth]{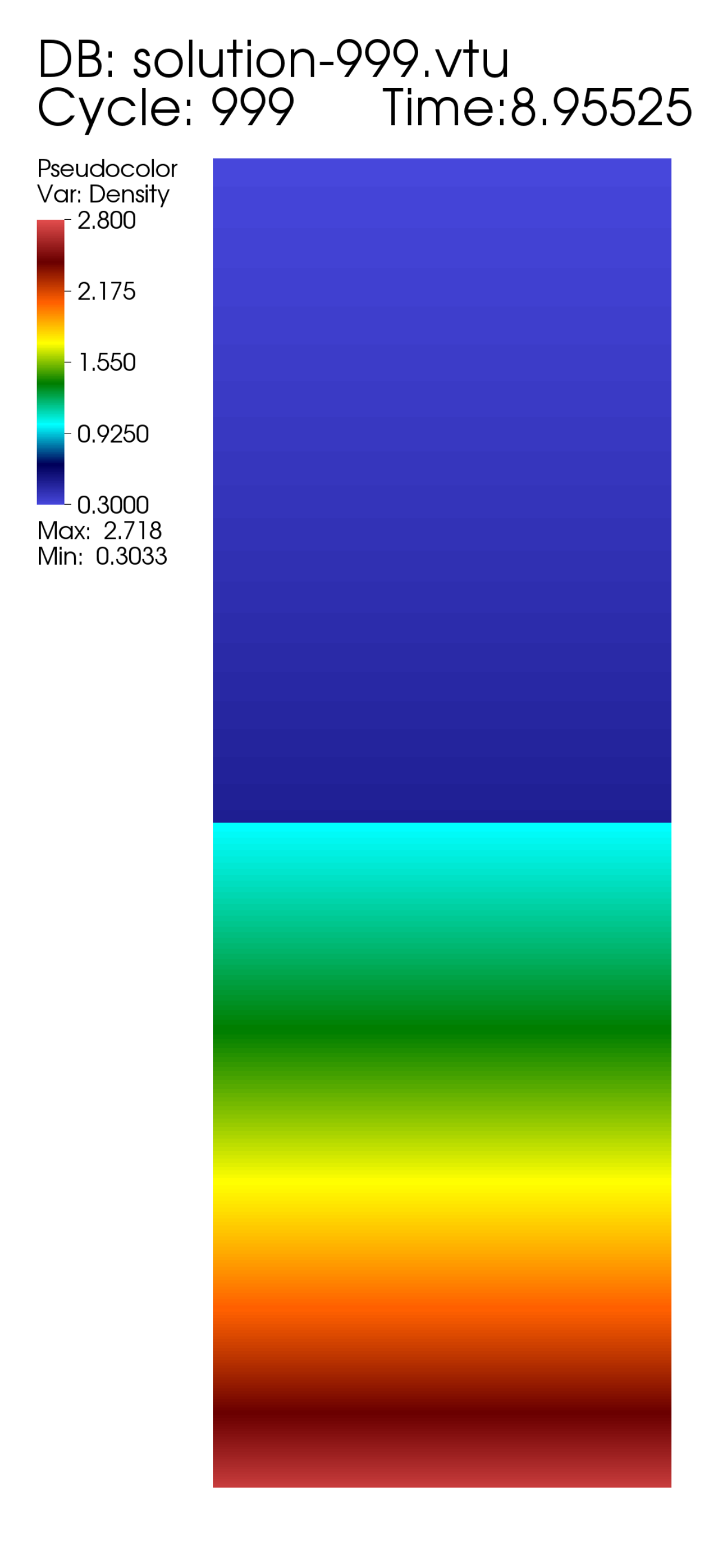}  \\
(a) & (b) & (c) & (d) \\
\includegraphics[width=0.24\textwidth]{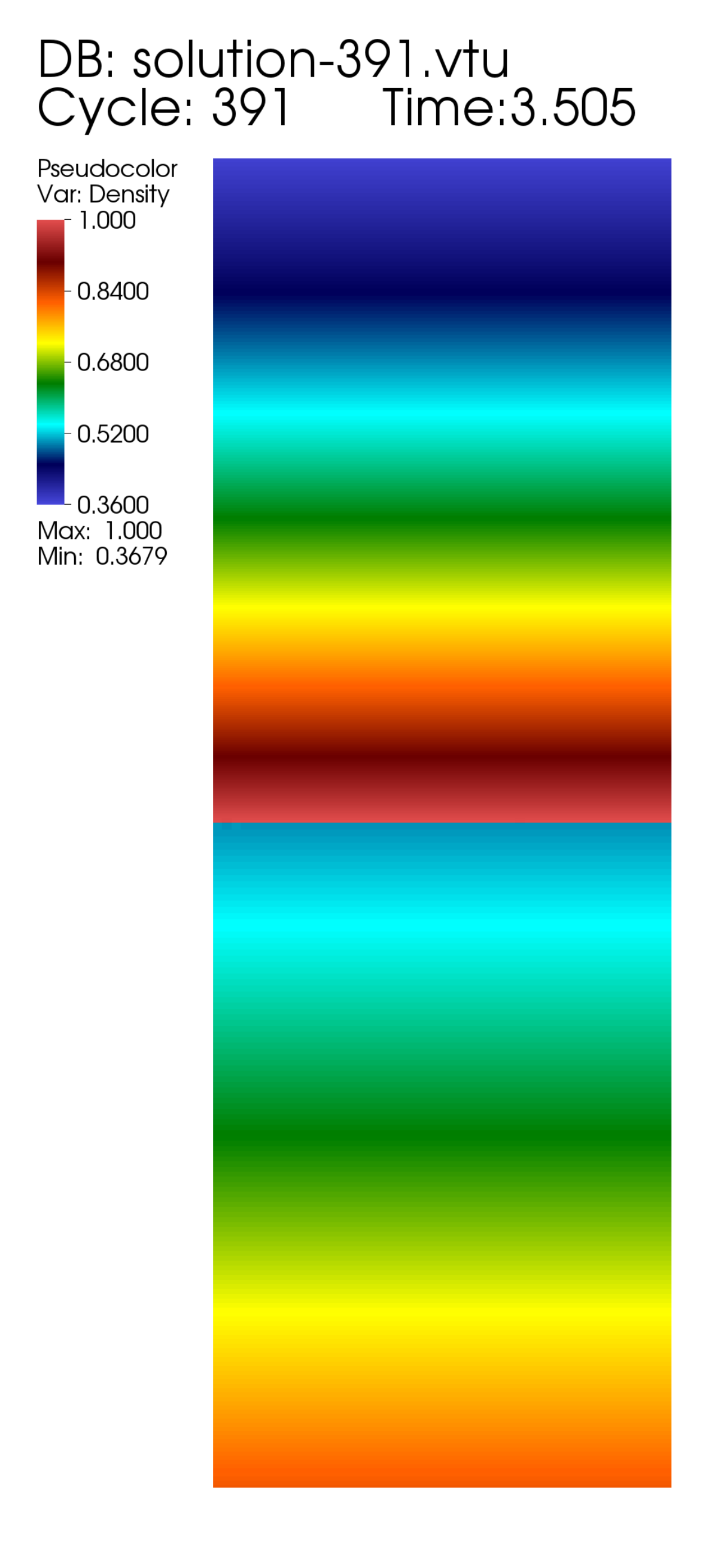} &
\includegraphics[width=0.24\textwidth]{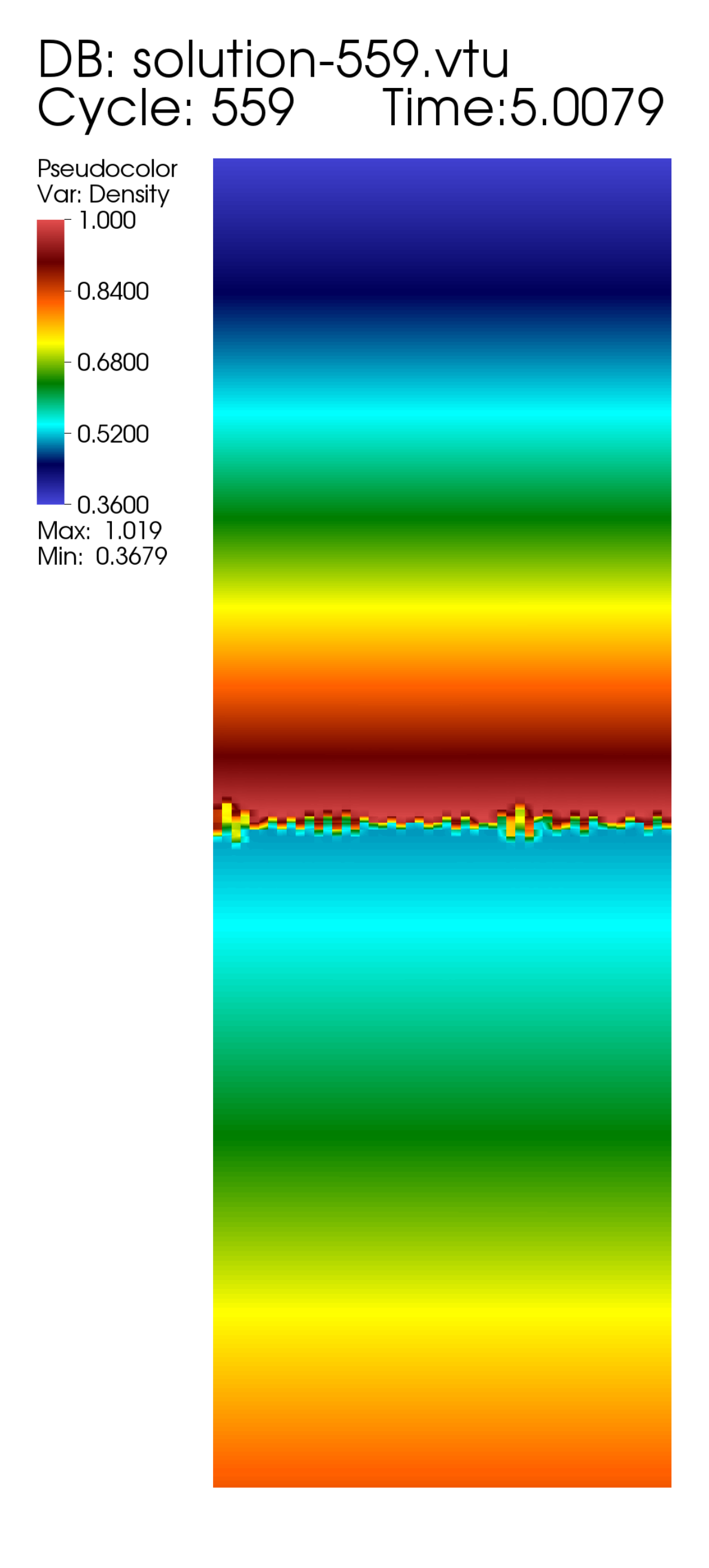} &
\includegraphics[width=0.24\textwidth]{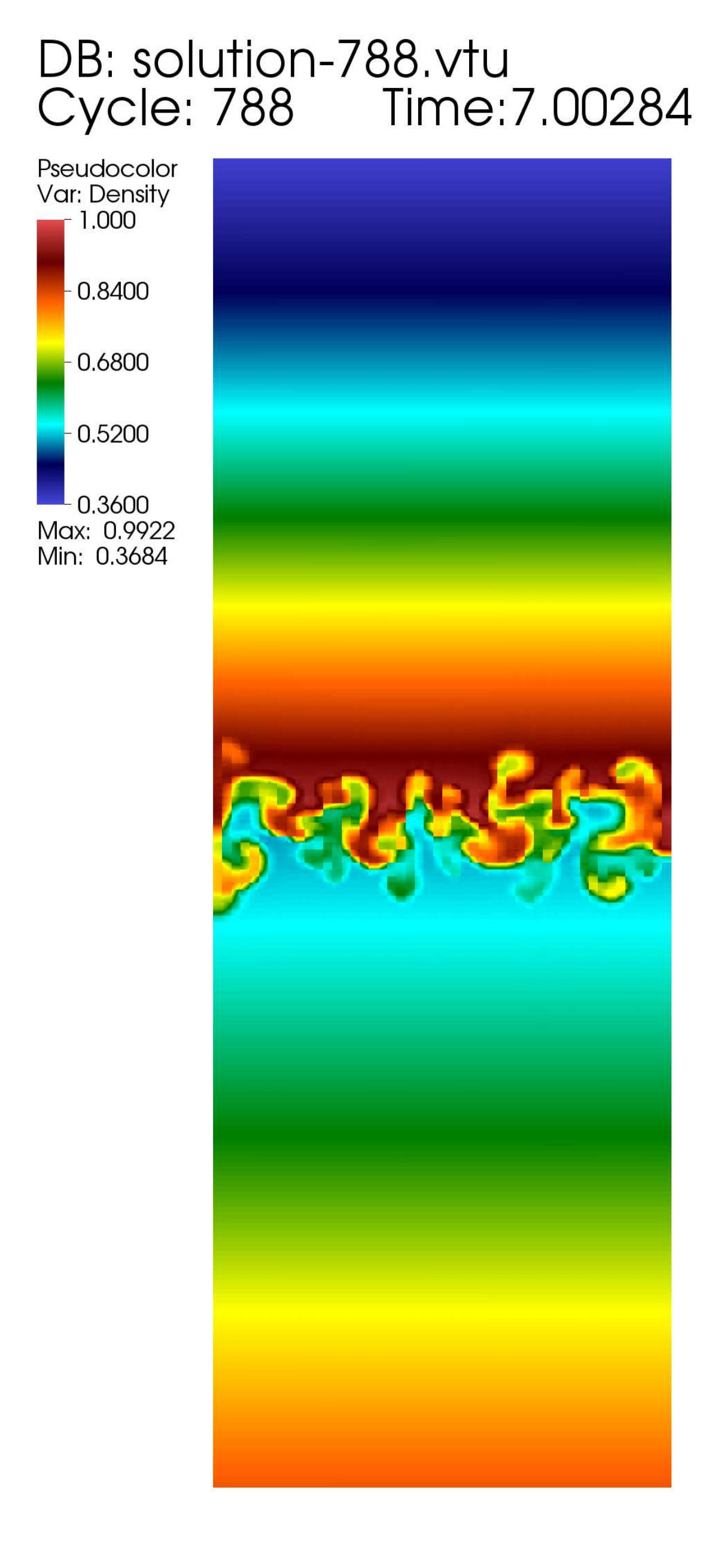} &
\includegraphics[width=0.24\textwidth]{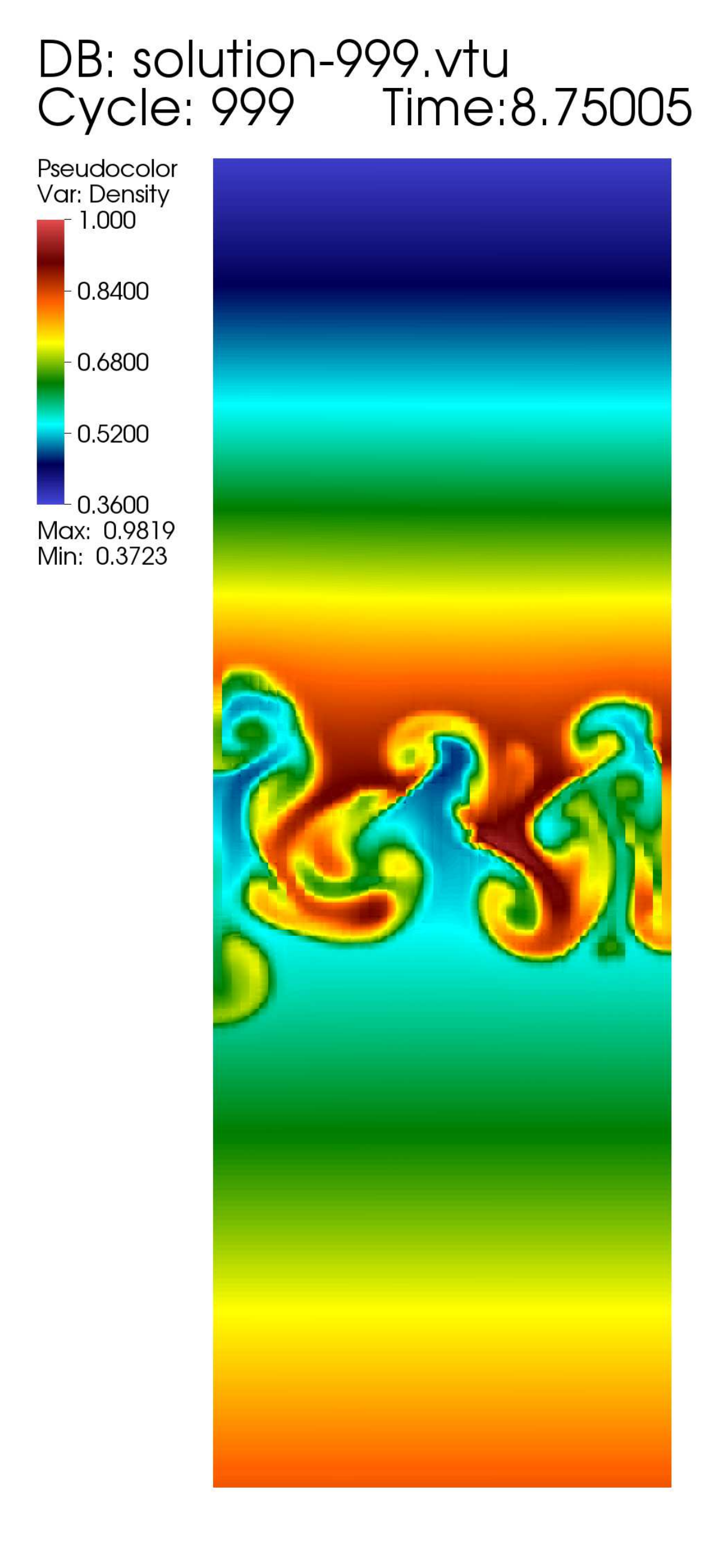} \\
(e) & (f) & (g) & (h)
\end{tabular}
\end{center}
\caption{Rayleigh-Taylor problem. (a)-(d) Case 1, (e)-(f) Case 2}
\label{fig:rtsol}
\end{figure}

The configuration in case 2 is however physically unstable since the heavier fluid is on top. The small round-off errors will eventually grow with time and lead to the Rayleigh-Taylor instability. In figure~(\ref{fig:rtsol}), we show the density at large times and observe that the solution does not change for case 1 which is a stable configuration. At sufficiently large times, the configuration in case 2 develops instabilities near the density jump but it remains well-balanced away from the interface.
%-------------------------------------------------------------------------------------
\subsection{Order of accuracy study}
This test case is taken from~\cite{Xing:2013:HOW:2434597.2434626}. An exact solution of the Euler equations with gravity given by
\[
\rho = 1 + 0.2 \sin(\pi(x+y-t(u_0+v_0))), \qquad u = u_0, \qquad v = v_0
\]
\[
p = p_0 + t(u_0 + v_0) -x - y + 0.2 \cos(\pi(x+y -t(u_0 + v_0)))/\pi
\]
is used to study the order of accuracy. The gravitational potential in this case is given by $\pot(x,y) =x+y$. For the parameters, we take $u_0 = v_0 = 1$, $p_0 = 4.5$ as in~\cite{Xing:2013:HOW:2434597.2434626}. The above initial condition is evolved in time using the well-balanced DG scheme upto a final time of $t=0.1$ units. On all parts of the boundary the numerical flux (HLLC) is used to calculate the flux by making use of the exact solution. Tables~(\ref{tab:conv1}), (\ref{tab:conv2}) show the error measured in $L^2$ norm for different mesh sizes and polynomial degrees on a Cartesian mesh. We observe that the error converges at the rate of $N+1$ for the polynomial degree $N$ in all the variables.

\begin{table}
\begin{center}
\begin{tabular}{|c|c|c|c|c|c|c|c|c|}
\hline
$1/h$ & \multicolumn{2}{|c|}{$\rho u$} & \multicolumn{2}{|c|}{$\rho v$} & \multicolumn{2}{|c|}{$\rho$} & \multicolumn{2}{|c|}{$E$} \\
\hline 
  & Error & Rate & Error & Rate & Error & Rate & Error & Rate \\
\hline
50 & 0.00134154 & -- & 0.00134154 & -- & 0.0012837 & -- & 0.00161287 & -- \\
100 & 0.000335446 & 1.99 & 0.000335446 & 1.99 & 0.00032044 & 2.00 & 0.000411141 & 1.97 \\
200 & 8.35627e-05 & 2.00 & 8.35627e-05 & 2.00 & 7.97842e-05 & 2.00 & 0.00010335 & 1.99 \\
400 & 2.08348e-05 & 2.00 & 2.08348e-05 & 2.00 & 1.98754e-05 & 2.00 & 2.58109e-05 & 2.00\\
\hline
\end{tabular}
\caption{Convergence of error for degree $N=1$}
\label{tab:conv1}
\end{center}
\end{table}

\begin{table}
\begin{center}
\begin{tabular}{|c|c|c|c|c|c|c|c|c|}
\hline
$1/h$ & \multicolumn{2}{|c|}{$\rho u$} & \multicolumn{2}{|c|}{$\rho v$} & \multicolumn{2}{|c|}{$\rho$} & \multicolumn{2}{|c|}{$E$} \\
\hline 
  & Error & Rate & Error & Rate & Error & Rate & Error & Rate \\
\hline
25  & 7.7019e-05  & -- & 7.7019e-05 & -- &  7.80868e-05 & -- & 9.32865e-05 & -- \\
50  & 9.68863e-06 & 2.99 &  9.68863e-06 & 2.99 & 9.76471e-06 & 2.99 & 1.16849e-05 & 2.99\\
100 & 1.21506e-06 & 2.99 &  1.21506e-06 & 2.99 & 1.22031e-06 & 3.00 & 1.46256e-06 & 2.99 \\
200 & 1.52134e-07 & 2.99 & 1.52134e-07 & 2.99 & 1.52503e-07 & 3.00 & 1.8247e-07 & 3.00 \\
\hline
\end{tabular}
\caption{Convergence of error for degree $N=2$}
\label{tab:conv2}
\end{center}
\end{table}
%-------------------------------------------------------------------------------------
\subsection{Radial Rayleigh-Taylor instability}
\begin{figure}
\begin{center}
\begin{tabular}{cc}
\includegraphics[width=0.48\textwidth]{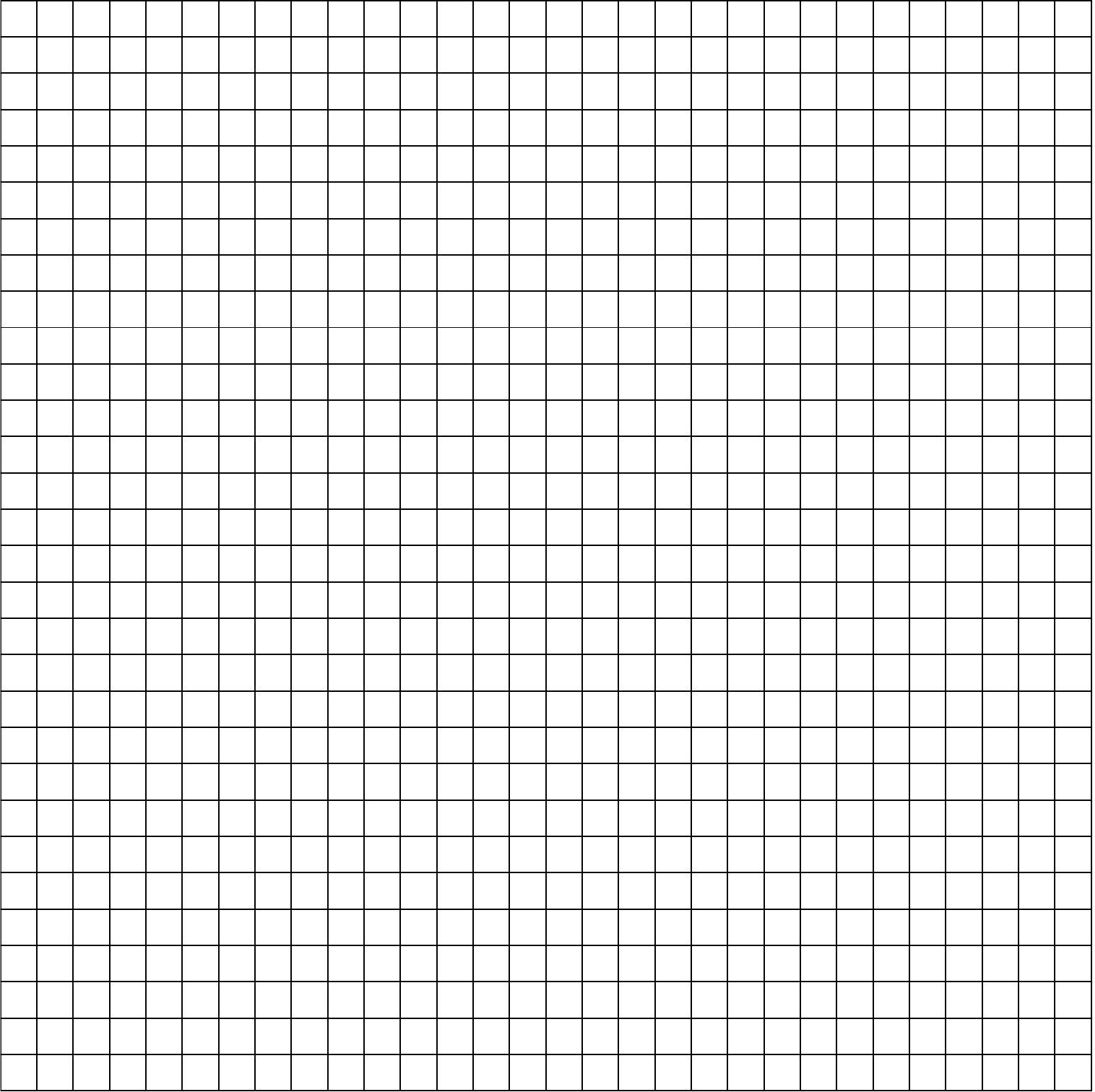} &
\includegraphics[width=0.48\textwidth]{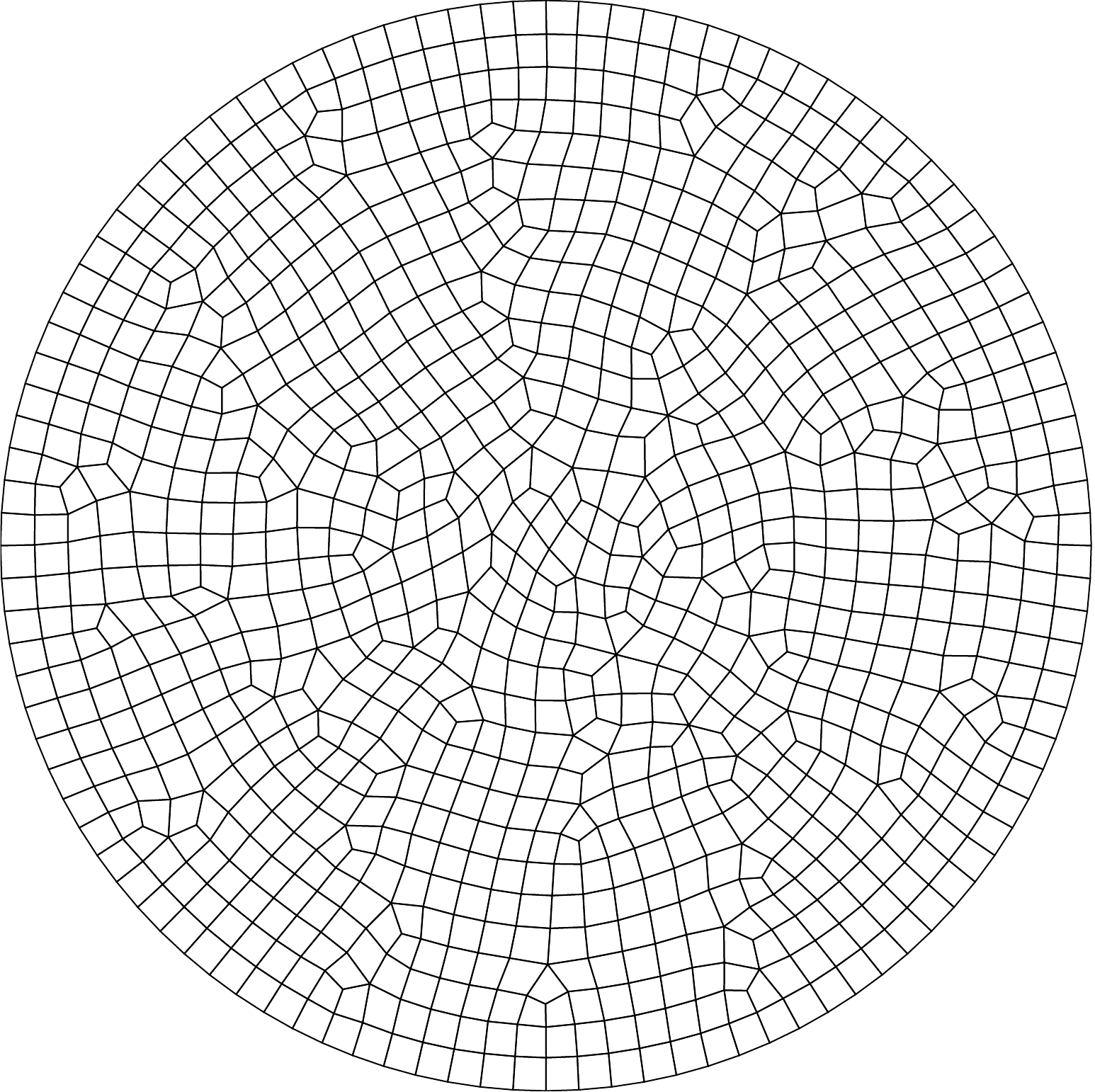}\\
(a) & (b)
\end{tabular}
\end{center}
\caption{Sample grids used for radial Rayleigh Taylor problem: (a) $30\times 30$ Cartesian cells, (b) unstructured mesh of 956 cells}
\label{fig:rrtgrid}
\end{figure}

Consider a radial potential given by $\pot = r$. The isothermal hydrostatic solution is given by
\[
\rho = p = \exp(-r)
\]
We compute the solution with above initial condition on Cartesian and unstructured meshes. The domain and mesh used are shown in figure~(\ref{fig:rrtgrid}) in case of the coarsest mesh used. Tables~(\ref{tab:rrtwbcart}), (\ref{tab:rrtwbuns}) show the error in solution at time $t=1$ on Cartesian and unstructured meshes, which shows the well-balanced property is satisfied close to machine precision.

\begin{table}
\begin{center}
\begin{tabular}{|l|c|c|c|c|}
\hline
  & $\rho u$ & $\rho v$ & $\rho$ & $E$ \\
\hline
$Q_1$, 30x30  &  6.08113e-13 & 6.06081e-13 & 2.11611e-14 & 4.13127e-14  \\
$Q_1$, 50x50  &  5.45634e-13 & 5.44973e-13 & 1.29319e-14 & 2.66376e-14  \\
$Q_1$, 100x100 & 5.4918e-13 & 5.49012e-13 & 9.63433e-15 & 2.11463e-14  \\
\hline
$Q_2$, 30x30  &  6.29776e-13 & 6.27278e-13 & 1.9777e-14 & 5.08689e-14  \\
$Q_2$, 50x50  &  5.5376e-13 & 5.52903e-13 & 1.5635e-14 & 3.85134e-14  \\
$Q_2$, 100x100 & 5.51645e-13 & 5.5142e-13 & 2.173e-14 & 5.25578e-14  \\
\hline
\end{tabular}
\caption{Well balanced test for radial Rayleigh-Taylor problem on Cartesian mesh}
\label{tab:rrtwbcart}
\end{center}
\end{table}

\begin{table}
\begin{center}
\begin{tabular}{|l|c|c|c|c|}
\hline
  & $\rho u$ & $\rho v$ & $\rho$ & $E$ \\
\hline
$Q_1$, 956   & 3.03033e-16 & 3.23738e-16 & 6.66245e-16 & 2.11449e-16  \\
$Q_1$, 2037  & 5.20653e-16 & 5.10865e-16 & 9.82565e-16 & 4.06402e-16  \\
$Q_1$, 10710 & 2.00037e-15 & 1.57078e-15 & 2.21284e-15 & 4.66515e-15  \\
\hline
$Q_2$, 956   & 2.69429e-14 & 3.31591e-14 & 4.50499e-14 & 1.61455e-13  \\
$Q_2$, 2037  & 7.68549e-14 & 1.1834e-13 & 1.01632e-13 & 3.84886e-13  \\
$Q_2$, 10710 & 2.92633e-13 & 2.44323e-13 & 2.9449e-13 & 4.10069e-13\\
\hline
\end{tabular}
\caption{Well balanced test for radial Rayleigh-Taylor problem on unstructured mesh}
\label{tab:rrtwbuns}
\end{center}
\end{table}

\begin{figure}
\begin{center}
\begin{tabular}{ccc}
\includegraphics[width=0.31\textwidth]{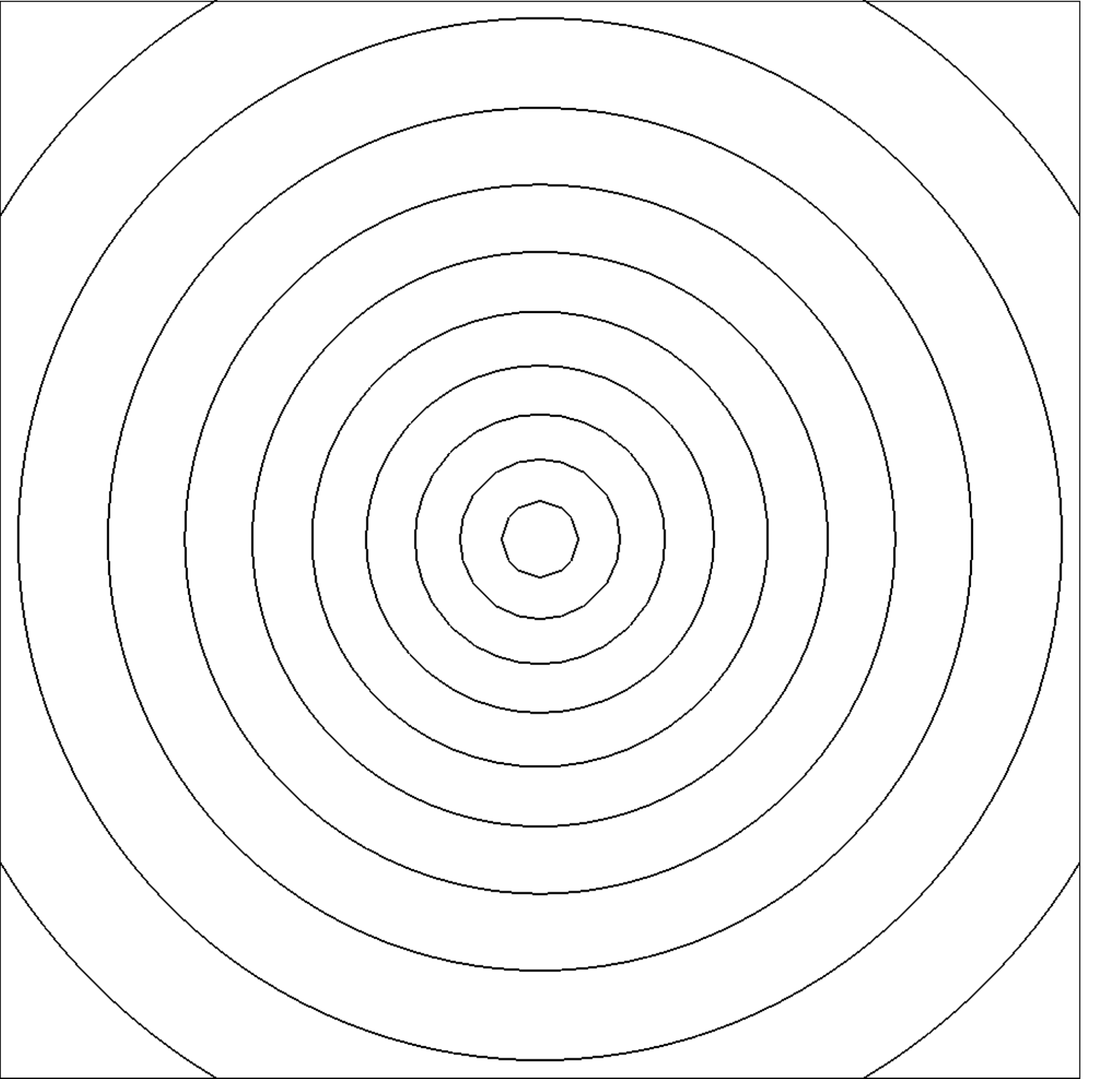} &
\includegraphics[width=0.31\textwidth]{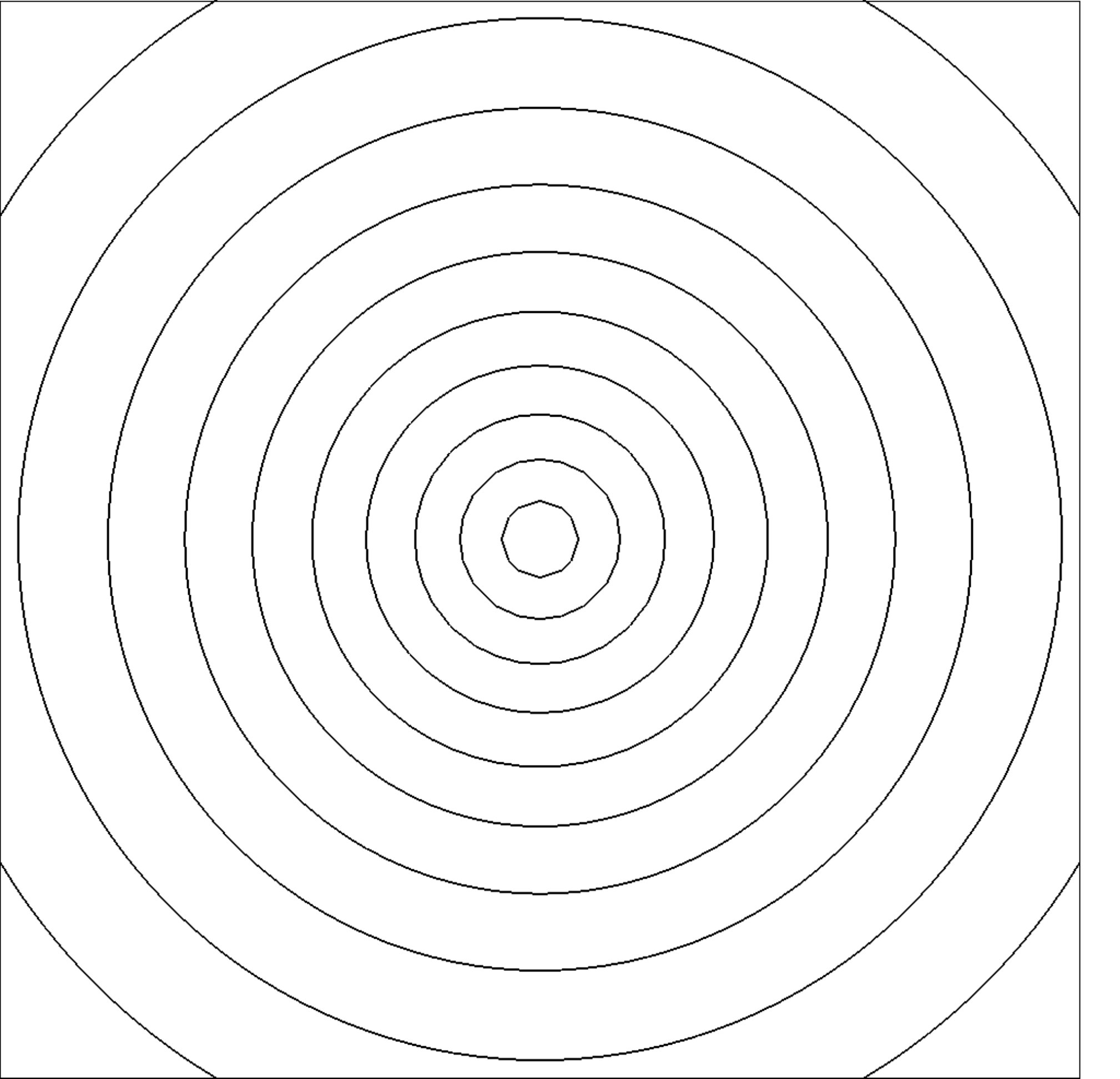} &
\includegraphics[width=0.31\textwidth]{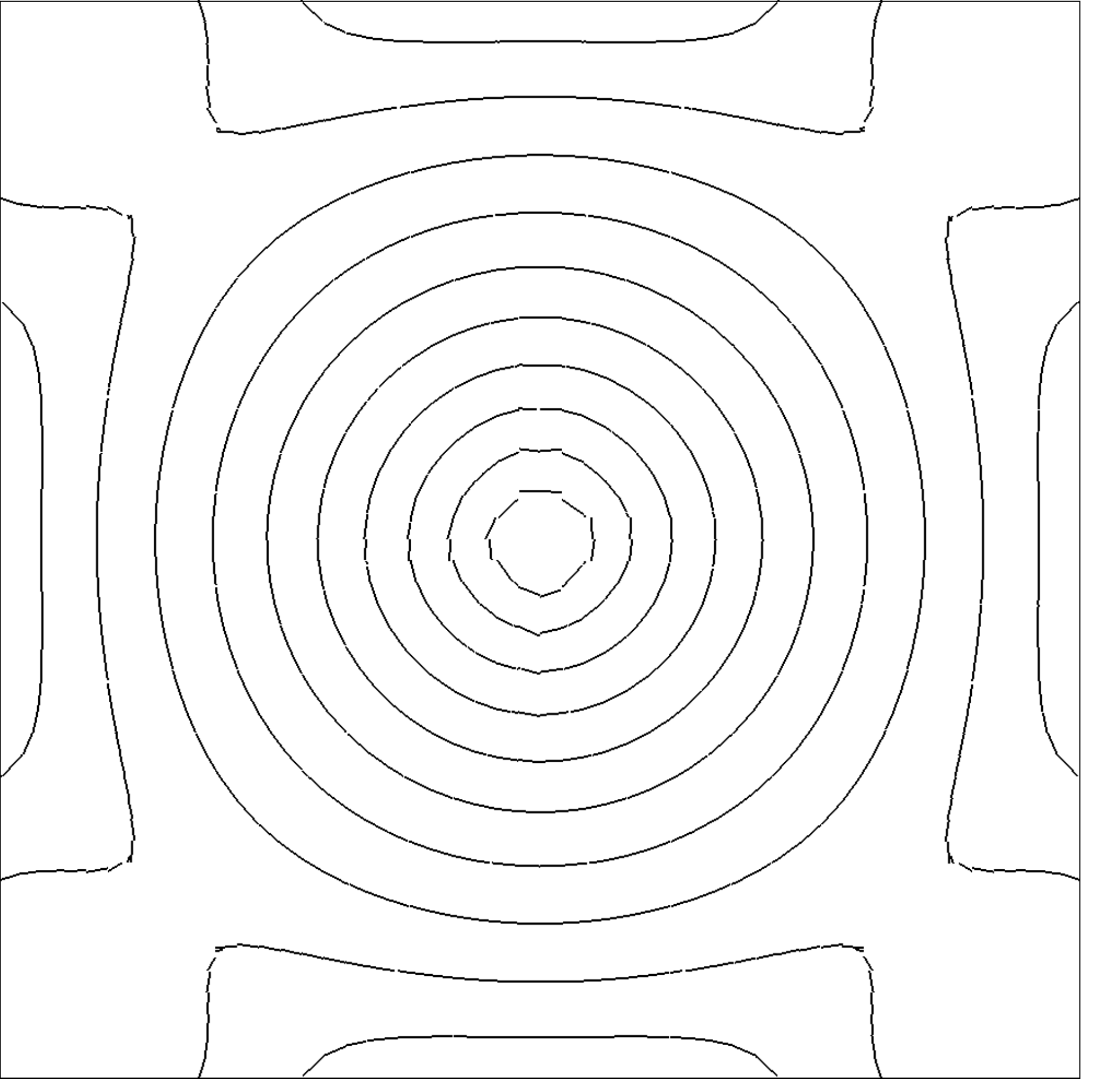} \\
(a) & (b) & (c)
\end{tabular}
\caption{Well balanced test for radial Rayleigh-Taylor problem on $50 \times 50$ mesh. (a) Initial density, (b) well-balanced scheme, density at $t=1.5$, (c) non well-balanced scheme, density at $t=1.5$}
\label{fig:rrtcompare}
\end{center}
\end{figure}

We next compute the above hydrostatic solution with a scheme which is not well balanced. The source terms in the non well-balanced scheme are computed using the exact value of the gravitational force $\nabla \pot$. On the outer boundary, which is an artificial boundary, the flux is computed from the interior solution. We compute the solution until time $t=1.5$ on a mesh of $50 \times 50$ using our well-balanced scheme and the non well-balanced scheme. The density contours plotted in figure~(\ref{fig:rrtcompare}) show that the well-balanced scheme retains the initial condition, while the non well-balanced scheme generates large errors. If we continue the computation for longer times, then the non well-balanced scheme fails at some point since the artificial boundary conditions cannot be made stable and accurate when the solution does not remain stationary near the outer boundary.

\begin{figure}
\begin{center}
\begin{tabular}{cc}
\includegraphics[width=0.48\textwidth]{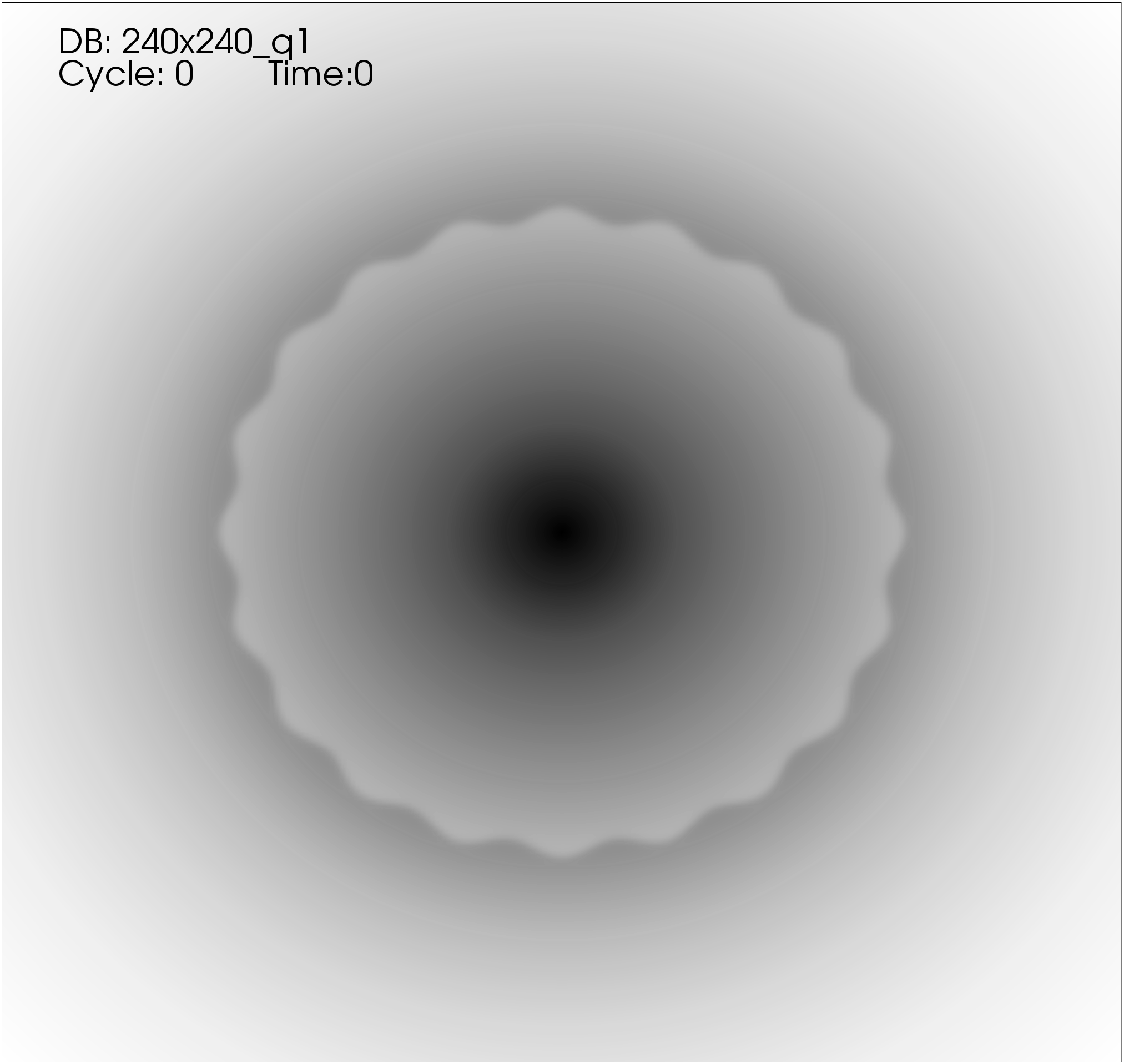} &
\includegraphics[width=0.48\textwidth]{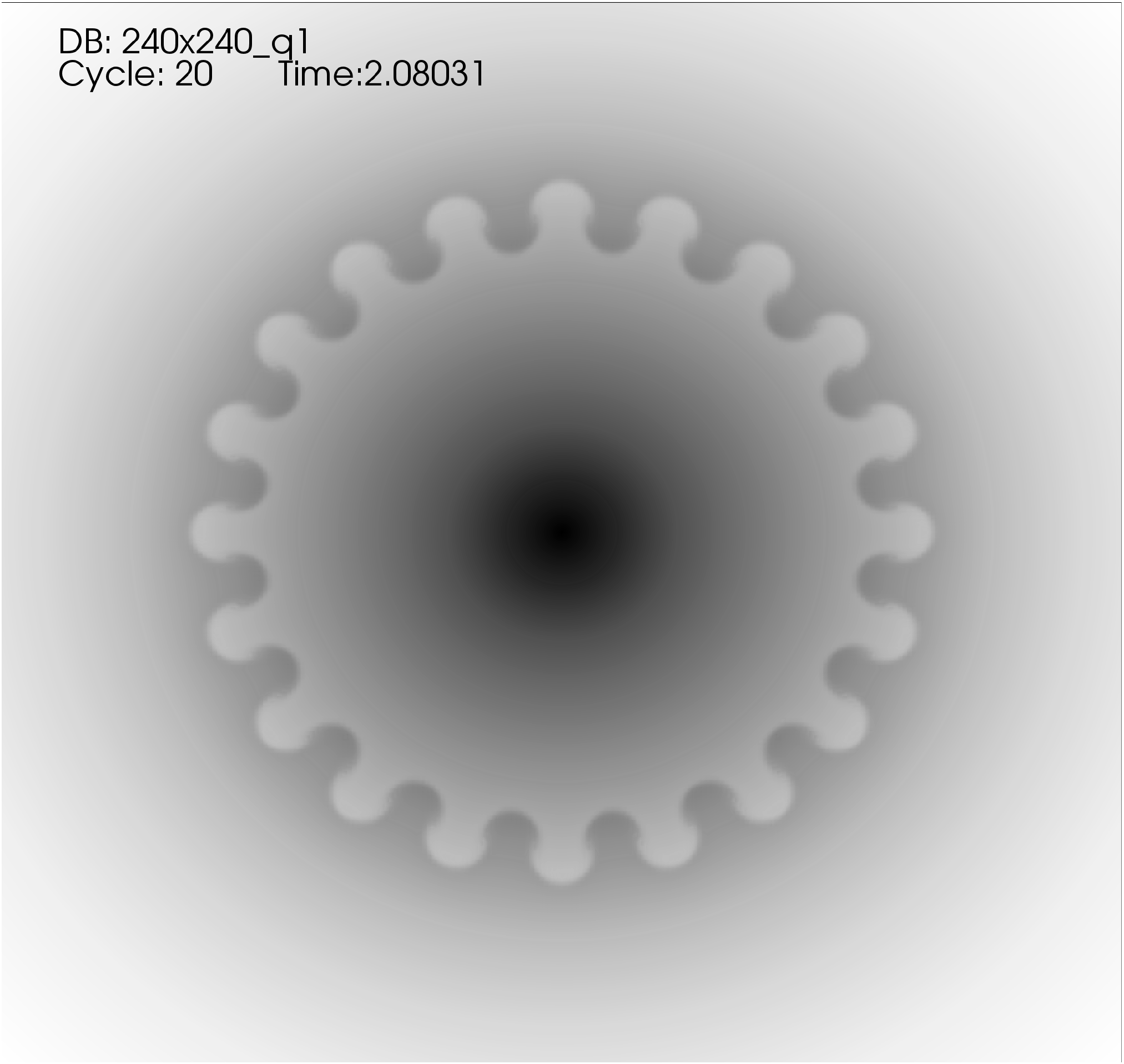} \\
(a) & (b) \\
\includegraphics[width=0.48\textwidth]{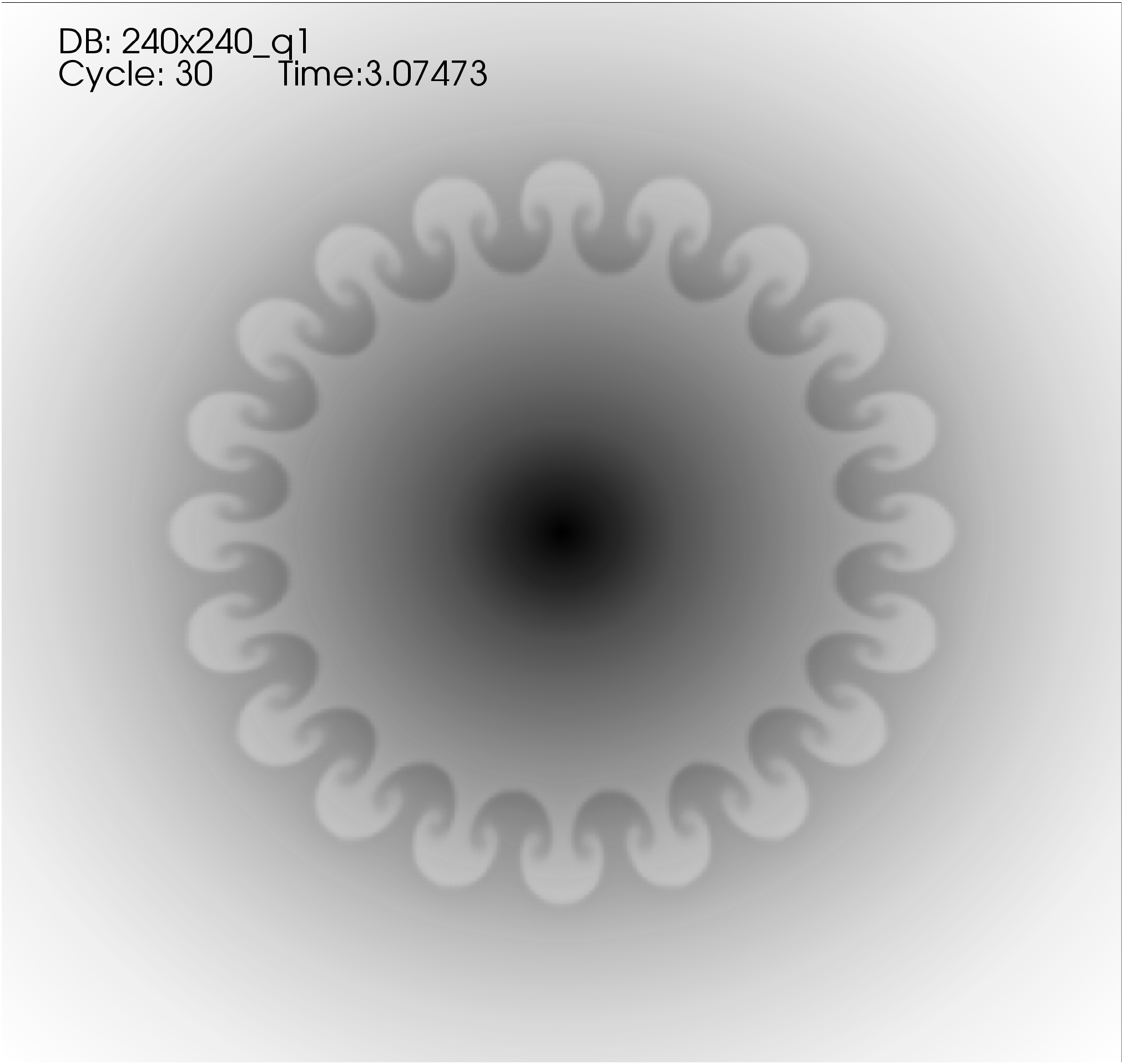} &
\includegraphics[width=0.48\textwidth]{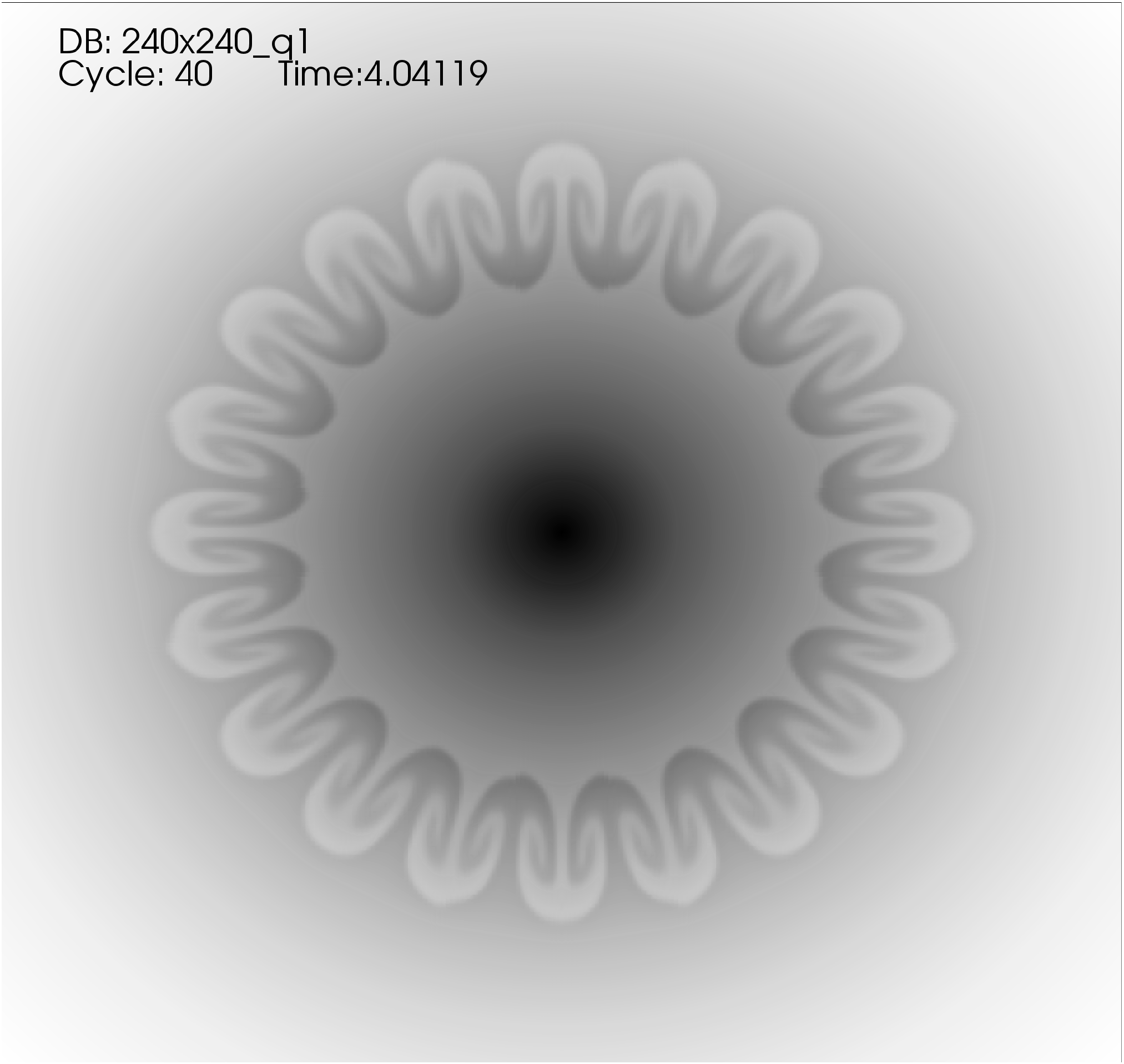} \\
(c) & (d)
\end{tabular}
\caption{Radial Rayleigh-Taylor problem on Cartesian mesh of $240 \times 240$ and $Q_1$ polynomials, showing density at different times}
\label{fig:rrt240q1}
\end{center}
\end{figure}

In the next test, we put a perturbation in density over the isothermal radial     solution. The initial pressure and density are given by
\[
p = \begin{cases}
\ee^{-r} & r \le r_0 \\
\ee^{-\frac{r}{\alpha} + r_0\frac{(1-\alpha)}{\alpha}} & r > r_0
\end{cases}, \qquad \rho = \begin{cases}
\ee^{-r} & r \le r_i(\theta) \\
\frac{1}{\alpha} \ee^{-\frac{r}{\alpha} + r_0\frac{(1-\alpha)}{\alpha}} & r >   r_i(\theta)
\end{cases}
\]
where $r_i(\theta) = r_0(1 + \eta \cos(k\theta))$ and $\alpha = \exp(-r_0)/(\exp(-r_0)+ \Delta_\rho)$. Hence the density jumps by an amount $\Delta_\rho > 0$ at the    interface defined by $r=r_i(\theta)$ whereas the pressure is continuous.                Following~\cite{levequebale}, we take $\Delta_\rho = 0.1$, $\eta=0.02$, $k=20$  and use a mesh of $240 \times 240$ cells on the domain $[-1,+1] \times [-1,+    1]$. In the regions $r < r_0(1-\eta)$ and $r > r_0(1+\eta)$ the initial         condition is in stable equilibrium but due to the discontinuous density, a Rayleigh-Taylor instability develops at the interface defined by $r=r_i(\theta)$. The evolution of the density interface with time is shown in figure~(\ref{fig:rrt240q1}). The disturbance is seen to be localized around the initial interface and the solution remains unchanged in other regions. The characteristic plume like structures are clearly visible in the pictures. This level of accuracy is not possible with a scheme that is not well-balanced and such schemes can even be unstable, as discussed above, due to the inability to maintain the hydrostatic solution near the boundaries of the domain, which is an artificial boundary arising due to truncation of the computational domain.

%-------------------------------------------------------------------------------------
\subsection{Shock tube problem}
\begin{figure}
\begin{center}
\begin{tabular}{cc}
\includegraphics[width=0.48\textwidth]{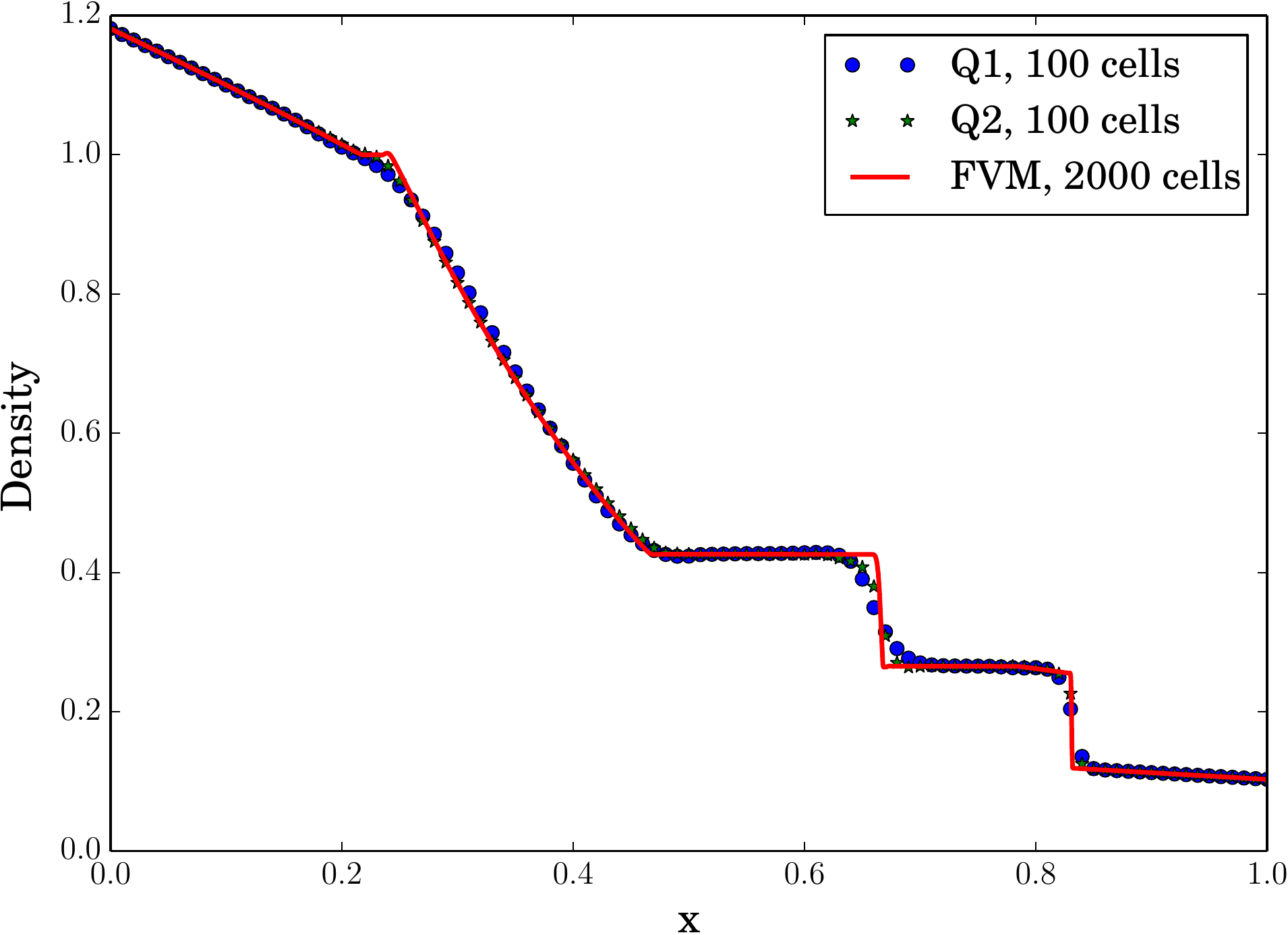} &
\includegraphics[width=0.48\textwidth]{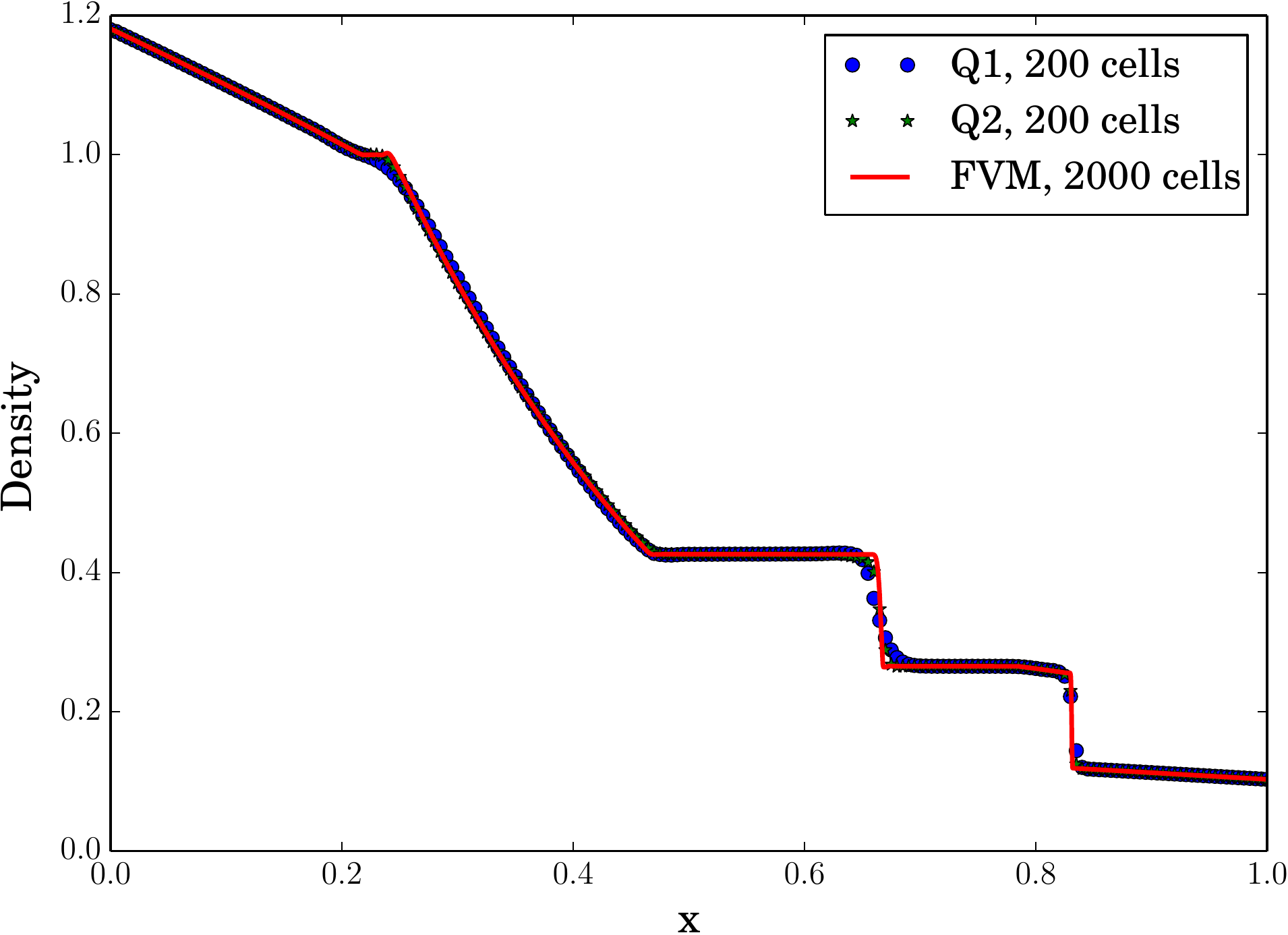} \\
(a) & (b)
\end{tabular}
\caption{Shock tube problem under gravitational field}
\label{fig:shocktube}
\end{center}
\end{figure}

We consider the standard Sod test case together with a gravitational field $\pot(x) = x$ as in~\cite{Xing:2013:HOW:2434597.2434626}. The domain is $[0,1]$ and the initial conditions are given by
\[
(\rho,u,p) = \begin{cases}
(1, 0, 1) &  x < \half \\
(0.125, 0, 0.1) & x > \half
\end{cases}
\]
together with solid wall boundary conditions. The solutions are obtained on 100 and 200 cells until a time of $t=0.2$ using the HLLC flux and are shown in figures~(\ref{fig:shocktube}). We see that the density increases near $x=0$ due to the gravitational force which is directed to the left. The coarse mesh of 100 cells is already able to resolve all the features in the solution and there are no spurious oscillations. This test indicates that the modifications in reconstruction scheme and source term are not destroying the non-oscillatory nature of the scheme.
%-------------------------------------------------------------------------------------
\section{Summary}
We have proposed a DG scheme for Euler equations with gravitational source terms which is well-balanced for isothermal and polytropic hydrostatic solutions. The scheme is based on nodal Lagrange basis functions using Gauss-Lobatto-Legendre points. The well-balanced property holds for arbitrary gravitational potentials and even on unstructured grids. Standard numerical flux functions can be used in this approach. The scheme is able to compute perturbations around the hydrostatic solution accurately even on coarse meshes. 
%-------------------------------------------------------------------------------------
\section*{Acknowledgements}
Praveen Chandrashekar thanks the Airbus Foundation Chair for Mathematics of Complex Systems at TIFR-CAM, Bangalore, for support in carrying out this work. Markus Zenk thanks the DAAD Passage to India program which supported his visit to Bangalore during which time part of this work was conducted.
%-------------------------------------------------------------------------------------

%\section*{References}
\bibliographystyle{siam}
\bibliography{bibdesk}

\begin{thebibliography}{10}

\bibitem{atkinsshu1998}
{\sc Harold~L. Atkins and Chi-Wang Shu}, {\em Quadrature-free implementation of
  discontinuous galerkin method for hyperbolic equations}, AIAA Journal, 36
  (1998), pp.~775--782.

\bibitem{BangerthHartmannKanschat2007}
{\sc W.~Bangerth, R.~Hartmann, and G.~Kanschat}, {\em {deal.II} -- a general
  purpose object oriented finite element library}, ACM Trans. Math. Softw., 33
  (2007), pp.~24/1--24/27.

\bibitem{BarJ89}
{\sc T.~J. Barth and D.~C. Jespersen}, {\em The design and analysis of upwind
  schemes on unstructured meshes}, no.~AIAA-89-0366, 1989.

\bibitem{doi:10.1137/140984373}
{\sc Praveen Chandrashekar and Christian Klingenberg}, {\em A second order
  well-balanced finite volume scheme for euler equations with gravity}, SIAM
  Journal on Scientific Computing, 37 (2015), pp.~B382--B402.

\bibitem{Cockburn:1998:RDG:287244.287254}
{\sc Bernardo Cockburn and Chi-Wang Shu}, {\em The {R}unge-{K}utta
  discontinuous {G}alerkin method for conservation laws {V}: {M}ultidimensional
  systems}, J. Comput. Phys., 141 (1998), pp.~199--224.

\bibitem{FLD:FLD1674}
{\sc A.~Ern, S.~Piperno, and K.~Djadel}, {\em A well-balanced runge--kutta
  discontinuous galerkin method for the shallow-water equations with flooding
  and drying}, International Journal for Numerical Methods in Fluids, 58
  (2008), pp.~1--25.

\bibitem{NME:NME2579}
{\sc Christophe Geuzaine and Jean-Fran{\c c}ois Remacle}, {\em Gmsh: A 3-d
  finite element mesh generator with built-in pre- and post-processing
  facilities}, International Journal for Numerical Methods in Engineering, 79
  (2009), pp.~1309--1331.

\bibitem{godlewski-raviart-2}
{\sc Edwige Godlewski and Pierre-Arnaud Raviart}, {\em Numerical Approximation
  of Hyperbolic Systems of Conservation Laws}, Springer, 1996.

\bibitem{hesthavenbook}
{\sc Jan~S. Hesthaven and Tim Warburton}, {\em Nodal Discontinuous Galerkin
  Methods: Algorithms, Analysis, and Applications}, Texts in Applied
  Mathematics, Springer, 2008.

\bibitem{Kappeli:2014:WSE:2567016.2567396}
{\sc R.~K\"{a}ppeli and S.~Mishra}, {\em Well-balanced schemes for the {E}uler
  equations with gravitation}, J. Comput. Phys., 259 (2014), pp.~199--219.

\bibitem{koprivagassner2010}
{\sc David~A. Kopriva and Gregor Gassner}, {\em On the quadrature and weak form
  choices in collocation type discontinuous galerkin spectral element methods},
  Journal of Scientific Computing, 44 (2010), pp.~136--155.

\bibitem{Leveque:2011:WPF:2004328.2004371}
{\sc Randall~J. Leveque}, {\em A well-balanced path-integral f-wave method for
  hyperbolic problems with source terms}, J. Sci. Comput., 48 (2011),
  pp.~209--226.

\bibitem{levequebale}
{\sc Randall.~J. LeVeque and Derek.~S. Bale}, {\em Wave propagation methods for
  conservation laws with source terms}, in Hyperbolic Problems: Theory,
  Numerics, Applications, Rolf Jeltsch and Michael Fey, eds., vol.~130 of
  International Series of Numerical Mathematics, Birkh{\"a}user Basel, 1999,
  pp.~609--618.

\bibitem{LiXing2015}
{\sc Gang Li and Yulong Xing}, {\em Well-balanced discontinuous galerkin
  methods for euler equations under gravitational fields}, J. Sci. Comput.,
  (2015 (in press)).

\bibitem{Roe1981357}
{\sc P.~L. Roe}, {\em Approximate {R}iemann solvers, parameter vectors, and
  difference schemes}, Journal of Computational Physics, 43 (1981), pp.~357 --
  372.

\bibitem{Shu1988439}
{\sc Chi-Wang Shu and Stanley Osher}, {\em Efficient implementation of
  essentially non-oscillatory shock-capturing schemes}, Journal of
  Computational Physics, 77 (1988), pp.~439 -- 471.

\bibitem{torobook}
{\sc Eleuterio~F. Toro}, {\em Riemann Solvers and Numerical Methods for Fluid
  Dynamics: A Practical Introduction}, Springer, 1999.

\bibitem{hllc}
{\sc Eleuterio~F. Toro, M.~Spruce, and W.~Speares}, {\em Restoration of the
  contact surface in the {HLL}-{R}iemann solver}, Shock Waves, 4 (1994),
  pp.~25--34.

\bibitem{Xing2014536}
{\sc Yulong Xing}, {\em Exactly well-balanced discontinuous galerkin methods
  for the shallow water equations with moving water equilibrium}, Journal of
  Computational Physics, 257, Part A (2014), pp.~536 -- 553.

\bibitem{Xing:2006:HOW:1140818.1140826}
{\sc Yulong Xing and Chi-Wang Shu}, {\em High order well-balanced finite volume
  weno schemes and discontinuous galerkin methods for a class of hyperbolic
  systems with source terms}, J. Comput. Phys., 214 (2006), pp.~567--598.

\bibitem{Xing:2013:HOW:2434597.2434626}
\leavevmode\vrule height 2pt depth -1.6pt width 23pt, {\em High order
  well-balanced {WENO} scheme for the gas dynamics equations under
  gravitational fields}, J. Sci. Comput., 54 (2013), pp.~645--662.

\end{thebibliography}

\end{document}